\documentclass[a4paper,11pt]{article}

\pdfoutput=1
\usepackage{jheppub}
\usepackage[utf8]{inputenc}
\usepackage{pdflscape}
\usepackage{tikz}
\usetikzlibrary{positioning}
\usetikzlibrary{arrows}
\usetikzlibrary{arrows.meta}

\usetikzlibrary{decorations.pathreplacing}
\usepackage{xcolor,colortbl}
\usepackage{mathtools}
\usepackage{ragged2e}
\newlength\ubwidth
\newcommand\parunderbrace[2]{\settowidth\ubwidth{$#1$}\underbrace{#1}_{\parbox{\ubwidth}{\scriptsize\RaggedRight#2}}}
\usetikzlibrary{calc}

\usepackage{booktabs}
\usepackage{paralist}
\usepackage[font={small,it},labelfont=bf]{caption} 
\usepackage{subcaption}

\usetikzlibrary{shapes}
\usepackage{braket}
\usepackage{MnSymbol}
\usepackage{extarrows}
\usepackage{csquotes}
\usepackage{array}
\usepackage{makecell}

\def\nodeAS#1#2{\overset{#1}{\underset{#2}{{~~~~\;~{\circ \supset^{\Lambda^2}}}}}}
\def\be{\begin{equation}}
\def\ee{\end{equation}}
\def\cb{\mathcal{M}_C}
\def\3d{$3d~\mathcal N=4$}
\def\node#1#2{\overset{#1}{\underset{#2}{\circ}}}
\def\sqnode#1#2{\overset{#1}{\underset{#2}{\scriptstyle\square}}}
\def\flav#1{\overset{\scriptstyle#1}{\overset{\square}{\scriptstyle\vert}}}
\def\squ#1{\overset{{\scriptstyle\square}\rlap{\,\,$\scriptstyle#1$}}{\scriptstyle\vert}}
\def\squl#1{\overset{\llap{$\scriptstyle#1$\,\,}{\scriptstyle\square}}{\scriptstyle\vert}}
\def\upnode#1{\overset{\scriptstyle#1}{\overset{\displaystyle\circ}{\scriptstyle\vert}}}
\def\topnodeDu#1{\overset{\scriptstyle#1}{\overset{\displaystyle\circ}{\scriptstyle\Uparrow}}}
\def\topnode#1#2{\overset{#1}{\overset{{\displaystyle\circ}\rlap{\,\,$\scriptstyle#2$}}{\scriptstyle\vert}}}
\def\wver#1#2{\overset{{\llap{$\scriptstyle#1$}\displaystyle{\square}{\rlap{$\scriptstyle#2$}}}}{\scriptstyle\vert}}
\def\fd#1#2{\overset{{\llap{$\scriptstyle#1$}\displaystyle{\square}{\rlap{$\scriptstyle#2$}}}}{\scriptstyle\downarrow}}
\def\fu#1#2{\overset{{\llap{$\scriptstyle#1$}\displaystyle{\square}{\rlap{$\scriptstyle#2$}}}}{\scriptstyle\uparrow}}
\def\fud#1#2{\overset{{\llap{$\scriptstyle#1$}\displaystyle{\square}{\rlap{$\scriptstyle#2$}}}}{\scriptstyle\updownarrows}}
\def\fdu#1#2{\overset{{\llap{$\scriptstyle#1$}\displaystyle{\square}{\rlap{$\scriptstyle#2$}}}}{\scriptstyle\downuparrows}}
\def\rarr{{\scriptstyle\rightarrow}}
\def\larr{{\scriptstyle\leftarrow}}
\def\lrarr{{\scriptstyle\leftrightarrows}}
\def\rlarr{{\scriptstyle\rightleftarrows}}

\def\nodesq#1#2{\overset{#1}{\underset{#2}{\boxcircle}}}
\usepackage{stmaryrd}

\usepackage{oubraces}
\usetikzlibrary{shapes.geometric}
\usetikzlibrary{shapes.misc}
\tikzset{gauge1/.style={draw=none,minimum size=0.4cm,fill=white,circle, draw}}
\tikzset{bluegauge/.style={draw=none,minimum size=0.4cm,fill=blue,circle, draw}}
\tikzset{redgauge/.style={draw=none,minimum size=0.4cm,fill=red,circle, draw}}
\tikzset{greenguage/.style={draw=none,minimum size=0.4cm,fill=green,circle, draw}}
\tikzset{cross/.style={path picture={ 
  \draw[black]
(path picture bounding box.south east) -- (path picture bounding box.north west) (path picture bounding box.south west) -- (path picture bounding box.north east);
}}}
\tikzset{cross2/.style={path picture={ 
  \draw[blue]
(path picture bounding box.south east) -- (path picture bounding box.north west) (path picture bounding box.south west) -- (path picture bounding box.north east);
}}}
\tikzset{cross3/.style={path picture={ 
  \draw[red]
(path picture bounding box.south east) -- (path picture bounding box.north west) (path picture bounding box.south west) -- (path picture bounding box.north east);
}}}
\tikzset{cross4/.style={path picture={ 
  \draw[magenta]
(path picture bounding box.south east) -- (path picture bounding box.north west) (path picture bounding box.south west) -- (path picture bounding box.north east);
}}}
\tikzset{orangegauge/.style={draw=none,minimum size=0.4cm,fill=orange,circle, draw}}
\tikzset{bluey/.style={draw=none,minimum size=0.4cm,fill=blue,circle, draw}}
\tikzset{magentagauge/.style={draw=none,minimum size=0.4cm,fill=magenta,circle, draw}}
\tikzset{cyangauge/.style={draw=none,minimum size=0.4cm,fill=cyan,circle, draw}}
\tikzset{purplee/.style={draw=violet,minimum size=0.4cm,fill=white,circle, draw=violet}}
\tikzset{dotsize/.style={draw=none,minimum size=0.6pt,fill=black,circle,inner sep=1pt, draw}}
\tikzset{gauge1/.style={draw=none,minimum size=0.4cm,fill=white,circle, draw}}
\tikzset{mini/.style={draw=none,minimum size=1pt,fill=white,circle,inner sep=3pt, draw}}
\tikzset{miniG/.style={draw=none,minimum size=1pt,fill=black,circle,inner sep=3pt, draw}}
\tikzset{cyangauge/.style={draw=none,minimum size=0.4cm,fill=cyan,circle, draw}}
\tikzset{blackgauge/.style={draw=none,minimum size=0.4cm,fill=black,circle, draw}}
\tikzset{flavour1/.style={draw=none,minimum size=0.5cm,fill=white, regular polygon,regular polygon sides=4,draw}}
\tikzset{purpleesquare/.style={draw=violet,minimum size=0.5cm,fill=white, regular polygon,regular polygon sides=4,draw}}
\tikzset{none/.style={draw=none}}
\tikzset{new edge style 1/.style={dashed}}
\tikzset{brace1/.style={decorate,decoration={brace,amplitude=5pt,mirror}}}
\tikzset{bluee/.style={line width=0.3mm,blue}}
\tikzset{greyline/.style={line width=0.5mm,gray}}
\tikzset{purpleline/.style={line width=0.3mm,violet}}
\tikzset{blueline/.style={line width=0.5mm,blue}}
\tikzset{orangee/.style={line width=0.3mm,orange}}
\tikzset{magentae/.style={line width=0.3mm,magenta}}
\tikzset{rede/.style={line width=0.3mm,red}}
\tikzset{greene/.style={line width=0.3mm,green}}
\tikzset{blackk/.style={line width=0.7mm,black}}
\tikzset{cyane/.style={line width=0.3mm,cyan}}
\tikzset{brace2/.style={decorate,decoration={brace,amplitude=5pt}}}
\tikzset{biggauge/.style={draw=none,minimum size=0.6cm,fill=white,circle, draw}}
\tikzset{largeflavor/.style={draw=none,minimum size=0.8cm,fill=white, regular polygon,regular polygon sides=4,draw}}
\pgfdeclarelayer{edgelayer}
\pgfdeclarelayer{nodelayer}
\usetikzlibrary{decorations.pathreplacing}
\pgfsetlayers{edgelayer,nodelayer,main}

\title{\boldmath The Higgs Mechanism -- Hasse Diagrams for Symplectic Singularities}

\author[\daleth]{Antoine Bourget}
\author[\daleth]{, Santiago Cabrera}
\author[\daleth]{, Julius F.\ Grimminger}
\author[\daleth,1]{, Amihay Hanany \note{ Imperial/TP/19/AH/02}}
\author[\beth]{, Marcus Sperling}
\author[\daleth]{, Anton Zajac}
\author[\daleth]{, and Zhenghao Zhong}
\affiliation[\daleth]{Theoretical Physics Group, The Blackett Laboratory, Imperial College London, Prince Consort Road
London, SW7 2AZ, UK}
\affiliation[\beth]{Yau Mathematical Sciences Center, Tsinghua University, Haidian District, Beijing, 100084, China}
\emailAdd{a.bourget@imperial.ac.uk}
\emailAdd{santiago.cabrera13@ic.ac.uk}
\emailAdd{julius.grimminger17@imperial.ac.uk}
\emailAdd{a.hanany@imperial.ac.uk}
\emailAdd{marcus.sperling@univie.ac.at}
\emailAdd{anton.zajac@imperial.ac.uk}
\emailAdd{zhenghao.zhong14@imperial.ac.uk}

\abstract{We explore the geometrical structure of Higgs branches of quantum field theories with 8 supercharges in  3, 4, 5 and 6 dimensions. They are symplectic singularities, and as such admit a decomposition (or \emph{foliation}) into so-called \emph{symplectic leaves}, which are related to each other by \emph{transverse slices}. We identify this foliation with the pattern of partial Higgs mechanism of the theory and, using brane systems and recently introduced notions of magnetic quivers and quiver subtraction, we formalise the rules to obtain the \emph{Hasse diagram} which encodes the structure of the foliation. While the unbroken gauge symmetry and the number of flat directions are obtainable by classical field theory analysis for Lagrangian theories, our approach allows us to characterise the geometry of the Higgs branch by a Hasse diagram with symplectic leaves and transverse slices, thus refining the analysis and extending it to non-Lagrangian theories. Most of the Hasse diagrams we obtain extend beyond the cases of nilpotent orbit closures known in the mathematics literature. The geometric analysis developed in this paper is applied to Higgs branches of several Lagrangian gauge theories, Argyres-Douglas theories, five dimensional SQCD theories at the conformal fixed point, and six dimensional SCFTs.}

\begin{document}

\maketitle

\section{Introduction}

\begin{figure}[h]
	\centering
	\begin{tabular}{c}
\begin{tikzpicture}[node distance=20pt]
	\tikzstyle{hasse} = [circle, fill,inner sep=2pt];
		\node at (-0.7,-0.5) [] (1a) [] {};
		\node [hasse] at (0,-0.5) [] (1b) [label=above:\footnotesize{$\{x,y\}$}] {};
		\node at (0.7,-0.5) [] (1c) [] {};
		\node [hasse] (2a) [label=left:\footnotesize{$\{x\}$},below of=1a] {};
		\node [] (2b) [below of=1b] {};
		\node [hasse] (2c) [label=right:\footnotesize{$\{y\}$},below of=1c] {};
		\node [] (3a) [below of=2a] {};
		\node [hasse]  (3b) [label=below:\footnotesize{$\{\}$},below of=2b] {};
		\node (3c) [below of=2c] {};
		\node (4a) [below of=3a] {};
		\draw (1b) edge [] node[] {} (2a)
		(1b) edge [] node[] {} (2c)
		(2a) edge [] node[] {} (3b)
		(2c) edge [] node[] {} (3b);
		\end{tikzpicture}
	\end{tabular}
	\caption{Hasse diagram encoding the poset structure of inclusions of the subsets of $\{x,y\}$.}\label{tab:hassesimpleexample}
\end{figure}
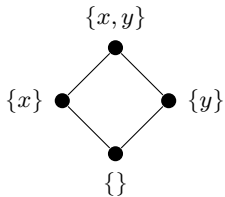

The Englert-Brout-Higgs-Guralnik-Hagen-Kibble mechanism for the Abelian case \cite{englert1964broken,higgs1964broken,guralnik1964global} and for the non-Abelian case \cite{kibble1967symmetry}, called \emph{Higgs mechanism} for short, by which a gauge group is broken to a subgroup by the vacuum expectation value (VEV) of scalar fields, has played a central role in theoretical physics in the last few decades. In the context of supersymmetric gauge theories with $8$ supercharges, the part of the moduli space called the Higgs branch refers to this old idea since it is parametrised by the VEVs of the scalar components which trigger this mechanism. For definiteness, let us adopt the language of 4d $\mathcal{N}=2$ theories -- even though we also consider 3d $\mathcal{N}=4$, 5d $\mathcal{N}=1$ and 6d $\mathcal{N}=(1,0)$ theories in this paper. 
The Higgs branch is parametrised by the scalars in the hypermultiplets. Under the assumption that there is enough matter in the theory, a generic VEV completely breaks the gauge group. However, on specific loci of positive codimension in the Higgs branch, the VEVs can leave a certain subgroup unbroken. Iterating this process, we see an interesting pattern of partial Higgsing, which can be characterised by the various subspaces of the Higgs branch. These subspaces are naturally partially ordered by inclusion of their closures, and as such can be arranged into a \emph{Hasse diagram}. A Hasse diagram simply represents a finite partially ordered set, in the form of a drawing of its transitive reduction. For illustration, Figure \ref{tab:hassesimpleexample} shows the Hasse diagram for the poset structure of a set $\{x,y\}$. 

Let us now consider the Higgs branch
from a purely geometric point of view. Supersymmetry implies that the Higgs branch is a hyper-K\"ahler cone \cite{Hitchin:1986ea}, or equivalently, a symplectic singularity \cite{beauville2000symplectic}. Symplectic spaces admit a natural foliation induced by the symplectic form, as recalled in Appendix \ref{SymplecticLeaves}. Concretely, this means that the Higgs branch can be partitioned into leaves which are partially ordered by inclusion of their closures, thus forming a Hasse diagram. This Hasse diagram reflects the symplectic structure (or more precisely the Poisson structure) of the Higgs branch. In addition, two adjacent leaves in the Hasse diagram define a transverse slice, which we call an \emph{elementary} Slodowy slice (or \emph{elementary slice} for short), whose dimension is equal to the codimension of the smaller leaf in the closure of the larger one \cite{Brieskorn:1970,Slodowy:1980,kraft1980minimal,Kraft1982,fu2017generic}. A transverse slice is a symplectic singularity as well. This means that the smaller leaf defines a set of points inside the closure of the larger leaf that are singular. The type of the singularity is precisely the same as the singularity found at the origin of the transverse slice. Therefore, the physical phenomenon of partial Higgsing is directly related to the mathematical structure of the moduli space, in particular, to the geometry of its singular points.

Before turning to the case of general symplectic singularities, let us mention a situation which may be familiar to the reader: closures of nilpotent orbits. These are prominent examples of symplectic singularities, and the symplectic foliation structure and the elementary slices have been worked out in the early 80s by Kraft and Procesi \cite{kraft1980minimal,Kraft1982}. 
Under the assumption that the closures of the slices are normal, any elementary slice belongs to one of the following two families: 
\begin{compactenum}[({S}1)]
    \item ADE singularities $\mathbb{C}^2/\Gamma$, where $\Gamma$ is a finite subgroup of $\mathrm{SU}(2)$, or
    \item Closures of minimal nilpotent orbits of simple Lie algebras. 
\end{compactenum}
(If the normality assumption is relaxed, more slices are allowed \cite{fu2017generic}.)
Therefore, the Hasse diagrams for nilpotent orbit closures are known.
The results of Kraft and Procesi have been reproduced in the physics of 3d $\mathcal{N}=4$ theories and their brane realisation in Type IIB string theory in \cite{Cabrera:2016vvv,Cabrera:2017njm}. Although nilpotent orbits play an important role in the present work, the Higgs branches considered here belong to a more general class. Approaching the general case means facing two challenges:
\begin{compactenum}[(i)]
    \item What are the Hasse diagrams for more general moduli spaces?
    \item How to compute these Hasse diagrams?
\end{compactenum}

To partly answer the first question, the following observation is crucial. Consider only theories whose Higgs branches are nilpotent orbit closures, hence the Hasse diagram is known. Remarkably, this \emph{Hasse diagram is reproduced by the pattern of partial Higgsing}, as we demonstrate below.
Therefore, one can elevate the partial Higgsing pattern to a guiding principle on how the Hasse diagram for the Higgs branch as symplectic singularity should look like in the case of a theories whose Higgs branch is no longer just a nilpotent orbit closure. In other words, we conjecture that \emph{partial Higgsing predicts the Hasse diagram of a classical Higgs branch} for a Lagrangian theory.

Next, the answer to the second question requires a recollection of some recent developments: 
Following the brane realisation \cite{Cabrera:2016vvv,Cabrera:2017njm} of the Kraft-Procesi transitions and understanding of the Hasse diagram in terms of Higgs and Coulomb branches for $3$d $\mathcal{N}=4$, the notion of \emph{quiver subtraction} has been explicitly introduced in \cite{Cabrera:2018ann}. Consequently, a transition of the type found by Kraft and Procesi is nothing but a subtraction of a quiver whose Coulomb branch is a minimal nilpotent orbit closure or a Kleinian singularity. In fact, precisely this subtraction process underlies the brane realisation \cite{Cabrera:2016vvv,Cabrera:2017njm}. However, in order to apply quiver subtraction, one requires that the Higgs branch in question has a realisation as a space of dressed monopole operators. This is realised in dimensions $3-6$ by computation of a \emph{magnetic quiver}, a concept already used in \cite{DelZotto:2014kka,Cremonesi:2015lsa,Ferlito:2017xdq} and explained from a brane perspective in $5$d and $6$d \cite{Cabrera:2018jxt,Cabrera:2019izd}, where it got its name. In 3 dimensions the magnetic quiver may be provided by 3d mirror symmetry. A mathematically rigorous construction of this space of dressed monopole operators is now available \cite{Nakajima:2015txa,Braverman:2016wma}.

Now, one can start computing Hasse diagrams for classical Higgs branches which are not nilpotent orbit closures. Suppose the magnetic quiver for the Higgs branch is known, then we conjecture that the \emph{Hasse diagram is obtained by subtracting only elementary slices}, as in the case of nilpotent orbit closures. Hence, one performs quiver subtraction on magnetic quivers. The crucial consistency check is then to compare the computed Hasse diagram with the prediction from partial Higgsing. Remarkably, we find agreement, as demonstrated in this paper. This agreement serves as indication that elementary slices for Coulomb branches of quivers with unitary nodes still seem to keep falling into the same two families (S1) and (S2), although we are not aware of a proof of this fact. For more general symplectic singularities, for which no magnetic quiver is known, our work suggests that other elementary slices can appear, see section \ref{subsection6d}. 

To summarise the conceptual steps: Hasse diagrams for nilpotent orbit closures are known from the mathematicians; elementary slices are classified. Remarkably, we find that the pattern of partial Higgsing of the theories whose Higgs branches are nilpotent orbit closures matches the Hasse diagram. Therefore, we use the partial Higgsing to \emph{predict} the Hasse diagram for classical Higgs branches which are more general symplectic singularities.
Equipped with the concepts of magnetic quivers and quiver subtraction,  we \emph{conjecture} that elementary slices for generic symplectic singularities are exactly the same as for nilpotent orbit closures. With this assumption, we can \emph{compute} Hasse diagrams, we \emph{compare} them against the prediction from the Higgs mechanism, and we find agreement.
Having established and verified this approach, the next logical step is to derive Hasse diagrams of Higgs branch at infinite coupling and give \emph{predictions} of the singularity structure of these moduli spaces.

It should also be mentioned that notions of Hasse diagrams have appeared in  recent physics literature. For instance, the relation between Hasse diagrams of nilpotent orbits and 6d SCFTs has been studied in \cite{Heckman:2016ssk,Heckman:2018jxk,Hassler:2019eso}. 
In addition, brane systems have been used in \cite{Rogers:2018dez,Rogers:2019pqe} to obtain Hasse diagrams corresponding to circular $3$d $\mathcal{N}=4$ quiver gauge theories. 

The remainder of this paper is organised as follows: 
In Section \ref{higgsrecap} we demonstrate how the predictions from classical Higgs mechanism match with the Hasse diagram obtained by assuming the KP transitions. Specifically, in Sections \ref{higgsforsu3} and \ref{branesclass} the example of $\mathrm{SU}(3)$ with $6$ fundamental hypermultiplets is carefully analysed. Thereafter, we consider a $\mathrm{SU}(4)$ gauge theory with one $2$nd rank antisymmetric hypermultiplet and $12$ fundamental hypermultiplets in Sections \ref{higgsforsu4} and \ref{magQuivforsu4}.

In Section \ref{sec:Hasse} the Hasse diagram is elaborated on in more detail. Next, the application of the techniques to Higgs branches of infinite gauge coupling are demonstrated in Section \ref{sec:Higgs_infinite} by considering the same examples as in Section \ref{higgsrecap}.

In Section \ref{sectionResults} we provide many examples of Hasse diagrams for the Higgs branch of Lagrangian theories with a simple gauge group, $5$d $\mathcal{N}=1$ and $6$d $\mathcal{N}=(1,0)$ SCFTs, which are obtained as UV fixed points corresponding to the infinite coupling limit of gauge theories, and selected $4$d $\mathcal{N}=2$ Argyres-Douglas theories. 

Appendix \ref{app:Branes_Quivers} provides background material on relevant brane configurations, magnetic quivers, and quiver subtraction. A brief reminder of symplectic leaves is given in Appendix \ref{SymplecticLeaves}, while Appendix \ref{full5d} provides all the Hasse diagrams for Higgs branches at infinite coupling of 5d SQCD.  

\section{Classical Higgs Mechanism and Hasse Diagrams}
\label{higgsrecap}

In this section the classical Higgs effect in gauge theories with $8$ supercharges is recalled and phrased in a new perspective using Hasse diagrams. 
By giving a non-zero VEV to hypermultiplet scalars of the theory, gauge bosons may become massive, breaking the gauge group to one of its subgroups. If the gauge group is fully broken one speaks of \emph{complete Higgsing}; however, this is not always possible as some unbroken part of the gauge group may remain even if a maximal number of the hypermultiplet scalars acquire a non-zero VEV. This case is termed \emph{maximal Higgsing}. Between the unbroken and the maximally broken case there exist different theories that can be obtained via \emph{partial Higgsing}, where the gauge group is broken to one of its subgroups. Representation theory  can be used to find the matter content of the Higgsed theory or to show that a specific partial Higgsing is impossible. To do this one decomposes the adjoint representation of the gauge group into representations of the unbroken group. All vector bosons in representations which are not the adjoint of the unbroken group become massive. This is only possible if there exist hypermultiplet scalars in the very same representations which are available to be \emph{eaten up}. If these representations are not in the decomposition of the matter content then no such Higgsing exists. If they are, then these scalars acquire a mass, and so does the entire multiplet. A representation theoretic approach to this problem is the subject of Section \ref{higgsforsu3}.

In Section \ref{branesclass} we will employ a modern method of using brane constructions and the tools of magnetic quivers and quiver subtraction to not only derive the same results as in Section \ref{higgsforsu3}, but to uncover more information about the Higgs branch, i.e.\ its structure as a symplectic singularity foliated by symplectic leaves and the transverse slices between them.

\subsection{\texorpdfstring{Partial Higgsing for $\mathrm{SU}(3)$ with $6$ Fundamental Hypermultiplets via Representation Theory}{Partial Higgsing for SU(3) with 6 Fundamental Hypermultiplets through Representation Theory}}
\label{higgsforsu3}
Consider an $\mathrm{SU}(3)$ theory with 6 fundamental hypermultiplets. The continuous subgroups of $\mathrm{SU}(3)$ are $\mathrm{SU}(2)\times \mathrm{U}(1)$, $\mathrm{SU}(2)$, $\mathrm{U}(1)\times \mathrm{U}(1)$, $\mathrm{U}(1)$ and the trivial group $\{1\}$. 

\paragraph{Higgsing $\mathrm{SU}(3)$ to $\mathrm{SU}(2)$.}
The $\mathrm{SU}(3)$ representations decompose as follows:
\begin{equation}
    \begin{split}
    &[1,0]_{A_2}\mapsto [1]_{A_1}+[0]_{A_1} \,,\\
    &[1,1]_{A_2}\mapsto [2]_{A_1}+\underbrace{2[1]_{A_1}+[0]_{A_1}}_{\textnormal{acquire mass}} \,.
    \end{split}
\end{equation}
The W-bosons need to acquire mass from a suitable component of the hypermultiplets. In terms of the representations, one computes
\begin{equation}
    6([1,0]_{A_2}+[0,1]_{A_2})-2(2[1]_{A_1}+[0]_{A_1})    =4([1]_{A_1}+[1]_{A_1})+10([0]_{A_1}) \,.
    \label{PHRC}
\end{equation}
This result implies two facts: (i) the remaining effective gauge group is $\mathrm{SU}(2)$ with $4$ hypermultiplets in the fundamental representation, 
and (ii) the quaternionic dimension of the subspace of the Higgs branch where the gauge group is broken to $\mathrm{SU}(2)$ is $\frac{10}{2}=5$, as seen from the number of massless hypermultiplets that transform as singlets in \eqref{PHRC}.

\paragraph{Higgsing $\mathrm{SU}(3)$ to $\mathrm{U}(1) \times \mathrm{SU}(2)$.}
Here, the representations decompose as follows:
\begin{equation}
    \begin{split}
        &[1,0]_{A_2}\mapsto q^{-2}[1]_{A_1}+q[0]_{A_1} \,, \qquad 
        [0,1]_{A_2}\mapsto q^{2}[1]_{A_1}+q^{-1}[0]_{A_1} \,,\\
        &[1,1]_{A_2}
        \mapsto 
        [2]_{A_1}+\underbrace{q^{-3}[1]_{A_1}+q^{3}[1]_{A_1}+[0]}_{\textnormal{should acquire mass}}  \,.
    \end{split}
\end{equation}
Repeating the computation of \eqref{PHRC}, one observes that the terms do not cancel because of the charge assignments. Hence, Higgsing $\mathrm{SU}(3)$ to $\mathrm{U}(1)\times \mathrm{SU}(2)$ is impossible. 
Analogously, the same statement can be shown for the sub-groups $\mathrm{U}(1) \times \mathrm{U}(1)$ and $\mathrm{U}(1)$.
\paragraph{Higgsing $\mathrm{SU}(2)$ to $\{1\}$.}
The Higgsing of the $\mathrm{SU}(2)$ theory to $\{1\}$, i.e.\ giving mass to all remaining gauge bosons, is unobstructed. One can straightforwardly deduce that the number of massless moduli is $10$, giving a 5 dimensional singular locus in quaternionic dimensions.

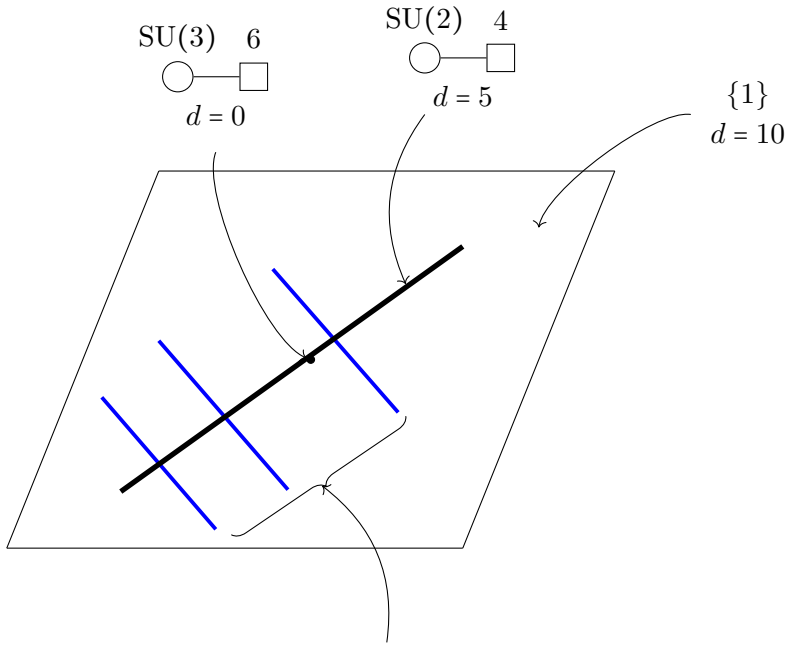
\begin{figure}[t]
    \centering
\begin{tikzpicture}
	\begin{pgfonlayer}{nodelayer}
		\node [style=none] (0) at (-1, 2) {};
		\node [style=none] (1) at (5, 2) {};
		\node [style=none] (2) at (-3, -3) {};
		\node [style=none] (3) at (3, -3) {};
		\node [style=none] (4) at (-1.5, -2.25) {};
		\node [style=none] (5) at (3, 1) {};
		\node [style=dotsize] (6) at (1, -0.5) {};
		\node [style=none] (7) at (6, 2.75) {};
		\node [style=none] (8) at (4, 1.25) {};
		\node [style=none] (9) at (2.25, 0.5) {};
		\node [style=none] (10) at (2.5, 2.75) {};
		\node [style=none] (11) at (-0.25, 2.25) {};
		\node [style=none] (14) at (3, 3) {$d=5$};
		\node [style=none] (16) at (6.75, 3) {$\{1\}$};
		\node [style=none] (17) at (6.75, 2.5) {$d=10$};
		\node [style=gauge1] (18) at (2.5, 3.5) {};
		\node [style=flavour1] (19) at (3.5, 3.5) {};
		\node [style=none] (20) at (2.5, 4) {$\mathrm{SU}(2)$};
		\node [style=none] (21) at (3.5, 4) {$4$};
		\node [style=none] (22) at (-0.25, 2.75) {$d=0$};
		\node [style=gauge1] (23) at (-0.75, 3.25) {};
		\node [style=flavour1] (24) at (0.25, 3.25) {};
		\node [style=none] (25) at (-0.75, 3.75) {$\mathrm{SU}(3)$};
		\node [style=none] (26) at (0.25, 3.75) {$6$};
		\node [style=none] (27) at (-1, -0.25) {};
		\node [style=none] (28) at (0.7, -2.225) {};
		\node [style=none] (29) at (0.5, 0.7) {};
		\node [style=none] (30) at (2.15, -1.2) {};
		\node [style=none] (31) at (-1.75, -1) {};
		\node [style=none] (32) at (-0.25, -2.75) {};
		\node [style=none] (33) at (-1.2, -1.7) {};
		\node [style=none] (35) at (-0.8, -1.725) {};
		\node [style=none] (36) at (-0.3, -1.05) {};
		\node [style=none] (38) at (0.1, -1.1) {};
		\node [style=none] (39) at (1.125, 0) {};
		\node [style=none] (41) at (1.55, -0.05) {};
		\node [style=none] (42) at (2, -4.25) {};
		\node [style=none] (43) at (1.15, -2.175) {};
		\node [style=none] (44) at (2, -4.75) {Transverse slices of dimension $d=5$};
		\node [style=none] (45) at (2.25, -1.25) {};
		\node [style=none] (46) at (-0.05, -2.825) {};
	\end{pgfonlayer}
	\begin{pgfonlayer}{edgelayer}
		\draw [in=180, out=0] (0.center) to (1.center);
		\draw (1.center) to (3.center);
		\draw (0.center) to (2.center);
		\draw (2.center) to (3.center);
		\draw [style=->, bend left=315, looseness=0.50] (7.center) to (8.center);
		\draw [style=->, bend right] (10.center) to (9.center);
		\draw [style=->, bend right=45, looseness=0.50] (11.center) to (6);
		\draw (18) to (19);
		\draw (23) to (24);
		\draw [style=->, bend right] (42.center) to (43.center);
		\draw [style=blueline] (31.center) to (32.center);
		\draw [style=blueline] (27.center) to (28.center);
		\draw [style=blueline] (29.center) to (30.center);
		\draw [style=blackk] (4.center) to (5.center);
		\draw [style=brace2] (45.center) to (46.center);
	\end{pgfonlayer}
\end{tikzpicture}
    \caption{A schematic representation of the Higgs branch of $\mathrm{SU}(3)$ with $6$ fundamental hypermultiplets. At the $0$-dimensional origin the theory is unbroken. On the $5$-dimensional symplectic leaf, represented by the black line without the origin, the theory is broken to $\mathrm{SU}(2)$ with $4$ fundamental hypermultiplets and 5 neutral hypermultiplets. On the 10-dimensional symplectic leaf, represented by the plane without the black line and the origin, the gauge group is completely broken and the effective theory contains only 10 neutral hypermultiplets. The blue lines represent the transverse space of a point on the 5-dimensional symplectic leaf inside the 10-dimensional symplectic leaf. Locally, each individual transverse space looks like the Higgs branch of $\mathrm{SU}(2)$ with $4$ fundamental hypermultiplets.}
    \label{HiggsSimplest}
\end{figure}

\paragraph{Higgs branch.}
The original $\mathrm{SU}(3)$ theory with $6$ fundamental flavours has a Higgs branch of quaternionic dimension 10, while the $\mathrm{SU}(2)$ with 4 fundamentals has Higgs branch of dimension $5$.
In anticipation of the later sections, the locus where the gauge group is consistently broken to a particular subgroup is called a \emph{symplectic leaf} of the full Higgs branch of $\mathrm{SU}(3)$ with 6 flavours, which is the total space. 
In this paper, the total space is understood as the closure of the largest leaf.
See Appendix \ref{SymplecticLeaves} for details on symplectic leaves. Over any point in a specific leaf there exists a residual gauge theory with the unbroken gauge group and some matter content. The massless moduli that appeared in the above analysis parametrise the leaf. The Higgs branch is schematically shown in Figure \ref{HiggsSimplest}.

\begin{figure}
    \centering
\begin{tikzpicture}
	\begin{pgfonlayer}{nodelayer}
		\node [style=miniG] (0) at (0, 3) {};
		\node [style=miniG] (1) at (0, 0) {};
		\node [style=miniG] (2) at (0, -3) {};
		\node [style=none] (3) at (2.75, 3) {};
		\node [style=none] (4) at (3.25, 3) {};
		\node [style=none] (7) at (3.25, 0.15) {};
		\node [style=none] (10) at (2.75, 0.15) {};
		\node [style=none] (11) at (-0.75, 1.5) {5};
		\node [style=none] (12) at (-0.75, -1.5) {5};
		\node [style=gauge1] (13) at (4, 1) {};
		\node [style=flavour1] (14) at (4, 2.5) {};
		\node [style=none] (17) at (4, 0.5) {$\mathrm{SU}(2)$};
		\node [style=none] (18) at (4, 3) {4};
		\node [style=none] (21) at (4.75, 3) {};
		\node [style=none] (22) at (5.25, 3) {};
		\node [style=none] (23) at (5.25, -3.025) {};
		\node [style=none] (24) at (4.75, -3.025) {};
		\node [style=gauge1] (25) at (6, -0.75) {};
		\node [style=flavour1] (26) at (6, 0.75) {};
		\node [style=none] (27) at (6, -1.25) {$\mathrm{SU}(3)$};
		\node [style=none] (28) at (6, 1.25) {6};
		\node [style=none] (29) at (1.25, 3.125) {};
		\node [style=none] (30) at (1.6, 3.125) {};
		\node [style=none] (31) at (1.6, 2.875) {};
		\node [style=none] (32) at (1.25, 2.875) {};
		\node [style=none] (33) at (2, 3) {$\{1\}$};
		\node [style=none] (34) at (0, 4) {Hasse diagram};
		\node [style=none] (35) at (3.25, 4) {Effective theory};
		\node [style=none] (36) at (0.5, 3) {10};
		\node [style=none] (37) at (0.5, 0) {5};
		\node [style=none] (38) at (0.5, -3) {0};
	\end{pgfonlayer}
	\begin{pgfonlayer}{edgelayer}
		\draw (3.center) to (4.center);
		\draw (3.center) to (4.center);
		\draw (4.center) to (7.center);
		\draw (7.center) to (10.center);
		\draw (0) to (1);
		\draw (1) to (2);
		\draw (13) to (14);
		\draw (21.center) to (22.center);
		\draw (21.center) to (22.center);
		\draw (22.center) to (23.center);
		\draw (23.center) to (24.center);
		\draw (25) to (26);
		\draw (29.center) to (30.center);
		\draw (30.center) to (31.center);
		\draw (31.center) to (32.center);
	\end{pgfonlayer}
\end{tikzpicture}
	\caption{Hasse diagram obtained from partial Higgsing of $\mathrm{SU}(3)$ with 6 fundamentals. 
	The effective gauge theory on each symplectic leaf is given by the quiver in the corresponding bracket and an extra number $N$ of neutral hypermultiplets, where $N$ is the number labelling the node. Note that $N$ is also the quaternionic dimension of the leaf.
	}\label{tab:HiggsSimplestHasse}
\end{figure}
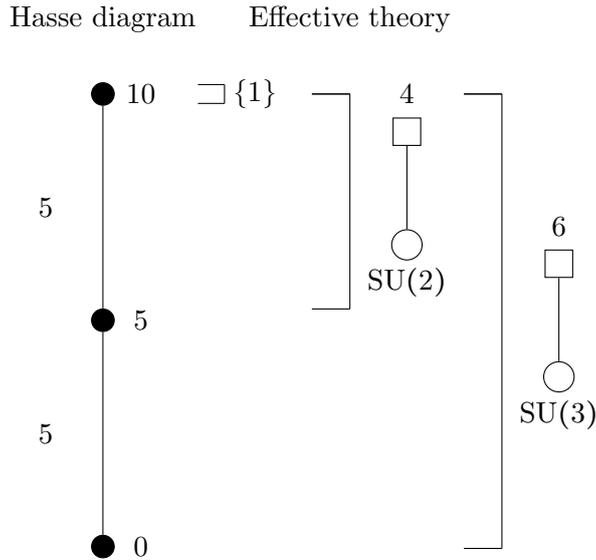

The information about partial Higgsing can be conveniently encoded diagrammatically in a \emph{Hasse diagram}. In the Hasse diagram of Figure \ref{tab:HiggsSimplestHasse}, the unbroken $\mathrm{SU}(3)$ corresponds to the bottom node, the residual $\mathrm{SU}(2)$ to the middle node, and the completely broken trivial group to the top node. 
The bottom node represents the locus where $\mathrm{SU}(3)$ is unbroken, which is equal to the smallest symplectic leaf (identical to its own closure), the Higgs branch of $\mathrm{SU}(3)$ is the space from the top node to the bottom node. Or put differently, the Higgs branch of $\mathrm{SU}(3)$ is the transverse slice between the trivial symplectic leaf and the  total space.
Next, the closure of the singular locus where $\mathrm{SU}(3)$ is broken to $\mathrm{SU}(2)$, i.e.\ the closure of another symplectic leaf, is given by the space from the bottom node to the middle, while the Higgs branch of the residual $\mathrm{SU}(2)$ theory is the space from the top to the middle node. 
Note in particular, that the closure of this leaf contains both, the singular locus where $\mathrm{SU}(3)$ is unbroken and broken to $\mathrm{SU}(2)$.
On the other hand, the Higgs branch of $\mathrm{SU}(2)$ is the transverse slice between the symplectic leaf, where $\mathrm{SU}(3)$ breaks to $\mathrm{SU}(2)$, and the total space, which is the Higgs branch of $\mathrm{SU}(3)$. 
Lastly, the closure of the singular locus for the Higgsing of $\mathrm{SU}(3)$ to $\{1\}$ is the space from the bottom to the top node. Hence, it is equal to the Higgs branch of $\mathrm{SU}(3)$ itself, and the transverse slice to the total space is the trivial space.

\subsection{\texorpdfstring{Gauge Enhancement of $\mathrm{SU}(3)$ with $6$ Fundamental Hypermultiplets via Brane Webs, Magnetic Quivers and the Kraft-Procesi Transition}{Gauge Enhancement of SU(3) with 6 fundamental hypermultiplets via Brane Webs and Magnetic Quivers}}
\label{branesclass}
Now the objective is to match the predictions from partial Higgsing from Section \ref{higgsforsu3} using the techniques of magnetic quivers of 5-brane webs \cite{Cabrera:2018jxt} and Kraft-Procesi transitions \cite{Cabrera:2016vvv,Cabrera:2017njm,Cabrera:2018ann}, as reviewed in Appendix \ref{app:Branes_Quivers}. Note that these two techniques had never been used together before. Hence, the discussion in the present section constitutes a new result. For the $5$d $\mathrm{SU}(3)$ theory with $6$ fundamental hypermultiplets one employs a 5-brane web construction. However, unlike in Section \ref{higgsforsu3} the emphasis is placed on \emph{partial gauge enhancement} (colloquially also known as un-Higgsing) of a maximally broken theory rather than the opposite view of partial Higgsing. Starting at a generic point on the Higgs branch, corresponding to the maximally broken phase, one realigns the light branes to move onto a singular locus of the Higgs branch to subsequently move onto a mixed branch of the theory.
Here, analogously to \cite{Cabrera:2016vvv,Cabrera:2017njm}, the assumption is that one needs to perform \emph{minimal} transitions in the brane configuration, cf.\  Appendix \ref{minimaltransitions}. These minimal transitions are called \emph{Kraft-Procesi transitions}.
Repeating this procedure allows to see all symplectic leaves and transverse spaces of the Higgs branch; hence, the Hasse diagram can be derived from the brane configuration. Using the notion of magnetic quivers \cite{Cabrera:2018jxt}, one can characterise the Hasse diagram in more detail than in Section \ref{higgsforsu3}.
Inspecting Figure \ref{fig:SU(3)6branesclass} reveals the following:
\begin{itemize}
    \item The left column depicts brane configurations for the different symplectic leaves. They are obtained by \emph{removing} elementary slices, one at a time. These are Kraft-Procesi transitions \cite{Cabrera:2016vvv,Cabrera:2017njm}, with the difference that they are performed in 5-brane webs.
    \item The central column provides the (electric) quivers, previously computed in Section \ref{higgsforsu3}, whose Higgs branch is the transverse slice between the corresponding symplectic leaf and the full space (the Higgs branch of $\mathrm{SU}(3)$ with 6 flavours).
    \item The right column contains the magnetic quivers which describe each symplectic leaf. The precise relationship is that the $3d~\mathcal N=4$ Coulomb branch of the magnetic quiver is the closure of the symplectic leaf.
    \item The arrows between the different 5-brane webs in the left column denote the sub-webs which realise the Kraft-Procesi transitions. The Higgs branch of such 5-brane webs are the elementary slices.
    \item The arrows in the right column denote the magnetic quivers corresponding to the elementary slices. The precise relationship is that the $3d~\mathcal N=4$ Coulomb branch of the magnetic quiver is the elementary slice.
\end{itemize}

\begin{figure}[t]
	\centering
	\begin{tabular}{m{2.6cm} m{2.8cm}}
	Hasse diagram  & Magnetic quiver 
	 \\ 
  	\begin{tikzpicture}
		\tikzstyle{hasse} = [circle, fill,inner sep=2pt];
		\node [hasse] (1) [label=right:\footnotesize{$10$}] {};
		\node (2) [below of=1] {};
		\node [hasse] (3) [label=right:\footnotesize{$5$},below of=2] {};
		\node  (4) [below of=3] {};
		\node[hasse] (5) [label=right:\footnotesize{$0$},below of=4] {};
		\draw (1) edge [] node[label=left:\footnotesize{$\;\;\;\;\;d_4$}] {} (3)
			(3) edge [] node[label=left:\footnotesize{$a_5$}] {} (5);
	\end{tikzpicture}
	& \thead{ $\node{}1-\node{}2-\node{\overset{\displaystyle\overset 1 \circ~~} \diagdown  \overset{\displaystyle\overset{~~1} {~~ \circ}} \diagup}3 -\node{}2-\node{}1$ \\ \\ \\ \\
	$\overset{\overset{\displaystyle\overset 1 \circ}{\diagup~~~~~~~~~~\diagdown}}{\node{}{1}-\node{}1-\node{}1-\node{}1-\node{}{1}}$ \\ \\ \\ 
	 $\node{}1$ }
	\end{tabular}
	\caption{Hasse diagram with magnetic quivers representing the closures of symplectic leaves for the classical Higgs branch of $\mathrm{SU}(3)$ with $6$ fundamentals. The transverse slices between neighbouring symplectic leaves have been added in the edges of the Hasse diagram. This is a generalisation of the results by \cite{kraft1980minimal,Kraft1982} to a set of spaces different from nilpotent orbits.}\label{tab:SU(3)6HasseSub}
\end{figure}
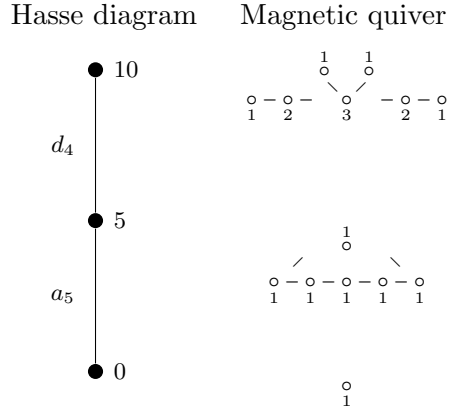

From the brane analysis of Kraft-Procesi transitions in Figure \ref{fig:SU(3)6branesclass}, the Hasse diagram of the Higgs branch of $\mathrm{SU}(3)$ with 6 flavours can be obtained. It is depicted in Figure \ref{tab:SU(3)6HasseSub}. This deserves some comments. To each node a magnetic quiver is associated, such that its space of dressed monopole operators describes the closure of the symplectic leaf. In other words, if the magnetic quiver is considered as an auxiliary $3$d $\mathcal N=4$ theory, its Coulomb branch is the closure of the symplectic leaf.  This is in marked contrast to the Hasse diagram of Figure \ref{tab:HiggsSimplestHasse}, where no explicit quiver description was available for the closures of the symplectic leaves. In addition, the links between connected points have been labelled by elementary slices, here $d_4$ and $a_5$. The way to see these is from the magnetic quivers that are obtained during the transition, see Figure \ref{fig:SU(3)6branesclass}. The space of dressed monopole operators of the magnetic quivers (alternatively their $3$d $\mathcal N=4$ Coulomb branches) are either the closure of the minimal nilpotent orbit of $A_5$, hence denoted as an $a_5$ Kraft-Procesi transition, or the minimal nilpotent orbit closure of $D_4$, which is called a $d_4$ Kraft-Procesi transition \cite{Kraft1982}. By construction, the minimal Kraft-Procesi transitions are the transverse slices between two neighbouring points in the Hasse diagram. Again, this understanding of the Hasse diagram reflects the refinement obtained due to the use of magnetic quivers and 5-brane webs compared to the classical Higgs mechanism analysis.

The present brane realisation of Kraft-Procesi transitions can be translated to an operation between the magnetic quivers. This is analogous to the \emph{quiver subtraction} \cite{Cabrera:2018ann} that was proposed to generalise the brane realisations of the Kraft-Procesi transitions for closures of nilpotent orbits \cite{Cabrera:2016vvv,Cabrera:2017njm}. A new formulation of the operation of \emph{quiver subtractions} can be found in Appendix \ref{AppendixSubtraction}. This new formulation covers the new Kraft-Procesi transitions found in the 5-brane webs, and it is fully consistent with the previous definition. It also achieves to unify the quiver subtraction operation as defined in \cite{Cabrera:2018ann} with the subtractions of $e_8$ Kraft-Procesi transitions utilised in \cite{Hanany:2018uhm,Cabrera:2019izd}.

\begin{landscape}
\begin{figure}[t]
\centering
\scalebox{0.6}{\begin{tikzpicture}
	\begin{pgfonlayer}{nodelayer}
		\node [style=none] (0) at (-6, 5) {\huge Brane Webs};
		\node [style=none] (1) at (2.25, 5) {};
		\node [style=none] (2) at (2.25, -18.5) {};
		\node [style=none] (3) at (4.75, 5) {\huge Electric Quivers};
		\node [style=none] (4) at (7.25, 5) {};
		\node [style=none] (5) at (7.25, -18) {};
		\node [style=none] (6) at (13, 5) {\huge Magnetic Quivers};
		\node [style=gauge1] (7) at (-8, 2) {};
		\node [style=gauge1] (8) at (-9.5, 2) {};
		\node [style=gauge1] (9) at (-11, 2) {};
		\node [style=gauge1] (10) at (-7, 4) {};
		\node [style=gauge1] (11) at (-5, 4) {};
		\node [style=gauge1] (12) at (-7, 0) {};
		\node [style=gauge1] (13) at (-5, 0) {};
		\node [style=gauge1] (14) at (-4, 2) {};
		\node [style=gauge1] (15) at (-2.5, 2) {};
		\node [style=gauge1] (16) at (-1, 2) {};
		\node [style=none] (17) at (-9.5, 2.125) {};
		\node [style=none] (18) at (-9.5, 1.875) {};
		\node [style=none] (19) at (-8, 2.125) {};
		\node [style=none] (20) at (-8, 1.875) {};
		\node [style=none] (21) at (-8, 2.125) {};
		\node [style=none] (22) at (-8, 1.875) {};
		\node [style=none] (23) at (-8, 2) {};
		\node [style=none] (24) at (-4, 2.125) {};
		\node [style=none] (25) at (-4, 2) {};
		\node [style=none] (26) at (-4, 1.875) {};
		\node [style=none] (27) at (-4, 2.125) {};
		\node [style=none] (28) at (-4, 1.875) {};
		\node [style=none] (29) at (-2.5, 2.125) {};
		\node [style=none] (30) at (-2.5, 1.875) {};
		\node [style=gauge1] (31) at (-3, -2.5) {};
		\node [style=gauge1] (32) at (-4, -2.5) {};
		\node [style=gauge1] (34) at (-2.5, -1.5) {};
		\node [style=gauge1] (35) at (-1.75, -1.5) {};
		\node [style=gauge1] (36) at (-2.5, -3.5) {};
		\node [style=gauge1] (37) at (-1.75, -3.5) {};
		\node [style=gauge1] (38) at (-1.25, -2.5) {};
		\node [style=gauge1] (39) at (-0.25, -2.5) {};
		\node [style=none] (45) at (-3, -2.375) {};
		\node [style=none] (46) at (-3, -2.625) {};
		\node [style=none] (51) at (-1.25, -2.375) {};
		\node [style=none] (52) at (-1.25, -2.625) {};
		\node [style=none] (53) at (-5, -1) {};
		\node [style=none] (54) at (-5, -4) {};
		\node [style=gauge1] (55) at (-8, -7) {};
		\node [style=gauge1] (56) at (-9.5, -7) {};
		\node [style=gauge1] (57) at (-11, -7) {};
		\node [style=gauge1] (58) at (-7, -5) {};
		\node [style=gauge1] (59) at (-5, -5) {};
		\node [style=gauge1] (60) at (-7, -9) {};
		\node [style=gauge1] (61) at (-5, -9) {};
		\node [style=gauge1] (62) at (-4, -7) {};
		\node [style=gauge1] (63) at (-2.75, -7) {};
		\node [style=gauge1] (64) at (-1.25, -7) {};
		\node [style=none] (65) at (-9.5, -6.875) {};
		\node [style=none] (66) at (-9.5, -7.125) {};
		\node [style=none] (67) at (-8, -6.875) {};
		\node [style=none] (68) at (-8, -7.125) {};
		\node [style=none] (69) at (-8, -6.875) {};
		\node [style=none] (70) at (-8, -7.125) {};
		\node [style=none] (71) at (-8, -7) {};
		\node [style=none] (72) at (-4, -6.875) {};
		\node [style=none] (73) at (-4, -7) {};
		\node [style=none] (74) at (-4, -7.125) {};
		\node [style=none] (75) at (-4, -6.875) {};
		\node [style=none] (76) at (-4, -7.125) {};
		\node [style=none] (77) at (-2.75, -6.875) {};
		\node [style=none] (78) at (-2.75, -7.125) {};
		\node [style=none] (80) at (-7, -6) {};
		\node [style=none] (81) at (-5, -6) {};
		\node [style=none] (82) at (-4.5, -6.875) {};
		\node [style=none] (83) at (-4.5, -7.125) {};
		\node [style=none] (84) at (-5, -8) {};
		\node [style=none] (85) at (-7.5, -7.125) {};
		\node [style=none] (86) at (-7, -8) {};
		\node [style=none] (87) at (-7.5, -6.875) {};
		\node [style=gauge1] (88) at (-8, -14.75) {};
		\node [style=gauge1] (89) at (-9, -14.75) {};
		\node [style=gauge1] (90) at (-10, -14.75) {};
		\node [style=gauge1] (91) at (-6.5, -12.75) {};
		\node [style=gauge1] (92) at (-5.5, -12.75) {};
		\node [style=gauge1] (93) at (-6.75, -16.75) {};
		\node [style=gauge1] (94) at (-5.25, -16.75) {};
		\node [style=gauge1] (95) at (-4, -14.75) {};
		\node [style=gauge1] (96) at (-3, -14.75) {};
		\node [style=gauge1] (97) at (-2, -14.75) {};
		\node [style=none] (98) at (-9, -14.7) {};
		\node [style=none] (99) at (-9, -14.8) {};
		\node [style=none] (102) at (-8, -14.625) {};
		\node [style=none] (103) at (-8, -14.875) {};
		\node [style=none] (104) at (-8, -14.75) {};
		\node [style=none] (105) at (-4.025, -14.625) {};
		\node [style=none] (107) at (-4.025, -14.875) {};
		\node [style=none] (110) at (-3, -14.7) {};
		\node [style=none] (111) at (-3, -14.8) {};
		\node [style=none] (112) at (-6.5, -13.25) {};
		\node [style=none] (113) at (-5.5, -13.25) {};
		\node [style=none] (114) at (-6.75, -13.75) {};
		\node [style=none] (115) at (-5.25, -13.75) {};
		\node [style=none] (116) at (-7.5, -14.625) {};
		\node [style=none] (117) at (-4.525, -14.625) {};
		\node [style=none] (118) at (-6.75, -15.5) {};
		\node [style=none] (119) at (-5.25, -15.5) {};
		\node [style=none] (120) at (-4.525, -14.875) {};
		\node [style=none] (121) at (-7.5, -14.75) {};
		\node [style=none] (122) at (-7.5, -14.9) {};
		\node [style=none] (123) at (-4.525, -14.75) {};
		\node [style=none] (124) at (-4, -14.7) {};
		\node [style=none] (125) at (-4, -14.8) {};
		\node [style=none] (126) at (-8.025, -14.7) {};
		\node [style=none] (127) at (-8.025, -14.8) {};
		\node [style=none] (128) at (-4.75, -9.75) {};
		\node [style=none] (129) at (-4.75, -12.75) {};
		\node [style=gauge1] (130) at (-2.5, -11.25) {};
		\node [style=gauge1] (131) at (-3.25, -11.25) {};
		\node [style=gauge1] (132) at (-4, -11.25) {};
		\node [style=gauge1] (133) at (-1.5, -10.25) {};
		\node [style=gauge1] (134) at (-1, -10.25) {};
		\node [style=gauge1] (135) at (-1.5, -12.25) {};
		\node [style=gauge1] (136) at (-1, -12.25) {};
		\node [style=gauge1] (137) at (0, -11.25) {};
		\node [style=gauge1] (138) at (0.75, -11.25) {};
		\node [style=gauge1] (139) at (1.5, -11.25) {};
		\node [style=none] (140) at (-3.25, -11.125) {};
		\node [style=none] (141) at (-3.25, -11.375) {};
		\node [style=none] (142) at (-2.5, -11.125) {};
		\node [style=none] (143) at (-2.5, -11.375) {};
		\node [style=none] (144) at (-2.5, -11.125) {};
		\node [style=none] (145) at (-2.5, -11.375) {};
		\node [style=none] (146) at (-2.5, -11.375) {};
		\node [style=none] (147) at (0, -11.125) {};
		\node [style=none] (148) at (0, -11.25) {};
		\node [style=none] (149) at (0, -11.375) {};
		\node [style=none] (150) at (0, -11.125) {};
		\node [style=none] (151) at (0, -11.375) {};
		\node [style=none] (152) at (0.75, -11.125) {};
		\node [style=none] (153) at (0.75, -11.375) {};
		\node [style=none] (154) at (-1.5, -10.75) {};
		\node [style=none] (155) at (-1, -10.75) {};
		\node [style=none] (156) at (-0.5, -11.125) {};
		\node [style=none] (157) at (-0.5, -11.375) {};
		\node [style=none] (158) at (-1, -11.75) {};
		\node [style=none] (159) at (-2, -11.375) {};
		\node [style=none] (160) at (-1.5, -11.75) {};
		\node [style=none] (161) at (-2, -11.125) {};
		\node [style=none] (162) at (-2.5, -11.25) {};
		\node [style=none] (163) at (0, -11.25) {};
		\node [style=none] (164) at (4, 2) {$\{1\}$};
		\node [style=none] (167) at (4, -7.55) {$\mathrm{SU}(2)$};
		\node [style=none] (168) at (4, -5.025) {$4$};
		\node [style=none] (171) at (4, -15.3) {$\mathrm{SU}(3)$};
		\node [style=none] (172) at (4, -13.05) {$6$};
		\node [style=bluegauge] (173) at (11, 1) {};
		\node [style=bluegauge] (174) at (15, 1) {};
		\node [style=redgauge] (175) at (12, 1) {};
		\node [style=redgauge] (176) at (14, 1) {};
		\node [style=cyangauge] (177) at (13, 1) {};
		\node [style=orangegauge] (178) at (12.25, 2) {};
		\node [style=greenguage] (180) at (17.75, -2.25) {};
		\node [style=greenguage] (181) at (19.25, -2.25) {};
		\node [style=greenguage] (182) at (17.25, -3.25) {};
		\node [style=greenguage] (183) at (18.5, -3.25) {};
		\node [style=greenguage] (184) at (19.75, -3.25) {};
		\node [style=greenguage] (185) at (13, -6.75) {};
		\node [style=cyangauge] (186) at (13, -7.75) {};
		\node [style=redgauge] (187) at (12, -7.75) {};
		\node [style=redgauge] (188) at (14, -7.75) {};
		\node [style=bluegauge] (189) at (11, -7.75) {};
		\node [style=bluegauge] (190) at (15, -7.75) {};
		\node [style=greenguage] (192) at (16.5, -12.75) {};
		\node [style=greenguage] (193) at (17.5, -12.75) {};
		\node [style=greenguage] (194) at (18.5, -12.75) {};
		\node [style=greenguage] (195) at (19.5, -12.75) {};
		\node [style=greenguage] (196) at (20.5, -12.75) {};
		\node [style=greenguage] (197) at (13, -15) {};
		\node [style=none] (198) at (14.25, -1.5) {};
		\node [style=none] (199) at (14.25, -4.5) {};
		\node [style=none] (200) at (14.25, -10.75) {};
		\node [style=none] (201) at (14.25, -13.75) {};
		\node [style=none] (202) at (11, 0.5) {$1$};
		\node [style=none] (203) at (12, 0.5) {$2$};
		\node [style=none] (204) at (13, 0.5) {$3$};
		\node [style=none] (205) at (14, 0.5) {$2$};
		\node [style=none] (206) at (15, 0.5) {$1$};
		\node [style=none] (207) at (13.75, 2.5) {$1$};
		\node [style=none] (208) at (12.25, 2.5) {$1$};
		\node [style=none] (209) at (17.75, -1.75) {$1$};
		\node [style=none] (210) at (19.25, -1.75) {$1$};
		\node [style=none] (211) at (19.75, -3.75) {$1$};
		\node [style=none] (212) at (18.5, -3.75) {$2$};
		\node [style=none] (213) at (17.25, -3.75) {$1$};
		\node [style=none] (214) at (13, -6.25) {$1$};
		\node [style=none] (215) at (11, -8.25) {$1$};
		\node [style=none] (216) at (12, -8.25) {$1$};
		\node [style=none] (217) at (13, -8.25) {$1$};
		\node [style=none] (218) at (14, -8.25) {$1$};
		\node [style=none] (219) at (15, -8.25) {$1$};
		\node [style=none] (220) at (18.5, -11.25) {$1$};
		\node [style=none] (221) at (16.5, -13.25) {$1$};
		\node [style=none] (222) at (17.5, -13.25) {$1$};
		\node [style=none] (223) at (18.5, -13.25) {$1$};
		\node [style=none] (224) at (19.5, -13.25) {$1$};
		\node [style=none] (225) at (20.5, -13.25) {$1$};
		\node [style=none] (226) at (13, -15.5) {$1$};
		\node [style=blackgauge] (227) at (18.5, -11.75) {};
		\node [style=magentagauge] (443) at (13.75, 2) {};
		\node [style=none] (444) at (15.75, -2.75) {};
		\node [style=none] (445) at (16.5, -2.75) {};
		\node [style=none] (446) at (15.75, -12) {};
		\node [style=none] (447) at (16.5, -12) {};
		\node [style=purplee] (448) at (4, -7) {};
		\node [style=purpleesquare] (449) at (4, -5.5) {};
		\node [style=purplee] (450) at (4, -15) {};
		\node [style=purpleesquare] (451) at (4, -13.5) {};
	\end{pgfonlayer}
	\begin{pgfonlayer}{edgelayer}
		\draw [style=new edge style 1] (1.center) to (2.center);
		\draw [style=new edge style 1] (4.center) to (5.center);
		\draw [style=orangee] (10) to (12);
		\draw [style=cyane] (21.center) to (24.center);
		\draw [style=cyane] (25.center) to (23.center);
		\draw [style=cyane] (22.center) to (26.center);
		\draw [style=rede] (27.center) to (29.center);
		\draw [style=rede] (28.center) to (30.center);
		\draw [style=rede] (19.center) to (17.center);
		\draw [style=rede] (18.center) to (20.center);
		\draw [style=bluee] (15) to (16);
		\draw [style=bluee] (9) to (8);
		\draw [style=greene] (35) to (37);
		\draw [style=greene] (45.center) to (51.center);
		\draw [style=greene] (52.center) to (46.center);
		\draw [style=greene] (34) to (36);
		\draw [style=greene] (31) to (32);
		\draw [style=greene] (39) to (38);
		\draw [style=->, bend left=60, looseness=0.50] (53.center) to (54.center);
		\draw [style=cyane] (73.center) to (71.center);
		\draw [style=rede] (76.center) to (78.center);
		\draw [style=rede] (66.center) to (68.center);
		\draw [style=bluee] (63) to (64);
		\draw [style=bluee] (57) to (56);
		\draw [style=greene] (80.center) to (58);
		\draw [style=greene] (59) to (81.center);
		\draw [style=greene] (85.center) to (70.center);
		\draw [style=greene] (85.center) to (86.center);
		\draw [style=greene] (86.center) to (60);
		\draw [style=greene] (84.center) to (61);
		\draw [style=greene] (84.center) to (83.center);
		\draw [style=greene] (83.center) to (76.center);
		\draw [style=greene] (82.center) to (75.center);
		\draw [style=greene] (81.center) to (82.center);
		\draw [style=greene] (87.center) to (69.center);
		\draw [style=greene] (80.center) to (87.center);
		\draw [style=greene] (80.center) to (81.center);
		\draw [style=greene] (84.center) to (86.center);
		\draw [style=greene] (116.center) to (102.center);
		\draw [style=greene] (117.center) to (105.center);
		\draw [style=greene] (116.center) to (114.center);
		\draw [style=greene] (114.center) to (112.center);
		\draw [style=greene] (112.center) to (113.center);
		\draw [style=greene] (113.center) to (115.center);
		\draw [style=greene] (115.center) to (117.center);
		\draw [style=greene] (113.center) to (92);
		\draw [style=greene] (91) to (112.center);
		\draw [style=greene] (114.center) to (115.center);
		\draw [style=greene] (119.center) to (120.center);
		\draw [style=greene] (120.center) to (107.center);
		\draw [style=greene] (118.center) to (93);
		\draw [style=greene] (119.center) to (94);
		\draw [style=greene] (119.center) to (118.center);
		\draw [style=greene] (123.center) to (95);
		\draw [style=greene] (121.center) to (104.center);
		\draw [style=greene] (122.center) to (118.center);
		\draw [style=greene] (122.center) to (103.center);
		\draw [style=greene] (124.center) to (110.center);
		\draw [style=greene] (111.center) to (125.center);
		\draw [style=greene] (126.center) to (98.center);
		\draw [style=greene] (99.center) to (127.center);
		\draw [style=greene] (89) to (90);
		\draw [style=greene] (97) to (96);
		\draw [style=->, bend left=60, looseness=0.50] (128.center) to (129.center);
		\draw [style=greene] (145.center) to (141.center);
		\draw [style=greene] (153.center) to (151.center);
		\draw [style=greene] (139) to (138);
		\draw [style=greene] (132) to (131);
		\draw [style=greene] (162.center) to (163.center);
		\draw (178) to (177);
		\draw (177) to (176);
		\draw (176) to (174);
		\draw (173) to (175);
		\draw (180) to (183);
		\draw (183) to (181);
		\draw (183) to (184);
		\draw (183) to (182);
		\draw (185) to (190);
		\draw (190) to (188);
		\draw (188) to (186);
		\draw (186) to (187);
		\draw (187) to (189);
		\draw (189) to (185);
		\draw (192) to (193);
		\draw (193) to (194);
		\draw (194) to (195);
		\draw (195) to (196);
		\draw (177) to (175);
		\draw [style=->, bend left=60, looseness=0.50] (198.center) to (199.center);
		\draw [style=->, bend left=60, looseness=0.50] (200.center) to (201.center);
		\draw [style=greene] (69.center) to (65.center);
		\draw [style=greene] (75.center) to (77.center);
		\draw (133) to (154.center);
		\draw (134) to (155.center);
		\draw (155.center) to (156.center);
		\draw (156.center) to (150.center);
		\draw (154.center) to (161.center);
		\draw (157.center) to (151.center);
		\draw (157.center) to (158.center);
		\draw (158.center) to (136);
		\draw (160.center) to (135);
		\draw (160.center) to (158.center);
		\draw (160.center) to (159.center);
		\draw (140.center) to (144.center);
		\draw (152.center) to (150.center);
		\draw (154.center) to (155.center);
		\draw (227) to (192);
		\draw (227) to (196);
		\draw (161.center) to (144.center);
		\draw (159.center) to (146.center);
		\draw [style=magentae] (11) to (13);
		\draw (443) to (177);
		\draw (444.center) to (445.center);
		\draw (446.center) to (447.center);
		\draw [style=purpleline] (448) to (449);
		\draw [style=purpleline] (450) to (451);
	\end{pgfonlayer}
\end{tikzpicture}}
\caption{Depiction of the different 5-brane webs in the gauge enhancements up to $\mathrm{SU}(3)$ with $6$ fundamentals. The methods developed in \cite{Cabrera:2018jxt} allow us to read magnetic quivers for the closure of all symplectic leaves in the Higgs branch as well as the transverse slices. This process can be translated into an operation between the magnetic quivers, called quiver subtraction. Coloured branes are assumed to be on different positions along the 7-branes.}
\label{fig:SU(3)6branesclass}
\end{figure}
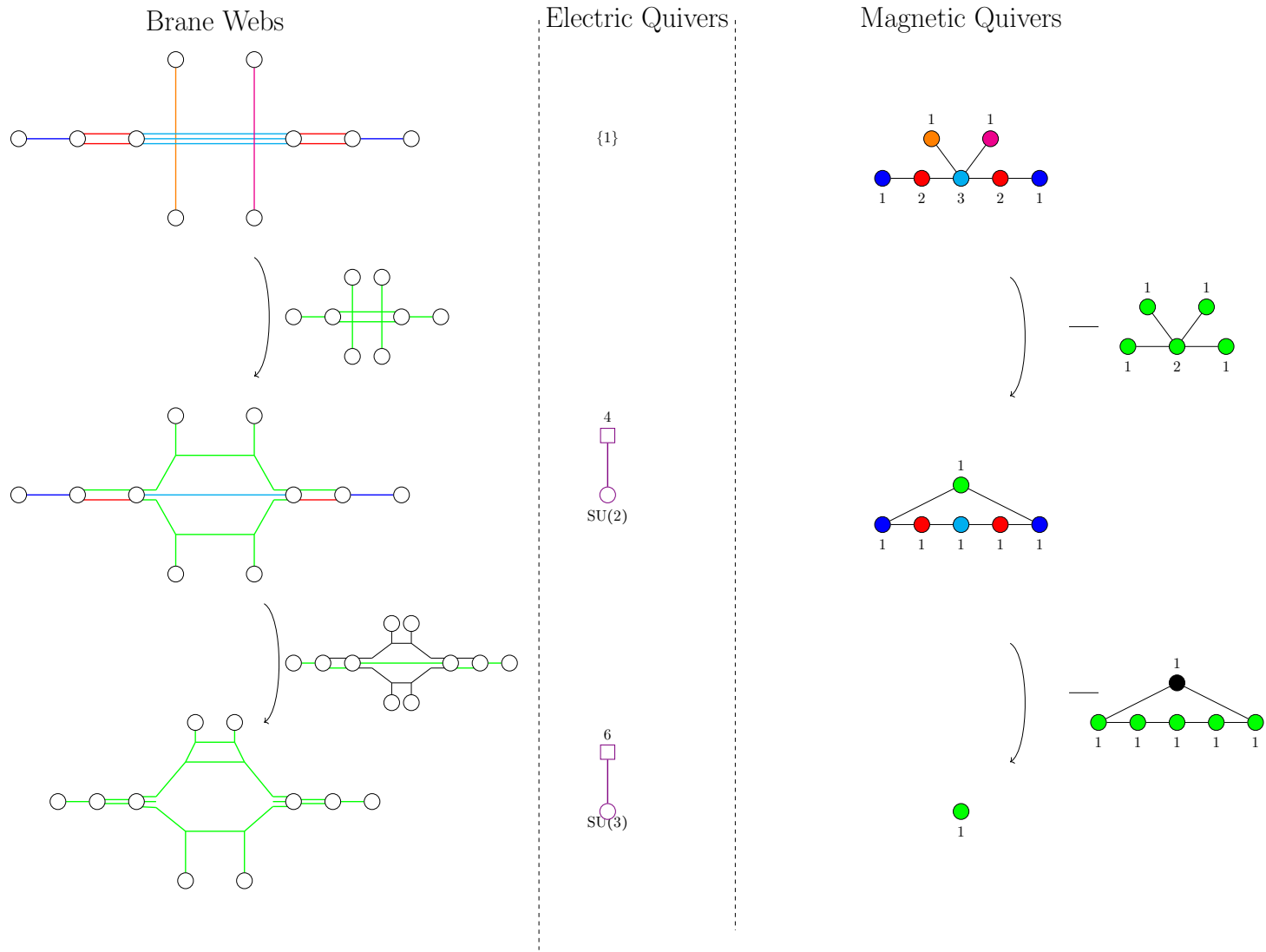
\end{landscape}

\subsection{\texorpdfstring{Partial Higgsing of $\mathrm{SU}(4)$ with a $2\mathrm{nd}$ rank antisymmetric $\Lambda^2$ and $12$ fundamental hypermultiplets via Representation Theory}{Partial Higgsing of SU(4) with a 2nd rank antisymmetric Lambda2 and 12 fundamental hypermultiplets via Representation Theory}}
\label{higgsforsu4}
We now proceed to repeat the analysis of Section \ref{higgsforsu3} for $\mathrm{SU}(4)$ with a $2\mathrm{nd}$ rank antisymmetric $\Lambda^2$ and $12$ fundamental hypermultiplets with the difference that we use a brane construction for the theory in 6d. As before, we first study the partial Higgs mechanism.
The subgroups of $\mathrm{SU}(4)$ reachable by partial Higgsing are $\mathrm{Sp}(2)$, $\mathrm{SU}(3)$, $\mathrm{SU}(2)$, and $\{1\}$. 
\paragraph{Higgsing $\mathrm{SU}(4)$ to $\mathrm{Sp}(2)$.}
To start with, the relevant representations decompose as follows:
\begin{equation}
    \begin{split}
    &[1,0,0]_{A_3}\mapsto [1,0]_{C_2} \,, \qquad \; [0,0,1]_{A_3}\mapsto [1,0]_{C_2} \,\\
    &[0,1,0]_{A_3}\mapsto [0,1]_{C_2}+[0,0]_{C_2} \,, \\
    &[1,0,1]_{A_3}\mapsto [2,0]_{C_2} +\underbrace{[0,1]_{C_2}}_{\textnormal{acquire mass}} \,.
    \end{split}
\end{equation}
With this one obtains the massless degrees of freedom after Higgsing via
\begin{equation}
        12([1,0,0]_{A_3}+[0,0,1]_{A_3})+2([0,1,0]_{A_3})-2([0,1]_{C_2})
        =24[1,0]_{C_2}+2([0,0]_{C_2})\,.
    \label{eq:breakSU4toSp2}
\end{equation}
Hence, the theory is broken to $\mathrm{Sp}(2)$ with $12$ fundamental hypermultiplets ($24$ half-hypers) on a symplectic leaf of quaternionic dimension $1$. 

\paragraph{Higgsing $\mathrm{Sp}(2)$ to $\mathrm{SU}(2)$.}
The $\mathrm{Sp}(2)$ can be further broken to $\mathrm{Sp}(1) \cong \mathrm{SU}(2)$. In detail,
\begin{equation}
    \begin{split}
        &[1,0]_{C_2}\mapsto [1]_{A_1}+2[0]_{A_1} \,,\\
        &[2,0]_{C_2}\mapsto [2]_{A_1}+\underbrace{2[1]_{A_1}+3[0]_{A_1}}_{\textnormal{acquire mass}} \,,
    \end{split}
\end{equation}
and the Higgs mechanism yields
\begin{equation}
    \begin{split}
        12(2[1,0]_{C_2})-2(2[1]_{A_1}+3[0]_{A_1})=20[1]_{A_1}+42[0]_{A_1}\,.
    \end{split}
\end{equation}
Therefore, the remaining effective theory is a $\mathrm{Sp}(1) \cong \mathrm{SU}(2)$ gauge theory with $10$ fundamental hypermultiplets. The symplectic leaf in the Higgs branch of the full $\mathrm{SU}(4)$ theory on which the theory is broken to $\mathrm{SU}(2)$ is of quaternionic dimension $\frac{42}{2}+1=22$, where the $+1$ comes from \eqref{eq:breakSU4toSp2}. The transverse slice to the leaf corresponding to $\mathrm{Sp}(2)$ into the closure of the leaf associated to $\mathrm{SU}(2)$ is of quaternionic dimension $\frac{42}{2}=21$.

\paragraph{Higgsing $\mathrm{SU}(4)$ to $\mathrm{SU}(3)$.}
As usual, one begins by decomposing representations
\begin{equation}
    \begin{split}
        &[1,0,0]_{A_3}\mapsto [1,0]_{A_2}+[0,0]_{A_2}\,, \\
        &[0,1,0]_{A_3}\mapsto [1,0]_{A_2}+[0,1]_{A_2}\,, \\
        &[1,0,1]_{A_3}\mapsto [1,1]_{A_2}+\underbrace{[1,0]_{A_2}+[0,1]_{A_2}+[0,0]_{A_2}}_{\textnormal{acquire mass}} \,,
    \end{split}
\end{equation}
and the Higgs mechanism proceeds via
\begin{equation}
    \begin{split}
        &12([1,0,0]_{A_3}+[0,0,1]_{A_3})+2[0,1,0]_{A_3}-2([1,0]_{A_2}+[0,1]_{A_2}+[0,0]_{A_2})\\
        &=12([1,0]_{A_2}+[0,1]_{A_2})+22[0,0]_{A_2}\,.
    \end{split}
\end{equation}
As a consequence, the residual theory is $\mathrm{SU}(3)$ with $12$ fundamental hypermultiplets. The appearing $22$ massless gauge singlets
parametrise a symplectic leaf of quaternionic dimension $11$. 
\paragraph{Higgsing $\mathrm{SU}(3)$ to $\mathrm{SU}(2)$.}
Next, proceed by Higgsing the theory to $\mathrm{SU}(2)$ via
\begin{equation}
    \begin{split}
        &[1,0]_{A_2}\mapsto [1]_{A_1}+[0]_{A_1}\\
        &[1,1]_{A_2}\mapsto [2]_{A_1}+\underbrace{2[1]_{A_1}+[0]_{A_1}}_{\textnormal{acquire mass}}
    \end{split}
\end{equation}
such that the Higgs mechanism produces
\begin{equation}
    \begin{split}
        12([1,0]_{A_2}+[0,1]_{A_2})-2(2[1]_{A_1}+[0]_{A_1})=20[1]_{A_1}+22[0]_{A_1} \,.
    \end{split}
\end{equation}
Hence, this is the same theory as obtained from Higgsing the $\mathrm{Sp}(2)$, i.e.\ $\mathrm{SU}(2)$ with $10$ flavours. However, the number of massless singlets is $22$, which is different compared to the Higgsing of $\mathrm{Sp}(2)$ to $\mathrm{SU}(2)$. Hence, the transverse slice to the leaf for $\mathrm{SU}(3)$ in the closure of the leaf for $\mathrm{SU}(2)$ is of quaternionic dimension $11$.
\paragraph{Higgsing $\mathrm{SU}(2)$ to $\{1\}$.}
One straightforwardly obtains $34$ massless moduli, giving rise to a transverse slice to the leaves of $\mathrm{SU}(2)$ in the closure of the leaf of $\{1\}$ of quaternionic dimension $17$.

\begin{figure}[t]
\centering
 \scalebox{0.8}{\begin{tikzpicture}
	\begin{pgfonlayer}{nodelayer}
		\node [style=miniG] (1) at (0, 4) {};
		\node [style=miniG] (2) at (0, 0) {};
		\node [style=miniG] (3) at (2, -4) {};
		\node [style=miniG] (4) at (0, -8) {};
		\node [style=miniG] (5) at (-1.5, -6.5) {};
		\node [style=none] (6) at (-1.5, -3.25) {21};
		\node [style=none] (7) at (1.625, -2.175) {11};
		\node [style=none] (8) at (1.5, -6.25) {11};
		\node [style=none] (9) at (-1.25, -7.5) {1};
		\node [style=none] (10) at (-0.5, 2) {17};
		\node [style=none] (11) at (3.25, 4) {};
		\node [style=none] (12) at (3.75, 4) {};
		\node [style=none] (13) at (3.75, 0.15) {};
		\node [style=none] (14) at (3.25, 0.15) {};
		\node [style=none] (23) at (5, 4) {};
		\node [style=none] (24) at (5.5, 4) {};
		\node [style=none] (25) at (5.5, -4) {};
		\node [style=none] (26) at (5, -4) {};
		\node [style=none] (31) at (6.75, 4) {};
		\node [style=none] (32) at (7.25, 4) {};
		\node [style=none] (33) at (7.25, -8) {};
		\node [style=none] (34) at (6.75, -8) {};
		\node [style=none] (43) at (-2.75, 4) {};
		\node [style=none] (44) at (-3.25, 4) {};
		\node [style=none] (45) at (-3.25, -6.525) {};
		\node [style=none] (46) at (-2.75, -6.525) {};
		\node [style=gauge1] (47) at (4.5, 1.5) {};
		\node [style=flavour1] (48) at (4.5, 3) {};
		\node [style=none] (49) at (4.5, 1) {$\mathrm{SU}(2)$};
		\node [style=none] (50) at (4.5, 3.5) {10};
		\node [style=gauge1] (59) at (6.25, -1) {};
		\node [style=flavour1] (60) at (6.25, 0.5) {};
		\node [style=none] (61) at (6.25, -1.5) {$\mathrm{SU}(3)$};
		\node [style=none] (62) at (6.25, 1) {12};
		\node [style=gauge1] (67) at (8.25, -2.5) {};
		\node [style=flavour1] (68) at (8.25, -1) {};
		\node [style=none] (69) at (9, -2.5) {$\mathrm{SU}(4)$};
		\node [style=none] (70) at (8.25, -0.5) {12};
		\node [style=gauge1] (79) at (-4.25, -1.75) {};
		\node [style=flavour1] (80) at (-4.25, -0.25) {};
		\node [style=none] (81) at (-4.25, -2.25) {$\mathrm{Sp}(2)$};
		\node [style=none] (82) at (-4.25, 0.25) {12};
		\node [style=none] (83) at (8.25, -3.5) {$\Lambda^2$};
		\node [style=none] (84) at (0.5, 4) {39};
		\node [style=none] (85) at (0.5, 0) {22};
		\node [style=none] (86) at (2.5, -4) {11};
		\node [style=none] (87) at (0.5, -8) {0};
		\node [style=none] (88) at (-2, -6.5) {1};
		\node [style=none] (89) at (1.75, 4.15) {};
		\node [style=none] (90) at (2.25, 4.15) {};
		\node [style=none] (91) at (1.75, 3.9) {};
		\node [style=none] (92) at (2.25, 3.9) {};
		\node [style=none] (93) at (2.5, 4) {$\{1\}$};
		\node [style=none] (94) at (0, 5.5) {Hasse diagram};
		\node [style=none] (95) at (5.25, 5.5) {Effective theory};
	\end{pgfonlayer}
	\begin{pgfonlayer}{edgelayer}
		\draw (1) to (2);
		\draw (2) to (3);
		\draw (2) to (5);
		\draw (5) to (4);
		\draw (4) to (3);
		\draw (11.center) to (12.center);
		\draw (11.center) to (12.center);
		\draw (12.center) to (13.center);
		\draw (13.center) to (14.center);
		\draw (23.center) to (24.center);
		\draw (23.center) to (24.center);
		\draw (24.center) to (25.center);
		\draw (25.center) to (26.center);
		\draw (31.center) to (32.center);
		\draw (31.center) to (32.center);
		\draw (32.center) to (33.center);
		\draw (33.center) to (34.center);
		\draw (43.center) to (44.center);
		\draw (43.center) to (44.center);
		\draw (44.center) to (45.center);
		\draw (45.center) to (46.center);
		\draw (47) to (48);
		\draw (59) to (60);
		\draw (67) to (68);
		\draw (79) to (80);
		\draw [in=-60, out=-120, loop] (67) to ();
		\draw (89.center) to (90.center);
		\draw (90.center) to (92.center);
		\draw (91.center) to (92.center);
	\end{pgfonlayer}
\end{tikzpicture}}
\caption{Hasse diagram obtained from partial Higgsing of $\mathrm{SU}(4)$ with one hypermultiplet in the $2$nd rank antisymmetric respresentation and $12$ fundamental hypermultiplets. The effective gauge theory on each symplectic leaf in the Higgs branch is denoted by brackets. Note that the $2$nd rank antisymmetric representation of $\mathrm{SU}(4)$ is denoted by $\supset^{\Lambda^2}$. In Figure \ref{tab:Hasse_SU4_finite} the dimensions of the symplectic leaves and transverse slices are replaced by associated magnetic quivers.}\label{tab:SU(4)HasseSimplest}
\end{figure}
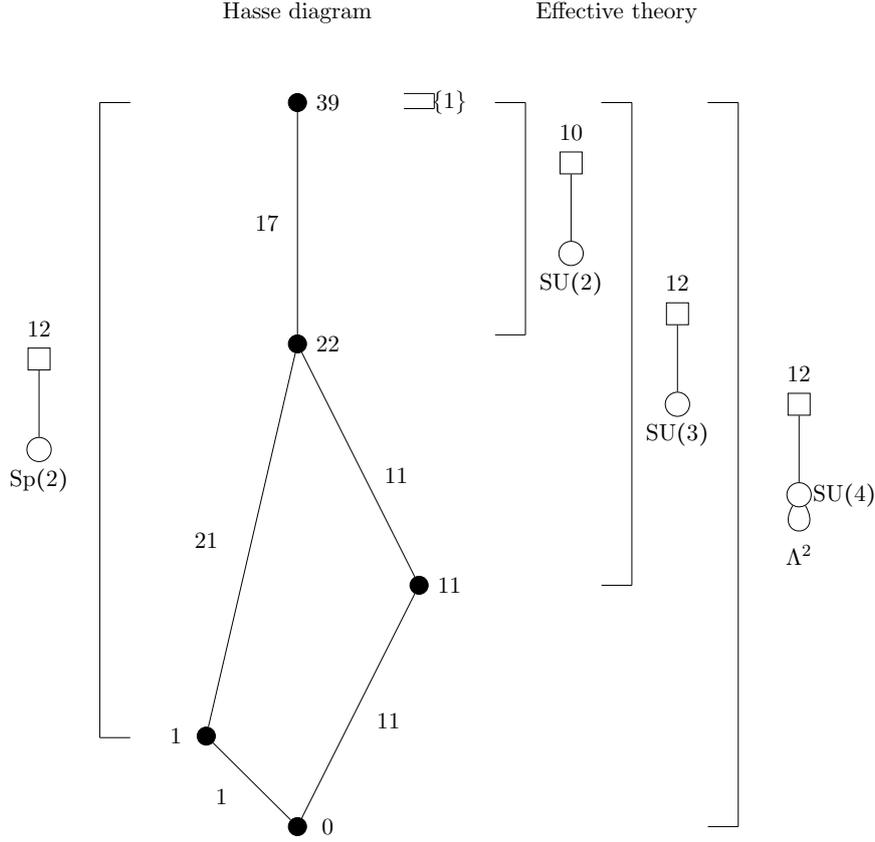

\paragraph{Higgs branch.}
The results can be summarised in a Hasse diagram, see Figure \ref{tab:SU(4)HasseSimplest}. The only conceptually new feature is the appearing bifurcation, which is a known phenomenon in Hasse diagrams, see for instance \cite{Hesselink:1976} for nilpotent orbits.
This is a consequence of the different ways one can Higgs the theory to different unbroken gauge groups. Two subsequent nodes are connected, if they can be related via a partial Higgsing.

Again, the Higgs branches of the theories obtained only describe transverse slices to any symplectic leaf in the total space. The geometry of other more general transverse slices are not accessible via this method, except for their dimensions.

\subsection{\texorpdfstring{Gauge Enhancement of $\mathrm{SU}(4)$ with a $2\mathrm{nd}$ rank antisymmetric $\Lambda^2$ and $12$ Fundamentals via Brane Construction and Magnetic Quivers}{Gauge Enhancement of SU(4) with a 2nd rank antisymmetric Lambda2 and 12 Fundamentals via Brane Construction, Magnetic Quivers and Kraft-Procesi Transitions}}\label{magQuivforsu4}

\begin{figure}[t]
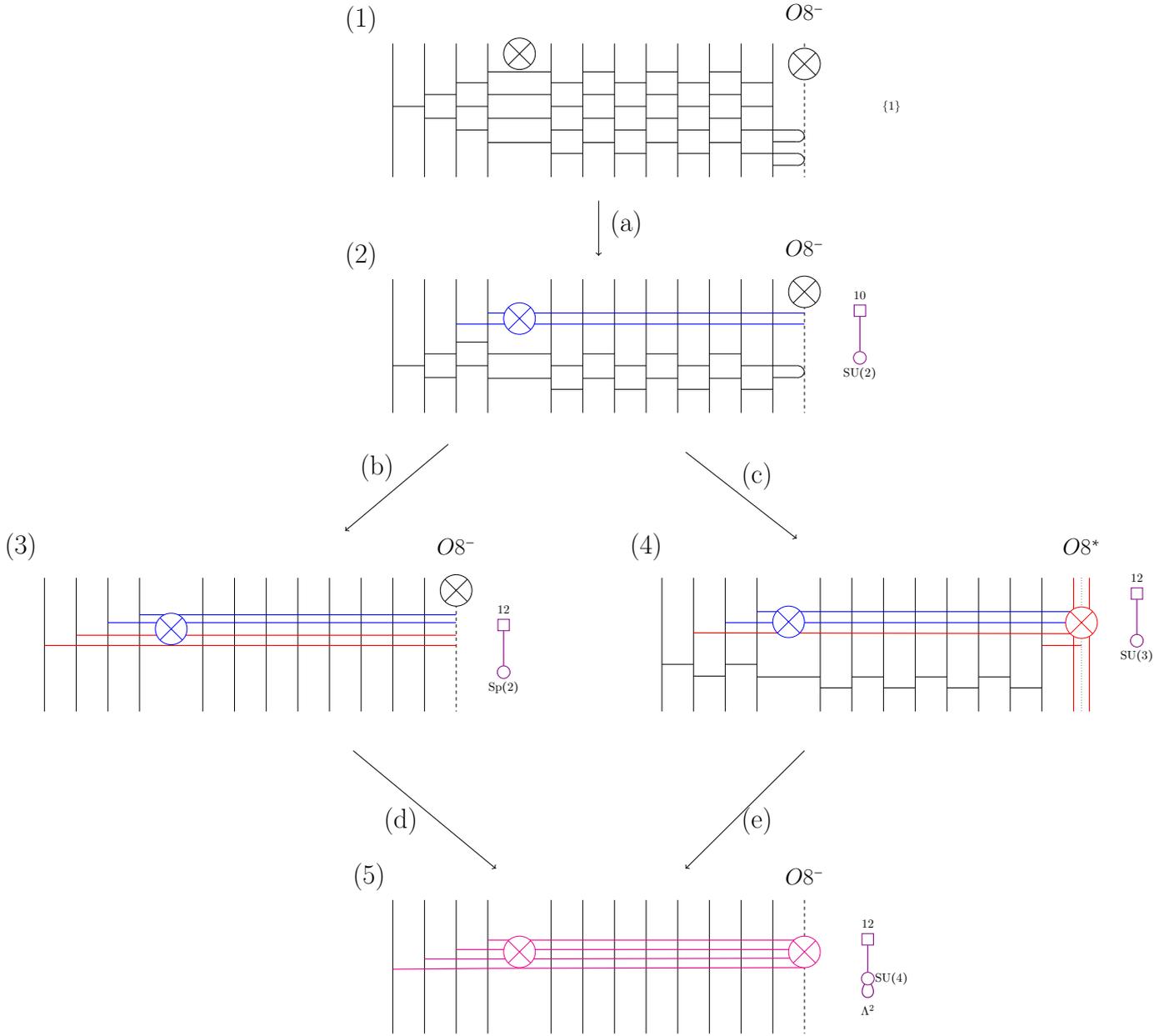

 \makebox[\textwidth][c]{
\scalebox{0.5}{
}}
\caption{Type IIA brane configurations for the different transitions in the Hasse diagram of $\mathrm{SU}(4)$ with one $2$nd rank antisymmetric and 12 fundamentals. The effective theories at each phase in the brane system are given in purple. Coloured branes are assumed to be stuck together.}
\label{fig:SU(4)12Branes}
\end{figure}

\begin{figure}[t]
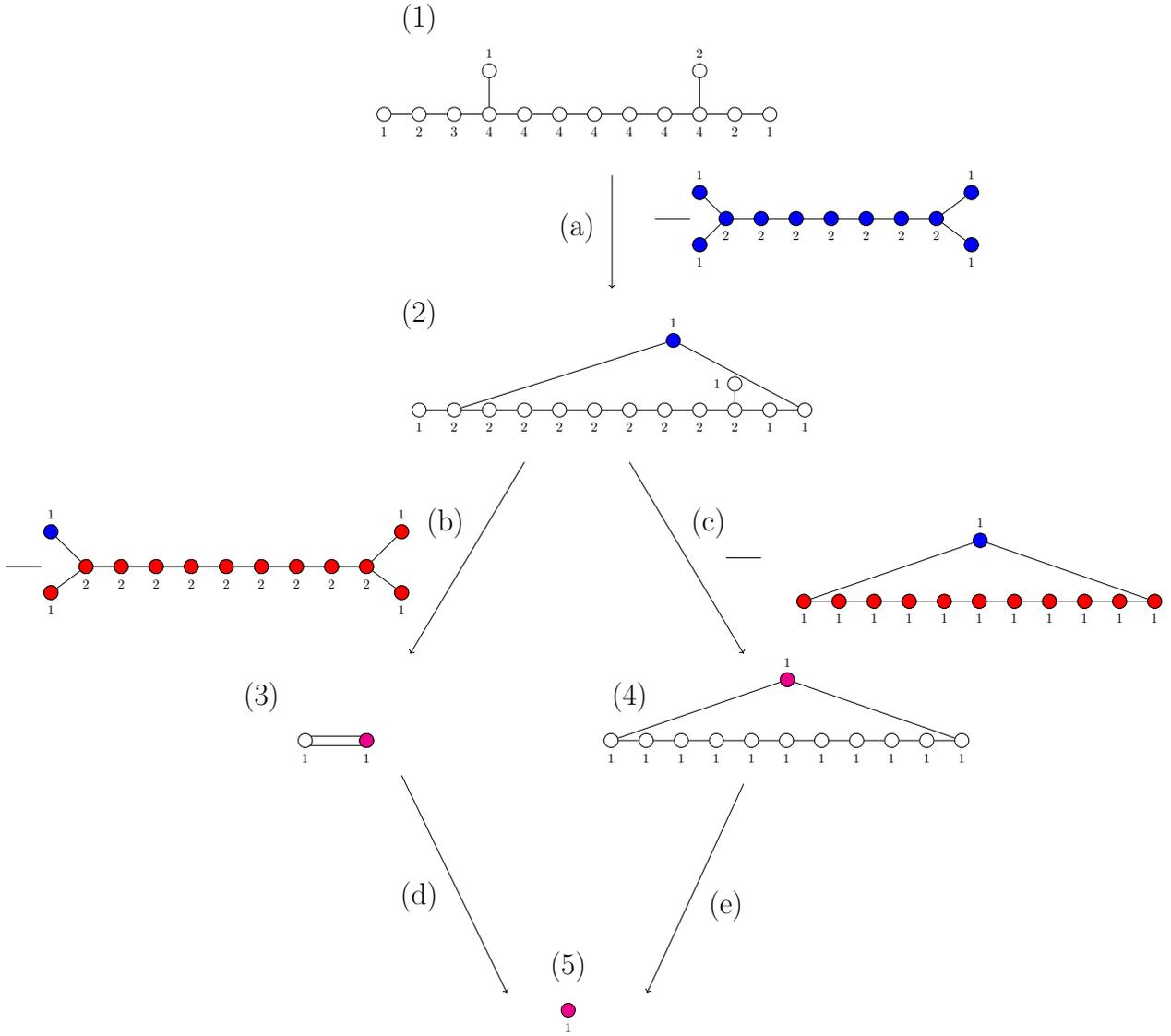

     \makebox[\textwidth][c]{
\scalebox{0.5}{%
}}
    \caption{The magnetic quivers for the different transitions of $\mathrm{SU}(4)$ with one 2nd rank antisymmetric and 12 fundamentals derived from the brane configurations. The colours in the quivers correspond to the coloured branes in Figure \ref{fig:SU(4)12Branes}. The method of reading magnetic quivers developed in \cite{Cabrera:2019izd} can be used to obtain every displayed quiver. Equivalently, one can use quiver subtraction.
    }
    \label{fig:SU(4)Magnetic}
\end{figure}

This theory can be seen as world-volume theory from a Type IIA involving D$6$, D$8$, and NS$5$ branes in the presence of an O$8^-$ plane. The brane configurations corresponding to the various mixed branches of the theory are depicted in Figure \ref{fig:SU(4)12Branes}. Subsequently, the magnetic quiver perspective is summarised in Figure \ref{fig:SU(4)Magnetic}. Some explanations are required for Figure \ref{fig:SU(4)12Branes} and \ref{fig:SU(4)Magnetic}.
\begin{itemize}
    \item The brane diagram labelled (1) in Figure \ref{fig:SU(4)12Branes} represents a generic point on the Higgs branch, i.e.\ at the largest symplectic leaf. All D$6$ branes are suspended between D$8$ branes; hence, the gauge group is completely broken. The corresponding magnetic quiver is displayed as (1) in Figure \ref{fig:SU(4)Magnetic}.
    \item In the brane system (2) of Figure \ref{fig:SU(4)12Branes} the blue D6 branes and the NS5 branes are aligned and the gauge group becomes enhanced to $\mathrm{SU}(2)$. The new effective theory has 10 hypermultiplets charged under the fundamental representation of $\mathrm{SU}(2)$ and 22 remaining neutral hypermultiplets, given by the D6s and the NS5 that are not aligned. In analogy with Sections \ref{higgsforsu3} and \ref{branesclass}, we conjecture that the 22 neutral hypermultiplets parametrise a symplectic leaf of quaternionic dimension 22. Similarly, we conjecture that the transverse slice between this symplectic leaf and the full space (the Higgs branch of $\mathrm{SU}(4)$ with a $2\mathrm{nd}$ rank antisymmetric $\Lambda^2$ and $12$ fundamentals) can be obtained as the Higgs branch of $\mathrm{SU}(2)$ with 10 flavours (the theory obtained from the aligned D6s and NS5).  The transverse slice is also obtained as the $3$d $\mathcal{N}=4$ Coulomb branch of the magnetic quiver depicted as (a) in Figure \ref{fig:SU(4)Magnetic}. This is therefore predicted to be a Kraft-Procesi transition of type $d_{10}$. Analogously, we propose that the 22 dimensional symplectic leaf can be characterized by giving its closure as the $3$d $\mathcal{N}=4$ Coulomb branch of another magnetic quiver. This is read off from the black D6 branes in (2) of Figure \ref{fig:SU(4)12Branes} with additionally counting the blue part as a single $\mathrm{U}(1)$ node. The magnetic quiver is shown as (2) in Figure \ref{fig:SU(4)Magnetic}. From this point in the brane system one can move to two different configurations representing different gauge enhancements.
    \item (3) in Figure \ref{fig:SU(4)12Branes} depicts the alignment of the remaining D6 branes (red) with the already aligned (blue) D6 branes. The gauge group is enhanced to $\mathrm{Sp}(2)$ and has 12 massless flavours. There is also 1 neutral hypermultiplet remaining. Therefore, we propose that this neutral hypermultiplet parametrizes a new symplectic leaf, of quaternionic dimension 1. The closure of the leaf is given as the $3d~\mathcal N=4$ Coulomb branch of the magnetic quiver (3) in Figure \ref{fig:SU(4)Magnetic}. The transverse slice between this symplectic leaf and the full space would be the Higgs branch of the $6d$ theory with $\mathrm{Sp}(2)$ gauge group and 12 flavours. The elementary slice (or transverse slice between the symplectic leaf (3) and the closure of symplectic leaf (2)) is given as the $3d~\mathcal N=4$ Coulomb branch of the magnetic quiver (b) in Figure \ref{fig:SU(4)Magnetic}. This is therefore a Kraft-Procesi transition of type $d_{12}$. Here, the part that was already aligned (blue) appears as a single $\mathrm{U}(1)$ and the part that is now being aligned (red) yields the rest of the magnetic quiver for the transverse slice.
    \item The arrow from (3) to (5) in Figure \ref{fig:SU(4)12Branes} shows the realignment of the black NS5 brane to make one piece. One is at the origin of the Higgs branch. The magnetic quivers corresponding to the elementary slice and the closure of the new symplectic leaf are (d) and (5) respectively. Therefore, this corresponds to a Kraft-Procesi transition of type $a_1$.
    \item From (2) in Figure \ref{fig:SU(4)12Branes} we could also realign the branes to realise a different pattern of gauge symmetry enhancement. Going to (4) in Figure \ref{fig:SU(4)12Branes}, one aligns the red D6 and NS5 branes with the already aligned blue branes. However, this process is rather non-trivial as it involves splitting the $O8^-$ brane into 2 half D8 branes and a $O8^*$. For details, the reader is referred to \cite{Cabrera:2019izd}. The elementary slice and the closure of the new symplectic leaf are given by magnetic quivers displayed as (c) and (4) in Figure \ref{fig:SU(4)Magnetic}, respectively. Therefore, this is a Kraft-Procesi transition of type $a_{11}$. The gauge group enhancement is to $\mathrm{SU}(3)$ with 12 flavours and 11 neutral hypermultiplets.
    \item Finally from (4) in Figure \ref{fig:SU(4)12Branes} one can move to (5) to go to the origin of the Higgs branch. The elementary slice is depicted as a magnetic quiver (e) in Figure \ref{fig:SU(4)Magnetic}. This corresponds to an $a_{11}$ Kraft-Procesi transition.
\end{itemize}

\clearpage

\clearpage

\begin{figure}[t]
	\centering
	\begin{tabular}{m{2.6cm} m{5.4cm}}
	Hasse diagram  & Magnetic quiver 
	 \\ 
\begin{tikzpicture}[node distance=20pt]
	\tikzstyle{hasse} = [circle, fill,inner sep=2pt];
		\node at (-0.7,-0.5) [] (1a) [] {};
		\node [hasse] at (0,-0.5) [] (1b) [label=left:\footnotesize{$39$}] {};
		\node at (0.7,-0.5) [] (1c) [] {};
		\node  (2a) [below of=1a] {};
		\node [] (2b) [below of=1b] {};
		\node (2c) [below of=1c] {};
		\node [] (3a) [below of=2a] {};
		\node  (3b) [below of=2b] {};
		\node (3c) [below of=2c] {};
		\node (4a) [below of=3a] {};
		\node [hasse] (4b)[label=left:\footnotesize{$22$},below of=3b] {};
		\node (4c) [below of=3c] {};
		\node [] (5a) [below of=4a] {};
		\node (5b) [below of=4b] {};
		\node  (5c) [below of=4c] {};
		\node  (6a) [below of=5a] {};
		\node  (6b) [below of=5b] {};
		\node (6c) [below of=5c] {};
		\node  (7a) [below of=6a] {};
		\node (7b) [below of=6b] {};
		\node (7c) [below of=6c] {}; 
		\node (8a) [below of=7a] {};
		\node  (8b) [below of=7b] {};
	        \node [hasse]  (8c) [label=right:\footnotesize{$11$},below of=7c] {};
	        \node (9a) [below of=8a] {};
		\node (9b) [below of=8b] {};
	        \node (9c) [below of=8c] {};
	        \node [hasse] (10a) [label=left:\footnotesize{$1$},below of=9a] {};
		\node (10b) [below of=9b] {};
	        \node (10c) [below of=9c] {};
	          \node (11a) [below of=10a] {};
		\node (11b) [below of=10b] {};
	        \node (11c) [below of=10c] {};
	         \node (12a) [below of=11a] {};
	         \node (13a) [below of=12a] {};
		         
		\node [hasse] (12b) [label=left:\footnotesize{$0$},below of=11b] {};
	        \node (12c) [below of=11c] {};
		\draw (1b) edge [] node[label=left:\footnotesize{$d_{10}$}] {} (4b)
		        (4b) edge [] node[label=right:\footnotesize{$a_{11}$}] {} (8c)
			(12b) edge [] node[label=right:\footnotesize{$a_{11}$}] {} (8c)
			(10a) edge [] node[label=left:\footnotesize{$c_1$}] {} (12b)
			(4b) edge [] node[label=left:\footnotesize{$d_{12}$}] {} (10a);
	\end{tikzpicture}
	&
\thead{ $\node{}1 -\node{}2 - \node{}3 -\node{\topnode{}1}4 -\node{}4-\node{}4-\node{}4-\node{}4-\node{}4-\node{\topnode{}2}4-\node{}2-\node{}1$ 
\\ \\ $\underbrace{\node{}1 -
 \overset{\overset{\displaystyle\overset 1 \circ}
	{\diagup~~~~~\diagdown}
	}
{\node{\topnode{}1}2  -\dots -\node{\topnode{}1}2}
-\node{}1}_{11}$  
 \\ \\ \\ $\underbrace{
 \overset{\overset{\displaystyle\overset 1 \circ}
	{\diagup~~~~~\diagdown}
	}
	{ \node{}{1}-\dots-\node{}{1} }
	}_{11}$ 
 \\ \\ $\node{}1 = \node{}1$ \\ \\
  $\node{}1$ \\ \\ \\}
	\end{tabular}
	\caption{Hasse diagram with magnetic quivers representing the closures of symplectic leaves for the classical Higgs branch of $\mathrm{SU}(4)$ with one 2nd rank antisymmetric and 12 fundamentals. The electric theories in Figure \ref{tab:SU(4)HasseSimplest} are suppressed. They reappear in Figure \ref{fig:subdiagramSU4}. 
	}\label{tab:Hasse_SU4_finite}
\end{figure}
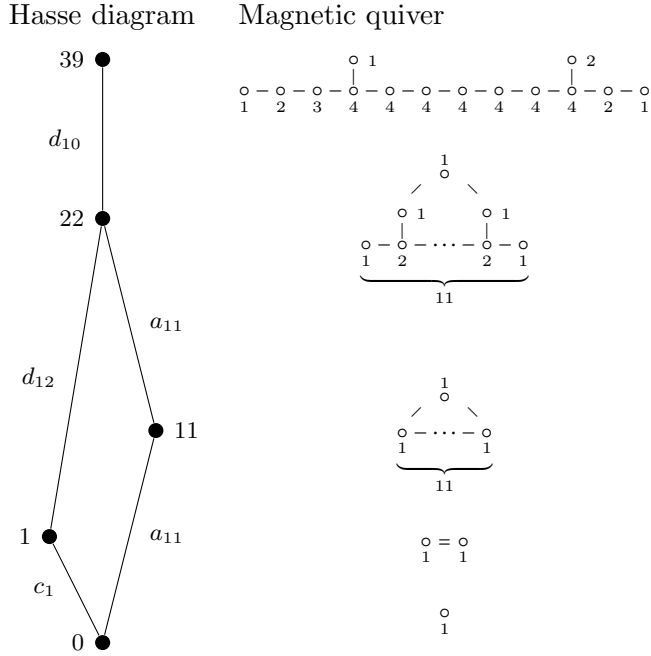

The Hasse diagram of Figure \ref{tab:Hasse_SU4_finite} now summarises all the insight gained from the brane and quiver perspective. Comparing to the prediction of the Higgs mechanism of Section \ref{higgsforsu4}, the Hasse diagrams agree, but refined version is found.
In detail, each point in the Hasse diagram is now associated with a magnetic quiver, whose space of monopole operators describes the closure of the corresponding symplectic leaf. Again, this is new information compared to the Higgsing approach in Figure \ref{tab:SU(4)HasseSimplest}, where only the dimensions had been determined.
Moreover, the working assumption of KP transitions allows to provide a magnetic quiver for each minimal transition such that the minimal transverse slices are geometrically describe as Coulomb branches. 

\section{Hasse diagrams: Generalisations and Comments}
\label{sec:Hasse}
In this section, the Hasse diagrams are explained in more detail.
The above section concluded by finding agreement between the \emph{predicted Hasse diagram} from partial Higgsing and the \emph{computed Hasse diagram} by assuming KP transitions only. The latter provides a refinement of the former.

Given a gauge group $G$ with enough fundamental hypermultiplets, the Higgs mechanism allows to partially break $G$ to some of its subgroups. The space of VEVs decomposes into symplectic leaves, where a leaf is defined as space of VEVs such that $G$ is broken to one particular subgroup. Choosing a point on a fixed leaf, the residual (massless) theory comprises the unbroken gauge group with some charged matter as well as some uncharged massless scalar moduli. These massless moduli parametrise the leaf, while the residual gauge theory, denoted as \emph{effective theory}, lives over any point in that particular leaf.  

Geometrically, the leaves themselves are mutually disjoint and posses no ordering. A partial ordering arises only upon taking closures of the leaves. Or put in physical terms, the closure of a leaf is given by the union of all leaves such that their residual gauge groups can be Higgsed to gauge group of the leaf one is taking the closure of. It is important to note that only the closure of a leaf is a symplectic singularity, while the leaf itself is not. This does not come as a surprise, given the known structure of nilpotent orbits and their closures.

Given a symplectic leaf, the effective or residual theory has a transparent meaning: the Higgs branch of the effective theory is the transverse slice of the leaf inside the closure of the total space. Hence, Figure \ref{tab:HiggsSimplestHasse} and \ref{tab:SU(4)HasseSimplest} the Higgs branch is seen the transverse slice that starts at the top of the Hasse diagram and extends all the way to the node labelled by the unbroken gauge group. Likewise, the closure of the leaf is everything from the bottom to that node.

However, the Higgs mechanism perspective provides only partial answers, because only transverse slices that start at the top are accessible. In order words, only transverse slices to the total space are given by Higgs branches of Higgsed theories. All other slices are not accessible. In addition, closures of symplectic leaves are accounted for by their dimension.

On the other hand, magnetic quivers and quiver subtraction allow to provide all details of the Hasse diagram. The starting point is slightly different as one begins with the magnetic quiver for the full Higgs branch and proceeds to trace out the Hasse diagram via minimal transitions. Thus, in Figure \ref{tab:SU(3)6HasseSub} and \ref{tab:Hasse_SU4_finite} the magnetic quiver on top describes the entire moduli space, which equals the full Higgs branch of the un-Higgsed theory. Proceeding with minimal subtractions generates a magnetic quiver for each node whose Coulomb branch describes the closure of the corresponding symplectic leaf, i.e.\ the space from the bottom of the Hasse diagram to the node.
Moreover, the minimal transitions have magnetic quivers associated to them, such that, the Coulomb branch thereof matches the transverse slice between the two nodes. Hence, the geometry of the minimal transitions is completely determined.

As demonstrated in the section below, one can now derive the geometric description of \emph{all} transverse slices in terms of a Higgs branch. Hence, completing the Hasse diagram.

\subsection{Transverse Slices}
Given a Hasse diagram, the magnetic quiver approach provides all the minimal transitions. As any minimal transition has an associated magnetic quiver, one can equivalently describe the transverse slice as Higgs branch of an electric theory. Moreover, any transverse slice (meaning between a leaf $\mathcal{L}'$ and a closure of a larger leaf $\mathcal{L}$ containing it, i.e.\ $\mathcal{L}'\subset \overline{\mathcal{L}}$) has to be expressible via a Higgs branch. Here, the previous examples are considered again to inspect this in more detail.

\begin{figure}[t]
\centering
\begin{tikzpicture}
	\begin{pgfonlayer}{nodelayer}
		\node [style=miniG] (0) at (0, 3) {};
		\node [style=miniG] (1) at (0, 0) {};
		\node [style=miniG] (2) at (0, -3) {};
		\node [style=none] (3) at (0.5, 3) {};
		\node [style=none] (4) at (1, 3) {};
		\node [style=none] (5) at (1, -3) {};
		\node [style=none] (6) at (0.5, -3) {};
		\node [style=none] (7) at (1, 0.15) {};
		\node [style=none] (8) at (1, -0.125) {};
		\node [style=none] (9) at (0.5, -0.125) {};
		\node [style=none] (10) at (0.5, 0.15) {};
		\node [style=none] (11) at (-0.75, 1.5) {$d_4$};
		\node [style=none] (12) at (-0.75, -1.5) {$a_5$};
		\node [style=gauge1] (13) at (1.75, 1) {};
		\node [style=flavour1] (14) at (1.75, 2.5) {};
		\node [style=gauge1] (15) at (1.75, -2.25) {};
		\node [style=flavour1] (16) at (1.75, -0.75) {};
		\node [style=none] (17) at (1.75, 0.5) {$\mathrm{SU}(2)$};
		\node [style=none] (18) at (1.75, 3) {4};
		\node [style=none] (19) at (1.75, -2.75) {$\mathrm{U}(1)$};
		\node [style=none] (20) at (1.75, -0.25) {6};
		\node [style=none] (21) at (3, 3) {};
		\node [style=none] (22) at (3.5, 3) {};
		\node [style=none] (23) at (3.5, -3.025) {};
		\node [style=none] (24) at (3, -3.025) {};
		\node [style=gauge1] (25) at (4.25, 0) {};
		\node [style=flavour1] (26) at (4.25, 1.5) {};
		\node [style=none] (27) at (4.25, -0.5) {$\mathrm{SU}(3)$};
		\node [style=none] (28) at (4.25, 2) {6};
	\end{pgfonlayer}
	\begin{pgfonlayer}{edgelayer}
		\draw (3.center) to (4.center);
		\draw (3.center) to (4.center);
		\draw (4.center) to (7.center);
		\draw (7.center) to (10.center);
		\draw (9.center) to (8.center);
		\draw (8.center) to (5.center);
		\draw (5.center) to (6.center);
		\draw (0) to (1);
		\draw (1) to (2);
		\draw (13) to (14);
		\draw (16) to (15);
		\draw (21.center) to (22.center);
		\draw (21.center) to (22.center);
		\draw (22.center) to (23.center);
		\draw (23.center) to (24.center);
		\draw (25) to (26);
	\end{pgfonlayer}
\end{tikzpicture}
\caption{Hasse diagram for $\mathrm{SU}(3)$ with 6 fundamentals. Each subdiagram enclosed in a bracket corresponds to a symplectic singularity. It is expected that  the Higgs branches of these theories are the corresponding transverse slices. The elementary slices define Kraft-Procesi transitions that are labelled in the edges of the Hasse diagram.}
\label{fig:subdiagramSU3}
\end{figure}
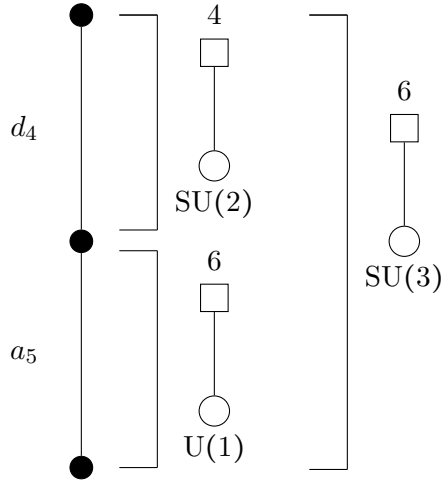

\begin{figure}[t]
\centering
\scalebox{0.8}{
\begin{tikzpicture}
	\begin{pgfonlayer}{nodelayer}
		\node [style=miniG] (1) at (0, 4) {};
		\node [style=miniG] (2) at (0, 0) {};
		\node [style=miniG] (3) at (2, -4) {};
		\node [style=miniG] (4) at (0, -8) {};
		\node [style=miniG] (5) at (-1.5, -6.5) {};
		\node [style=none] (6) at (-1.5, -3.25) {$d_{12}$};
		\node [style=none] (7) at (1.625, -2.175) {$a_{11}$};
		\node [style=none] (8) at (1.5, -6.25) {$a_{11}$};
		\node [style=none] (9) at (-1.25, -7.5) {$c_1$};
		\node [style=none] (10) at (0.5, 2) {$d_{10}$};
		\node [style=none] (11) at (2.75, 4) {};
		\node [style=none] (12) at (3.25, 4) {};
		\node [style=none] (13) at (3.25, 0.15) {};
		\node [style=none] (14) at (2.75, 0.15) {};
		\node [style=none] (15) at (2.75, 0) {};
		\node [style=none] (16) at (3.25, 0) {};
		\node [style=none] (17) at (3.25, -3.975) {};
		\node [style=none] (18) at (2.75, -3.975) {};
		\node [style=none] (19) at (2.75, -4.125) {};
		\node [style=none] (20) at (3.25, -4.125) {};
		\node [style=none] (21) at (3.25, -7.975) {};
		\node [style=none] (22) at (2.75, -7.975) {};
		\node [style=none] (23) at (5.5, 4) {};
		\node [style=none] (24) at (6, 4) {};
		\node [style=none] (25) at (6, -4) {};
		\node [style=none] (26) at (5.5, -4) {};
		\node [style=none] (27) at (6.5, 0) {};
		\node [style=none] (28) at (7, 0) {};
		\node [style=none] (29) at (7, -8) {};
		\node [style=none] (30) at (6.5, -8) {};
		\node [style=none] (31) at (8.25, 4) {};
		\node [style=none] (32) at (8.75, 4) {};
		\node [style=none] (33) at (8.75, -8) {};
		\node [style=none] (34) at (8.25, -8) {};
		\node [style=none] (35) at (-2.25, 0) {};
		\node [style=none] (36) at (-2.75, 0) {};
		\node [style=none] (37) at (-2.75, -6.275) {};
		\node [style=none] (38) at (-2.25, -6.275) {};
		\node [style=none] (39) at (-2.25, -6.5) {};
		\node [style=none] (40) at (-2.75, -6.5) {};
		\node [style=none] (41) at (-2.75, -8.025) {};
		\node [style=none] (42) at (-2.25, -8.025) {};
		\node [style=none] (43) at (-4.75, 4) {};
		\node [style=none] (44) at (-5.25, 4) {};
		\node [style=none] (45) at (-5.25, -6.275) {};
		\node [style=none] (46) at (-4.75, -6.275) {};
		\node [style=gauge1] (47) at (4, 1.5) {};
		\node [style=flavour1] (48) at (4, 3) {};
		\node [style=none] (49) at (4, 1) {$\mathrm{SU}(2)$};
		\node [style=none] (50) at (4, 3.5) {10};
		\node [style=gauge1] (51) at (4, -2.75) {};
		\node [style=flavour1] (52) at (4, -1.25) {};
		\node [style=none] (53) at (4, -3.25) {$\mathrm{U}(1)$};
		\node [style=none] (54) at (4, -0.75) {12};
		\node [style=gauge1] (55) at (4, -6.75) {};
		\node [style=flavour1] (56) at (4, -5.25) {};
		\node [style=none] (57) at (4, -7.25) {$\mathrm{U}(1)$};
		\node [style=none] (58) at (4, -4.75) {12};
		\node [style=gauge1] (59) at (6.75, 2) {};
		\node [style=flavour1] (60) at (6.75, 3.5) {};
		\node [style=none] (61) at (6.75, 1.5) {$\mathrm{SU}(3)$};
		\node [style=none] (62) at (6.75, 4) {12};
		\node [style=gauge1] (67) at (9.75, -2.25) {};
		\node [style=flavour1] (68) at (9.75, -0.75) {};
		\node [style=none] (69) at (10.5, -2.25) {$\mathrm{SU}(4)$};
		\node [style=none] (70) at (9.75, -0.25) {12};
		\node [style=gauge1] (71) at (-3.75, -3.75) {};
		\node [style=flavour1] (72) at (-3.75, -2.25) {};
		\node [style=none] (73) at (-3.75, -4.25) {$\mathrm{SU}(2)$};
		\node [style=none] (74) at (-3.75, -1.75) {12};
		\node [style=gauge1] (75) at (-3.75, -8) {};
		\node [style=flavour1] (76) at (-3.75, -6.75) {};
		\node [style=none] (77) at (-3.75, -8.5) {$\mathrm{O}(1)$};
		\node [style=none] (78) at (-3.75, -6.25) {$\mathrm{Sp}(1)$};
		\node [style=gauge1] (79) at (-6.25, -1.75) {};
		\node [style=flavour1] (80) at (-6.25, -0.25) {};
		\node [style=none] (81) at (-6.25, -2.25) {$\mathrm{Sp}(2)$};
		\node [style=none] (82) at (-6.25, 0.25) {12};
		\node [style=none] (83) at (9.75, -3.25) {$\Lambda^2$};
		\node [style=none] (84) at (8, -4.25) {$\mathcal{T}$};
	\end{pgfonlayer}
	\begin{pgfonlayer}{edgelayer}
		\draw (1) to (2);
		\draw (2) to (3);
		\draw (2) to (5);
		\draw (5) to (4);
		\draw (4) to (3);
		\draw (11.center) to (12.center);
		\draw (11.center) to (12.center);
		\draw (12.center) to (13.center);
		\draw (13.center) to (14.center);
		\draw (15.center) to (16.center);
		\draw (15.center) to (16.center);
		\draw (16.center) to (17.center);
		\draw (17.center) to (18.center);
		\draw (19.center) to (20.center);
		\draw (19.center) to (20.center);
		\draw (20.center) to (21.center);
		\draw (21.center) to (22.center);
		\draw (23.center) to (24.center);
		\draw (23.center) to (24.center);
		\draw (24.center) to (25.center);
		\draw (25.center) to (26.center);
		\draw (27.center) to (28.center);
		\draw (27.center) to (28.center);
		\draw (28.center) to (29.center);
		\draw (29.center) to (30.center);
		\draw (31.center) to (32.center);
		\draw (31.center) to (32.center);
		\draw (32.center) to (33.center);
		\draw (33.center) to (34.center);
		\draw (35.center) to (36.center);
		\draw (35.center) to (36.center);
		\draw (36.center) to (37.center);
		\draw (37.center) to (38.center);
		\draw (39.center) to (40.center);
		\draw (39.center) to (40.center);
		\draw (40.center) to (41.center);
		\draw (41.center) to (42.center);
		\draw (43.center) to (44.center);
		\draw (43.center) to (44.center);
		\draw (44.center) to (45.center);
		\draw (45.center) to (46.center);
		\draw (47) to (48);
		\draw (51) to (52);
		\draw (55) to (56);
		\draw (59) to (60);
		\draw (67) to (68);
		\draw (71) to (72);
		\draw (75) to (76);
		\draw (79) to (80);
		\draw [in=-60, out=-120, loop] (67) to ();
	\end{pgfonlayer}
\end{tikzpicture}}
\caption{Hasse diagram for 6d $\mathrm{SU}(4)$ with one 2nd rank antisymmetric and 12 fundamentals with electric quivers associated to every subdiagram between two points in the Hasse diagram. The $\mathcal{T}$ theory is a $U(2)$ gauge theory with twelve fundamentals of $\mathrm{SU}(2)$ with $\mathrm{U}(1)$ charge $1$ and two $\mathrm{SU}(2)$ singlets of $\mathrm{U}(1)$ charge $2$. }
\label{fig:subdiagramSU4}
\end{figure}
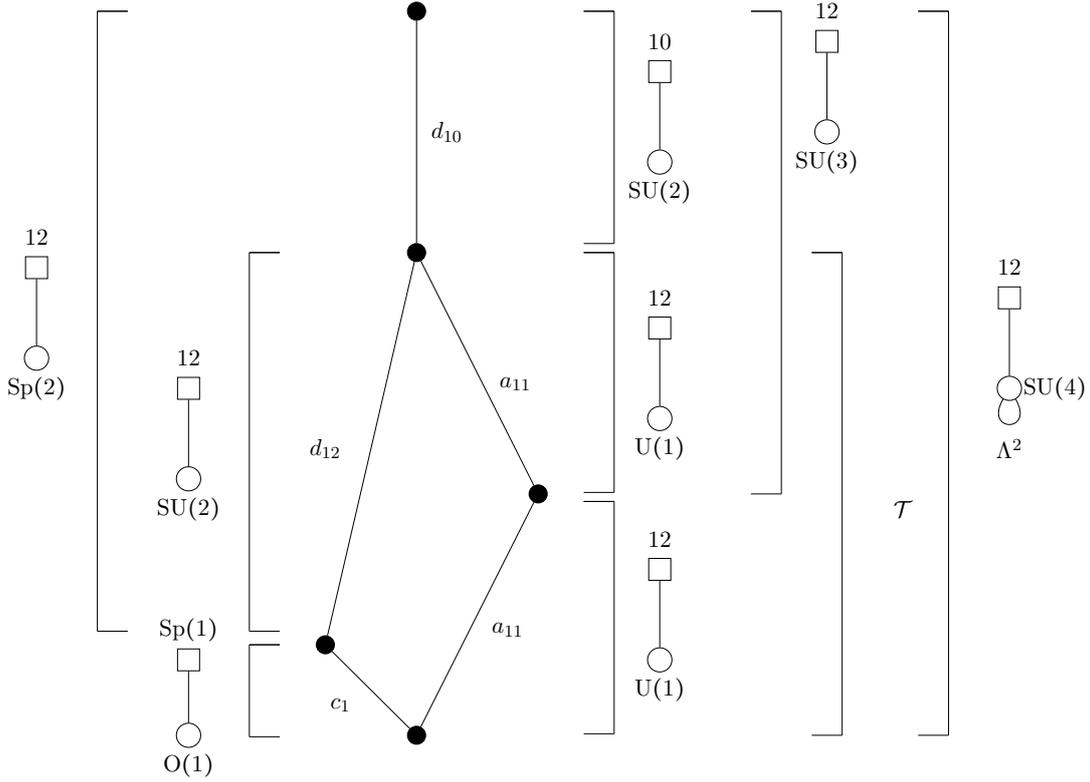

\paragraph{5d $\mathrm{SU}(3)$ with $6$ fundamentals.}
From the insights of the sections above, one associates different theories to the Hasse diagram in Figure \ref{fig:subdiagramSU3}. Trivially, the entire diagram corresponds to the original theory. While the diagram associated to the $d_4$ transition yields the $\mathrm{SU}(2)$ theory with $4$ fundamentals obtained via partial Higgsing, the theory for the $a_5$ transition cannot be obtained from a Higgsing of the original theory. Nevertheless, from the associated magnetic quiver it follows that the electric theory yields SQED with $5$ fundamentals. One notes in particular that the $\mathrm{U}(1)$ gauge group is also a consequence of the structure of the Hasse diagram, i.e.\ the $\mathrm{U}(1)$ is the commutant of $\mathrm{SU}(2)$ inside $\mathrm{SU}(3)$.
\paragraph{6d $\mathrm{SU}(4)$ with 2nd rank antisymmetric and 12 fundamentals.}
Returning to the Hasse diagram for the $\mathrm{SU}(4)$ example, it is straightforward to identify the theories that can be obtained from partial Higgsing of the original theory, see Figure \ref{tab:SU(4)HasseSimplest}. Starting from the top of Figure \ref{fig:subdiagramSU4}, the $d_{10}$ transition corresponds to the $\mathrm{SU}(2)$ theory with $10$ fundamentals that is obtained via partial Higgsing. Likewise, the combined transition $d_{10}+a_{11}$ is the Hasse diagram of $\mathrm{SU}(3)$ with $12$ fundamentals obtained via partial Higgsing. Next, the combination $d_{10}+d_{12}$ is the Hasse diagram of the $\mathrm{Sp}(2)$ theory with 12 flavours, which follows from partial Higgsing. Lastly, the entire graph returns the full $\mathrm{SU}(4)$ theory.

The transitions that do not start at the top are more difficult to extract as they are not obtainable via partial Higgsing. Nevertheless, each of the $a_{11}$ transitions is associated to SQED with 12 fundamentals. One way to see this is by focusing on the $a_{11}$ transition at the bottom and observe that the gauge group has to be the commutant of $\mathrm{SU}(3)$ inside $\mathrm{SU}(4)$, which yields a $\mathrm{U}(1)$. The number of flavours is then obtain from the dimensionality of the transition. Alternatively, one simply inspects the magnetic quiver for this minimal transition, which gives the same conclusion.

Similarly, the elementary slices $d_{12}$ and $c_1$ have known electric theories associated to them, which are read off from the associated magnetic quivers shown in Figure \ref{fig:SU(4)Magnetic}. Returning to the Hasse diagram in Figure \ref{fig:subdiagramSU4}, the theory corresponding to the combined $a_{11}+a_{11}$ or $c_1+d_{12}$ transition denoted $\mathcal{T}$ is a $\mathrm{U}(2)$ gauge theory (because the commutant of $\mathrm{SU}(2)$ inside $\mathrm{SU}(4)$ is $\mathrm{U}(2)$) with 12 fundamental hypermultiplets and 2 $\mathrm{SU}(2)$ singlets with charge 2 under the $\mathrm{U}(1)$.
\clearpage

\subsection{Global Symmetry}
\label{subsectionGlobalSymmetry}
We observe that the non-abelian part of the global symmetry of the Higgs branch always contains the global symmetry of the transverse slices to the origin. For instance, $\mathrm{SU}(3)$ with $6$ fundamentals has a non-abelian global symmetry of $\mathrm{SU}(6)$, which is reproduced by minimal $a_5$ transition to the origin in Figure \ref{tab:SU(3)6HasseSub}. Likewise, $\mathrm{SU}(4)$ with one $2$nd rank antisymmetric hypermultiplet and 12 fundamental hypermultiplets has a global $\mathrm{Sp}(1) \times \mathrm{SU}(12)$ symmetry, where the $\mathrm{Sp}(1)$ originates from the $2$nd antisymmetric representation of $\mathrm{SU}(4)$ which is real. The Hasse diagram of Figure \ref{tab:Hasse_SU4_finite} shows that there are two minimal transitions connected to the origin, one $c_1$ and one $a_{11}$; thus, matching the non-abelian symmetry.

Hence, if we know the global symmetry of the Higgs branch of a theory, we can use this information to give an upper bound to the number of lines and the type of the elementary slices at the bottom of the Hasse diagram. This is used for the computation of the Hasse diagram of $G_2$ in Section \ref{subsectionSimpleGaugeGroup} and the Hasse diagram in Section \ref{subsection6d}.

However, abelian $\mathrm{U}(1)$ factors are more cumbersome and it is currently not clear how to determine these from the Hasse diagram.

\subsection{Hasse Diagrams for Multiple Cones}
\label{subsectionMultiple}
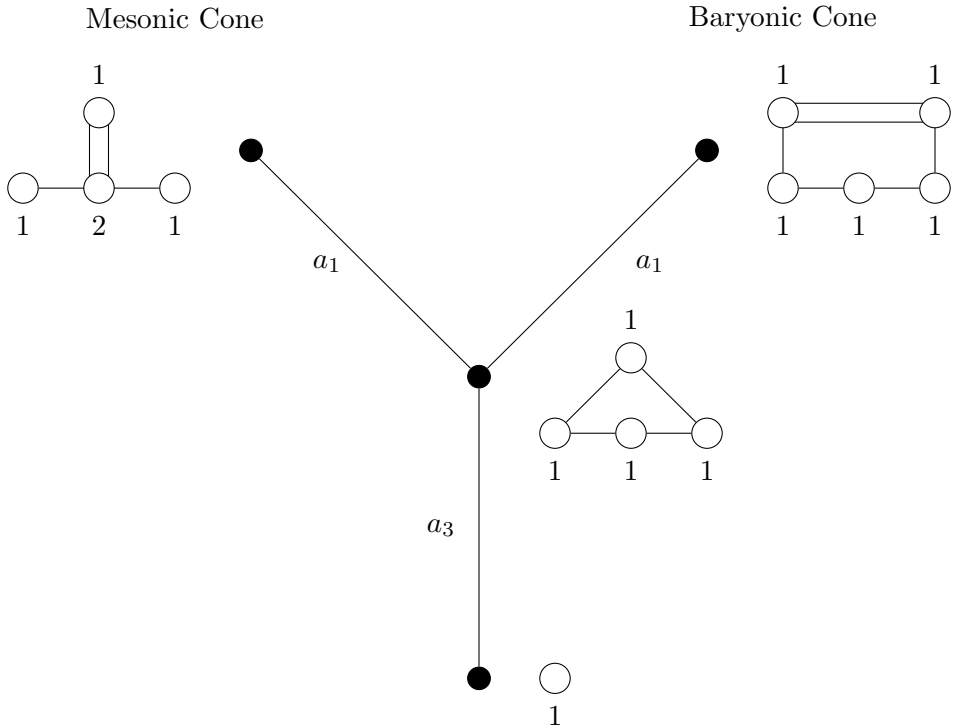
\begin{figure}[t]
\centering
\scalebox{1.0}{\begin{tikzpicture}
	\begin{pgfonlayer}{nodelayer}
		\node [style=gauge1] (3) at (-5, 3.5) {};
		\node [style=gauge1] (4) at (-5, 2.5) {};
		\node [style=gauge1] (5) at (-4, 2.5) {};
		\node [style=gauge1] (6) at (-6, 2.5) {};
		\node [style=none] (7) at (-5.125, 3.5) {};
		\node [style=none] (8) at (-4.875, 3.5) {};
		\node [style=none] (9) at (-5.125, 2.5) {};
		\node [style=none] (10) at (-4.875, 2.5) {};
		\node [style=none] (11) at (-6, 2) {1};
		\node [style=none] (12) at (-5, 2) {2};
		\node [style=none] (13) at (-4, 2) {1};
		\node [style=none] (14) at (-5, 4) {1};
		\node [style=gauge1] (15) at (4, 2.5) {};
		\node [style=gauge1] (16) at (5, 2.5) {};
		\node [style=gauge1] (17) at (6, 2.5) {};
		\node [style=gauge1] (18) at (6, 3.5) {};
		\node [style=gauge1] (19) at (4, 3.5) {};
		\node [style=none] (20) at (4, 3.625) {};
		\node [style=none] (21) at (4, 3.375) {};
		\node [style=none] (22) at (6, 3.625) {};
		\node [style=none] (23) at (6, 3.375) {};
		\node [style=none] (24) at (4, 2) {1};
		\node [style=none] (25) at (5, 2) {1};
		\node [style=none] (26) at (6, 2) {1};
		\node [style=none] (27) at (6, 4) {1};
		\node [style=none] (28) at (4, 4) {1};
		\node [style=gauge1] (29) at (1, -0.75) {};
		\node [style=gauge1] (30) at (2, -0.75) {};
		\node [style=gauge1] (31) at (3, -0.75) {};
		\node [style=gauge1] (33) at (2, 0.25) {};
		\node [style=none] (38) at (1, -1.25) {1};
		\node [style=none] (39) at (2, -1.25) {1};
		\node [style=none] (40) at (3, -1.25) {1};
		\node [style=none] (42) at (2, 0.75) {1};
		\node [style=miniG] (43) at (0, -4) {};
		\node [style=miniG] (44) at (0, 0) {};
		\node [style=miniG] (45) at (-3, 3) {};
		\node [style=miniG] (46) at (3, 3) {};
		\node [style=none] (47) at (-0.5, -2) {$a_3$};
		\node [style=none] (48) at (-2, 1.5) {$a_1$};
		\node [style=none] (49) at (2.25, 1.5) {$a_1$};
		\node [style=gauge1] (50) at (1, -4) {};
		\node [style=none] (51) at (1, -4.5) {1};
		\node [style=none] (52) at (-4, 4.75) {Mesonic Cone};
		\node [style=none] (53) at (4, 4.75) {Baryonic Cone};
	\end{pgfonlayer}
	\begin{pgfonlayer}{edgelayer}
		\draw (7.center) to (9.center);
		\draw (8.center) to (10.center);
		\draw (5) to (10.center);
		\draw (6) to (10.center);
		\draw (20.center) to (22.center);
		\draw (23.center) to (21.center);
		\draw (17) to (22.center);
		\draw (15) to (21.center);
		\draw (15) to (16);
		\draw (16) to (17);
		\draw (29) to (30);
		\draw (30) to (31);
		\draw (33) to (29);
		\draw (33) to (31);
		\draw (45) to (44);
		\draw (44) to (46);
		\draw (44) to (43);
	\end{pgfonlayer}
\end{tikzpicture}}
\caption{Hasse diagram of the Higgs branch of $\mathrm{SU}(3)$ with 4 fundamental hypers. The fact that the Higgs branch consists of a mesonic and a baryonic branch is represented in the Y shape of the Hasse diagram.}
\label{fig:two_cones}
\end{figure}

The Hasse diagrams mentioned in the previous sections represent symplectic singularities. However, a Higgs branch may be a union of several symplectic singularities with non-trivial intersection, see for instance \cite{Seiberg:1994aj} and also \cite{Ferlito:2016grh,Cabrera:2018jxt}. This structure can be encoded in a Hasse diagram. For example, the classical Higgs branch of $\mathrm{SU}(3)$ with 4 fundamental hypermultiplets consists of a mesonic branch and a baryonic branch. The mesonic branch is the next-to-minimal nilpotent orbit closure of $\mathrm{SU}(4)$, while the baryonic branch is an extension of the minimal nilpotent orbit closure of $\mathrm{SU}(4)$. The intersection between these two cones is the minimal nilpotent orbit closure of $\mathrm{SU}(4)$. Employing quiver subtraction one derives the Hasse diagram shown in Figure \ref{fig:two_cones}.
The structure of the two cones and their intersection is deduced from the \enquote{Y} shape of the Hasse diagram.

This concept is used abundantly in Section \ref{subsection5d}, where unions of up to three cones appear.

\section{Higgs Branches at Infinite Coupling}
\label{sec:Higgs_infinite}
Having discussed the Hasse diagrams that result from partial Higgsings of the gauge theory at finite coupling, we can employ the techniques of magnetic quivers and quiver subtraction to obtain the Hasse diagram for theories at infinite coupling. Working with brane configurations and magnetic quivers, Higgs branches of theories at infinite coupling do not represent a conceptual challenge. In fact, one essentially repeats the computation of Sections \ref{branesclass} and \ref{magQuivforsu4}.

\subsection{\texorpdfstring{5d $\mathrm{SU}(3)$ with 6 fundamentals and zero CS-level}{5d SU(3) with 6 fundamentals and zero CS-level}}
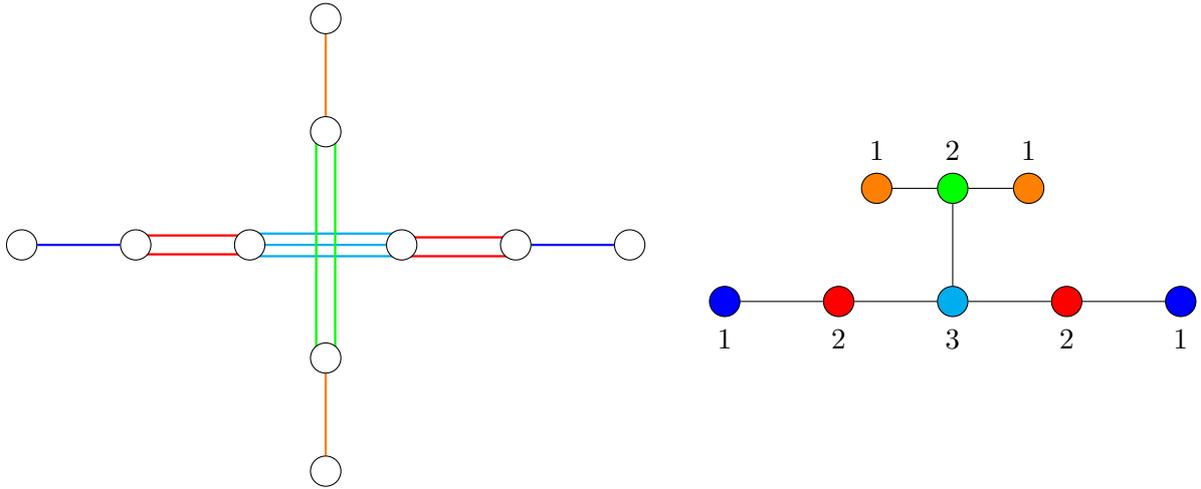
\begin{figure}[t]
\makebox[\textwidth][c]{
\begin{tikzpicture}
	\begin{pgfonlayer}{nodelayer}
		\node [style=gauge1] (0) at (-1, 0) {};
		\node [style=gauge1] (1) at (1, 0) {};
		\node [style=gauge1] (2) at (2.5, 0) {};
		\node [style=gauge1] (3) at (4, 0) {};
		\node [style=gauge1] (4) at (-2.5, 0) {};
		\node [style=gauge1] (5) at (-4, 0) {};
		\node [style=none] (6) at (-1, 0.15) {};
		\node [style=none] (7) at (-1, 0) {};
		\node [style=none] (8) at (-1, -0.15) {};
		\node [style=none] (9) at (1, 0.15) {};
		\node [style=none] (10) at (1, 0) {};
		\node [style=none] (11) at (1, -0.15) {};
		\node [style=none] (12) at (1, 0.1) {};
		\node [style=none] (13) at (1, 0) {};
		\node [style=none] (14) at (2.5, 0.1) {};
		\node [style=none] (15) at (2.5, -0.15) {};
		\node [style=none] (16) at (-1, -0.125) {};
		\node [style=none] (17) at (-1, 0.125) {};
		\node [style=none] (18) at (-2.5, 0.125) {};
		\node [style=none] (19) at (-2.5, -0.125) {};
		\node [style=none] (20) at (1, -0.15) {};
		\node [style=gauge1] (21) at (0, 1.5) {};
		\node [style=gauge1] (22) at (0, -1.5) {};
		\node [style=none] (23) at (-0.125, 1.5) {};
		\node [style=none] (24) at (0.125, 1.5) {};
		\node [style=none] (25) at (-0.125, -1.5) {};
		\node [style=none] (26) at (0.125, -1.5) {};
		\node [style=gauge1] (27) at (0, 3) {};
		\node [style=gauge1] (28) at (0, -3) {};
		\node [style=greenguage] (29) at (8.25, 0.75) {};
		\node [style=orangegauge] (30) at (9.25, 0.75) {};
		\node [style=orangegauge] (31) at (7.25, 0.75) {};
		\node [style=cyangauge] (32) at (8.25, -0.75) {};
		\node [style=redgauge] (33) at (9.75, -0.75) {};
		\node [style=redgauge] (34) at (6.75, -0.75) {};
		\node [style=bluegauge] (35) at (11.25, -0.75) {};
		\node [style=bluegauge] (36) at (5.25, -0.75) {};
		\node [style=none] (37) at (5.25, -1.25) {1};
		\node [style=none] (38) at (6.75, -1.25) {2};
		\node [style=none] (39) at (8.25, -1.25) {3};
		\node [style=none] (40) at (9.75, -1.25) {2};
		\node [style=none] (41) at (11.25, -1.25) {1};
		\node [style=none] (42) at (8.25, 1.25) {2};
		\node [style=none] (43) at (9.25, 1.25) {1};
		\node [style=none] (44) at (7.25, 1.25) {1};
	\end{pgfonlayer}
	\begin{pgfonlayer}{edgelayer}
		\draw [style=bluee] (3) to (2);
		\draw [style=bluee] (5) to (4);
		\draw [style=rede] (18.center) to (17.center);
		\draw [style=rede] (19.center) to (16.center);
		\draw [style=rede] (12.center) to (14.center);
		\draw [style=cyane] (6.center) to (9.center);
		\draw [style=cyane] (10.center) to (7.center);
		\draw [style=cyane] (8.center) to (11.center);
		\draw [style=rede] (15.center) to (20.center);
		\draw [style=greene] (24.center) to (26.center);
		\draw [style=greene] (25.center) to (23.center);
		\draw [style=orangee] (27) to (21);
		\draw [style=orangee] (22) to (28);
		\draw (31) to (29);
		\draw (29) to (30);
		\draw (29) to (32);
		\draw (32) to (33);
		\draw (33) to (35);
		\draw (34) to (32);
		\draw (36) to (34);
	\end{pgfonlayer}
\end{tikzpicture}}
\caption{5-brane web and magnetic quiver for 5d $\mathrm{SU}(3)$ gauge theory with $6$ fundamentals and vanishing CS-level on a generic point in the Higgs branch. Coloured branes are assumed to be on different positions along the 7-branes. The colours in the magnetic quiver reflect the way it was obtained from the brane web.}
\label{fig:SU3_infinite_coupling}
\end{figure}

Considering the infinite gauge coupling phase of the 5d $\mathrm{SU}(3)$ theory it is important to specify the Chern-Simons level, which is chosen to be zero here. The 5-brane web decomposition into sub-webs and the resulting magnetic quiver are shown in Figure \ref{fig:SU3_infinite_coupling}, see also \cite[Sec.\ 5]{Cabrera:2018jxt}. For convenience, different colours have been chosen to indicate independent sub-webs moving along D$7$ branes.

\begin{figure}[t]
\makebox[\textwidth][c]{
\begin{tikzpicture}
	\begin{pgfonlayer}{nodelayer}
		\node [style=miniG] (0) at (2, 4) {};
		\node [style=miniG] (1) at (2, 0.25) {};
		\node [style=miniG] (2) at (2, -4) {};
		\node [style=miniG] (3) at (8, -1) {};
		\node [style=none] (4) at (1.625, 1.25) {$d_5$};
		\node [style=none] (5) at (1.625, -2) {$a_5$};
		\node [style=none] (6) at (6, -2.5) {$a_1$};
		\node [style=none] (7) at (5.75, 1.5) {$e_6$};
		\node [style=gauge1] (8) at (9.25, -1) {};
		\node [style=gauge1] (9) at (10.25, -1) {};
		\node [style=none] (10) at (9.25, -0.9) {};
		\node [style=none] (11) at (9.25, -1.1) {};
		\node [style=none] (12) at (10.25, -0.9) {};
		\node [style=none] (13) at (10.25, -1.1) {};
		\node [style=gauge1] (14) at (0, 0.75) {};
		\node [style=gauge1] (15) at (-1.5, -0.25) {};
		\node [style=gauge1] (16) at (-0.75, -0.25) {};
		\node [style=gauge1] (17) at (0, -0.25) {};
		\node [style=gauge1] (18) at (0.75, -0.25) {};
		\node [style=gauge1] (19) at (1.5, -0.25) {};
		\node [style=gauge1] (20) at (5.725, 4.75) {};
		\node [style=gauge1] (21) at (4.725, 4.75) {};
		\node [style=gauge1] (22) at (6.725, 4.75) {};
		\node [style=gauge1] (23) at (5.725, 3.5) {};
		\node [style=gauge1] (24) at (6.725, 3.5) {};
		\node [style=gauge1] (25) at (7.725, 3.5) {};
		\node [style=gauge1] (26) at (4.725, 3.5) {};
		\node [style=gauge1] (27) at (3.725, 3.5) {};
		\node [style=none] (28) at (5.725, 5.25) {2};
		\node [style=none] (29) at (4.725, 5.25) {1};
		\node [style=none] (30) at (6.725, 5.25) {1};
		\node [style=none] (31) at (7.725, 3) {1};
		\node [style=none] (32) at (6.725, 3) {2};
		\node [style=none] (33) at (5.725, 3) {3};
		\node [style=none] (34) at (4.725, 3) {2};
		\node [style=none] (35) at (3.725, 3) {1};
		\node [style=none] (36) at (0, 1.25) {1};
		\node [style=none] (37) at (-1.5, -0.75) {1};
		\node [style=none] (38) at (-0.75, -0.75) {1};
		\node [style=none] (39) at (0, -0.75) {1};
		\node [style=none] (40) at (0.75, -0.75) {1};
		\node [style=none] (41) at (1.5, -0.75) {1};
		\node [style=none] (42) at (9.25, -1.5) {1};
		\node [style=none] (43) at (10.25, -1.5) {1};
		\node [style=gauge1] (44) at (3, -4) {};
		\node [style=none] (45) at (3, -4.5) {1};
		\node [style=gauge1] (47) at (-6.25, -1) {};
		\node [style=gauge1] (48) at (-5.25, -1) {};
		\node [style=none] (49) at (-6.25, -0.9) {};
		\node [style=none] (50) at (-6.25, -1.1) {};
		\node [style=none] (51) at (-5.25, -0.9) {};
		\node [style=none] (52) at (-5.25, -1.1) {};
		\node [style=none] (53) at (-6.25, -1.5) {1};
		\node [style=none] (54) at (-5.25, -1.5) {1};
		\node [style=none] (55) at (-2, -2.5) {$a_1$};
		\node [style=none] (56) at (-1.75, 1.5) {$e_6$};
		\node [style=miniG] (57) at (-4, -1) {};
	\end{pgfonlayer}
	\begin{pgfonlayer}{edgelayer}
		\draw (0) to (1);
		\draw (1) to (2);
		\draw (3) to (2);
		\draw (3) to (0);
		\draw (10.center) to (12.center);
		\draw (11.center) to (13.center);
		\draw (14) to (19);
		\draw (18) to (17);
		\draw (18) to (19);
		\draw (16) to (17);
		\draw (15) to (16);
		\draw (15) to (14);
		\draw (20) to (22);
		\draw (20) to (21);
		\draw (20) to (23);
		\draw (23) to (24);
		\draw (24) to (25);
		\draw (26) to (23);
		\draw (27) to (26);
		\draw (49.center) to (51.center);
		\draw (50.center) to (52.center);
		\draw (0) to (57);
		\draw (57) to (2);
	\end{pgfonlayer}
\end{tikzpicture}}
\caption{Hasse diagram for the Higgs branch of 5d $\mathrm{SU}(3)$ with 6 fundamentals and vanishing CS-level at infinite gauge coupling. Note that from the top quiver an $e_6$ can be subtracted in two different ways, related by action of the automorphism group of the top quiver, leading to the two $e_6$ lines and two $a_1$ lines. We can see that the Hasse diagram looks completely different from the finite coupling case of Figure \ref{tab:SU(3)6HasseSub} with exception of the $a_5$ line.}
\label{fig:Hasse_SU3_infinite}
\end{figure}
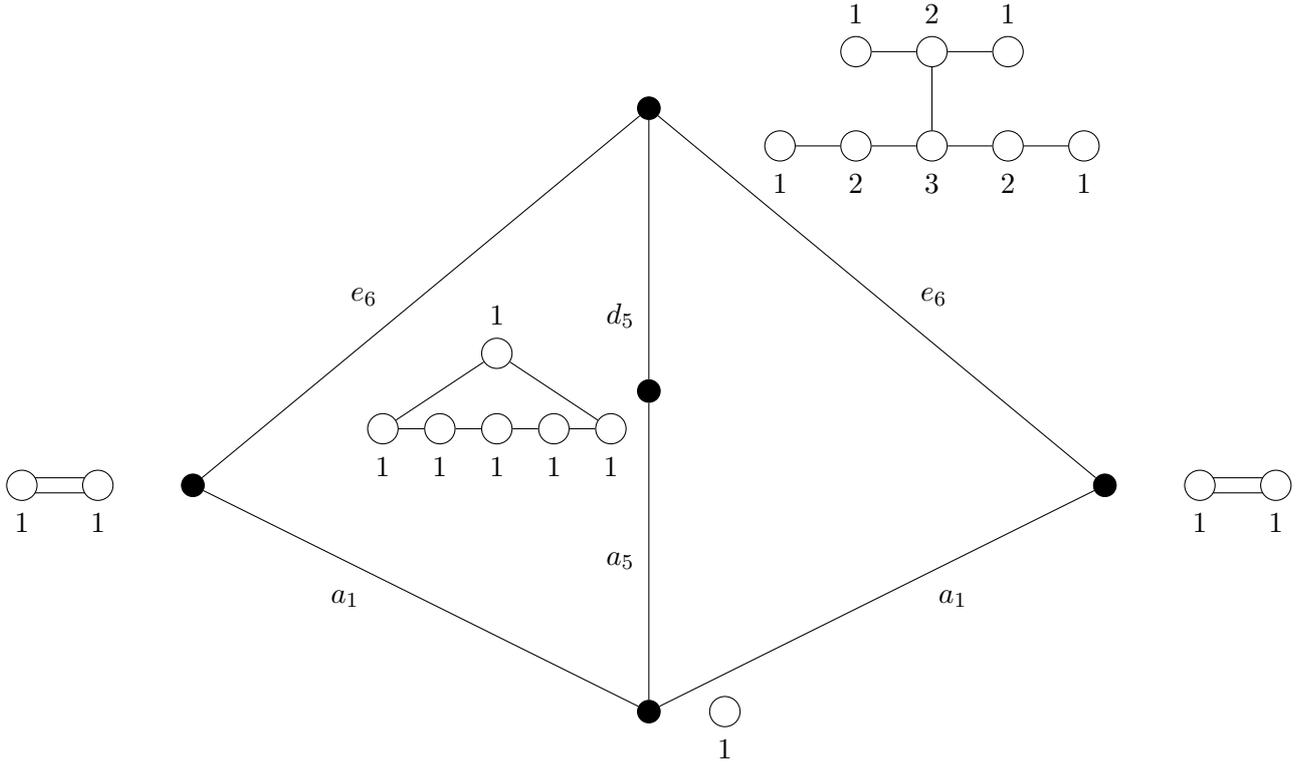

Taking the magnetic quiver of Figure \ref{fig:SU3_infinite_coupling} as a starting point, the process of quiver subtracting guided by elementary slices results in the Hasse diagram shown in Figure \ref{fig:Hasse_SU3_infinite}. Comparing this to the finite coupling Hasse diagram of Figure \ref{tab:SU(3)6HasseSub}, one realises that the structure has completely changed and that the finite coupling Hasse diagram is not contained in the one at infinite coupling. In particular, one notes the bifurcation and the appearance of the $e_6$ transition. In contrast to Figure \ref{tab:SU(3)6HasseSub}, the Hasse diagram of Figure \ref{fig:Hasse_SU3_infinite} cannot be obtained from partial Higgsing, because the theory is at infinite coupling where massless BPS objects contribute to the spectrum.

\subsection{\texorpdfstring{6d $\mathrm{SU}(4)$ with 2nd rank antisymmetric and $12$ fundamental hypers}{6d SU(4) with 2nd rank antisymmetric and 12 fundamental hypers}}\label{infinite_coupling_su4}

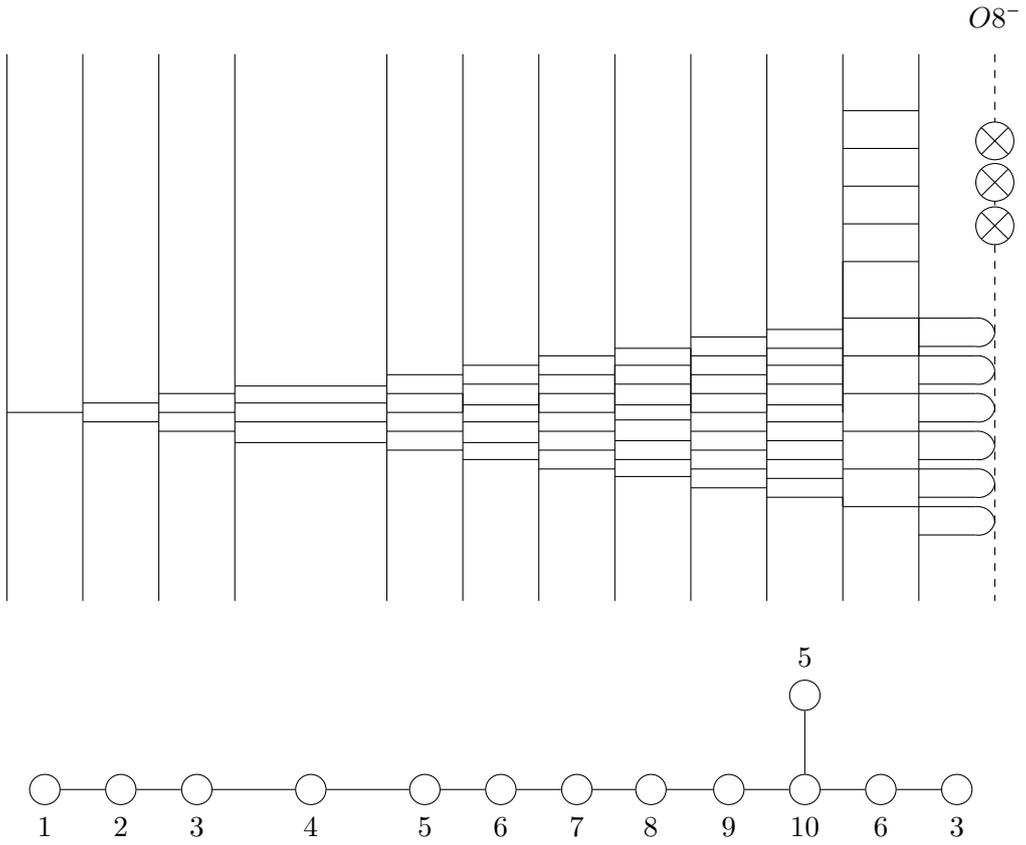
\begin{figure}[t]
\centering
\begin{tikzpicture}
	\begin{pgfonlayer}{nodelayer}
		\node [style=none] (6) at (1, 2) {};
		\node [style=none] (7) at (1, -5.25) {};
		\node [style=none] (22) at (9, 2) {};
		\node [style=none] (23) at (9, -5.25) {};
		\node [style=none] (32) at (1, -2.4) {};
		\node [style=none] (33) at (1, -2.625) {};
		\node [style=none] (34) at (1, -2.875) {};
		\node [style=none] (35) at (1, -3.15) {};
		\node [style=none] (76) at (1, -2.25) {};
		\node [style=none] (77) at (1, -2.5) {};
		\node [style=none] (78) at (1, -2.75) {};
		\node [style=none] (79) at (1, -3) {};
		\node [style=none] (80) at (2, 2) {};
		\node [style=none] (81) at (2, -2.75) {};
		\node [style=none] (82) at (2, -3.15) {};
		\node [style=none] (83) at (2, -2.125) {};
		\node [style=none] (84) at (2, -2.375) {};
		\node [style=none] (85) at (2, -2.65) {};
		\node [style=none] (86) at (2, -2.25) {};
		\node [style=none] (87) at (2, -2.5) {};
		\node [style=none] (88) at (2, -2.75) {};
		\node [style=none] (89) at (2, -3) {};
		\node [style=none] (90) at (3, 2) {};
		\node [style=none] (91) at (3, -2.75) {};
		\node [style=none] (92) at (3, -3.15) {};
		\node [style=none] (93) at (3, -2.125) {};
		\node [style=none] (94) at (3, -2.375) {};
		\node [style=none] (95) at (3, -2.65) {};
		\node [style=none] (96) at (3, -2.25) {};
		\node [style=none] (97) at (3, -2.5) {};
		\node [style=none] (98) at (3, -2.75) {};
		\node [style=none] (99) at (3, -3) {};
		\node [style=none] (100) at (4, 2) {};
		\node [style=none] (101) at (4, -2.75) {};
		\node [style=none] (102) at (4, -1.9) {};
		\node [style=none] (103) at (4, -2.125) {};
		\node [style=none] (104) at (4, -2.375) {};
		\node [style=none] (105) at (4, -2.65) {};
		\node [style=none] (106) at (4, -2.25) {};
		\node [style=none] (107) at (4, -2.5) {};
		\node [style=none] (108) at (4, -2.75) {};
		\node [style=none] (109) at (4, -3) {};
		\node [style=none] (110) at (5, 2) {};
		\node [style=none] (111) at (5, -2.75) {};
		\node [style=none] (112) at (5, -1.9) {};
		\node [style=none] (113) at (5, -2.125) {};
		\node [style=none] (114) at (5, -2.375) {};
		\node [style=none] (115) at (5, -2.65) {};
		\node [style=none] (116) at (5, -2.25) {};
		\node [style=none] (117) at (5, -2.5) {};
		\node [style=none] (118) at (5, -2.75) {};
		\node [style=none] (119) at (5, -3) {};
		\node [style=none] (120) at (6, 2) {};
		\node [style=none] (121) at (6, -2.75) {};
		\node [style=none] (122) at (6, -1.9) {};
		\node [style=none] (123) at (6, -2.125) {};
		\node [style=none] (124) at (6, -2.375) {};
		\node [style=none] (125) at (6, -2.65) {};
		\node [style=none] (126) at (6, -2.25) {};
		\node [style=none] (127) at (6, -2.5) {};
		\node [style=none] (128) at (6, -2.75) {};
		\node [style=none] (129) at (6, -3) {};
		\node [style=none] (130) at (7, 2) {};
		\node [style=none] (131) at (7, -2.75) {};
		\node [style=none] (132) at (7, -1.9) {};
		\node [style=none] (133) at (7, -2.125) {};
		\node [style=none] (134) at (7, -2.375) {};
		\node [style=none] (135) at (7, -2.65) {};
		\node [style=none] (136) at (7, -2.75) {};
		\node [style=none] (137) at (7, -2.25) {};
		\node [style=none] (138) at (7, -2.5) {};
		\node [style=none] (139) at (7, -3) {};
		\node [style=none] (140) at (8, 2) {};
		\node [style=none] (141) at (8, -2) {};
		\node [style=none] (142) at (8, -2.15) {};
		\node [style=none] (146) at (8, -3.5) {};
		\node [style=none] (150) at (-4, 2) {};
		\node [style=none] (151) at (-4, -5.25) {};
		\node [style=none] (152) at (-4, -0.9) {};
		\node [style=none] (153) at (-4, -1.625) {};
		\node [style=none] (154) at (-4, -2.375) {};
		\node [style=none] (155) at (-4, -3.15) {};
		\node [style=none] (156) at (-4, -1.25) {};
		\node [style=none] (157) at (-4, -2.75) {};
		\node [style=none] (158) at (-4, -2.75) {};
		\node [style=none] (159) at (-4, -5.25) {};
		\node [style=none] (160) at (-3, 2) {};
		\node [style=none] (161) at (-3, -5.25) {};
		\node [style=none] (162) at (-3, -0.9) {};
		\node [style=none] (163) at (-3, -2.625) {};
		\node [style=none] (164) at (-3, -2.875) {};
		\node [style=none] (165) at (-3, -3.15) {};
		\node [style=none] (166) at (-3, -1.25) {};
		\node [style=none] (167) at (-3, -2.75) {};
		\node [style=none] (168) at (-3, -2.75) {};
		\node [style=none] (169) at (-3, -5.25) {};
		\node [style=none] (170) at (-2, 2) {};
		\node [style=none] (171) at (-2, -5.25) {};
		\node [style=none] (172) at (-2, -0.9) {};
		\node [style=none] (173) at (-2, -2.625) {};
		\node [style=none] (174) at (-2, -2.875) {};
		\node [style=none] (175) at (-2, -3.15) {};
		\node [style=none] (176) at (-2, -3) {};
		\node [style=none] (177) at (-2, -2.5) {};
		\node [style=none] (178) at (-2, -2.75) {};
		\node [style=none] (179) at (-2, -5.25) {};
		\node [style=none] (180) at (-1, 2) {};
		\node [style=none] (181) at (-1, -5.25) {};
		\node [style=none] (182) at (-1, -2.4) {};
		\node [style=none] (183) at (-1, -2.625) {};
		\node [style=none] (184) at (-1, -2.875) {};
		\node [style=none] (185) at (-1, -3.15) {};
		\node [style=none] (186) at (-1, -3) {};
		\node [style=none] (187) at (-1, -2.5) {};
		\node [style=none] (188) at (-1, -2.75) {};
		\node [style=none] (189) at (-1, -5.25) {};
		\node [draw, circle, cross, minimum width=0.5 cm, fill=white] (190) at (9, 0.3) {};
		\node [style=none] (191) at (8.75, -3.5) {};
		\node [style=none] (193) at (8.75, -3.875) {};
		\node [style=none] (194) at (9, -3.7) {};
		\node [style=none] (196) at (8, -4) {};
		\node [style=none] (197) at (8.75, -4) {};
		\node [style=none] (198) at (8.75, -4.375) {};
		\node [style=none] (199) at (9, -4.2) {};
		\node [style=none] (608) at (9, 2.5) {$O8^-$};
		\node [style=none] (609) at (1, -3.25) {};
		\node [style=none] (610) at (2, -3.25) {};
		\node [style=none] (611) at (2, -2.875) {};
		\node [style=none] (612) at (2, -3.375) {};
		\node [style=none] (613) at (3, -2.875) {};
		\node [style=none] (614) at (3, -3.5) {};
		\node [style=none] (615) at (3, -3.375) {};
		\node [style=none] (616) at (3, -3.25) {};
		\node [style=none] (617) at (3, -2) {};
		\node [style=none] (618) at (4, -3.5) {};
		\node [style=none] (619) at (4, -2.85) {};
		\node [style=none] (620) at (4, -3.6) {};
		\node [style=none] (621) at (4, -2) {};
		\node [style=none] (622) at (4, -3.25) {};
		\node [style=none] (624) at (5, -2.85) {};
		\node [style=none] (625) at (5, -3.75) {};
		\node [style=none] (626) at (5, -3.6) {};
		\node [style=none] (627) at (5, -2) {};
		\node [style=none] (628) at (6, -2) {};
		\node [style=none] (630) at (6, -3.75) {};
		\node [style=none] (631) at (6, -1.65) {};
		\node [style=none] (633) at (7, -1.65) {};
		\node [style=none] (634) at (7, -2.5) {};
		\node [style=none] (635) at (7, -3.75) {};
		\node [style=none] (636) at (7, -3.5) {};
		\node [style=none] (637) at (8, -3.5) {};
		\node [style=none] (638) at (8, -3) {};
		\node [style=none] (639) at (8, -2.5) {};
		\node [style=none] (640) at (8, -1.75) {};
		\node [style=none] (641) at (6.75, -3) {};
		\node [style=none] (642) at (5, -3.375) {};
		\node [style=none] (643) at (4, -3.375) {};
		\node [style=none] (644) at (4, -3.125) {};
		\node [style=none] (645) at (5, -3.125) {};
		\node [style=none] (646) at (5, -1.75) {};
		\node [style=none] (647) at (6, -1.75) {};
		\node [style=none] (648) at (5, -3.25) {};
		\node [style=none] (649) at (6, -3.25) {};
		\node [style=none] (650) at (5, -3.5) {};
		\node [style=none] (651) at (6, -3.5) {};
		\node [style=none] (652) at (6, -2.875) {};
		\node [style=none] (653) at (7, -2.875) {};
		\node [style=none] (654) at (6, -3.375) {};
		\node [style=none] (655) at (7, -3.375) {};
		\node [style=none] (656) at (6, -3.125) {};
		\node [style=none] (657) at (7, -3.125) {};
		\node [style=none] (658) at (6, -3.625) {};
		\node [style=none] (659) at (7, -3.625) {};
		\node [style=none] (660) at (6, -3.875) {};
		\node [style=none] (661) at (7, -3.875) {};
		\node [style=none] (662) at (7, -1.5) {};
		\node [style=none] (663) at (8, -1.5) {};
		\node [style=none] (664) at (7, -2) {};
		\node [style=none] (665) at (8, -2) {};
		\node [style=none] (666) at (7, -3) {};
		\node [style=none] (667) at (8, -3) {};
		\node [style=none] (668) at (7, -4) {};
		\node [style=none] (669) at (8, -4) {};
		\node [style=none] (670) at (7, -1.725) {};
		\node [style=none] (672) at (8, -3.875) {};
		\node [style=none] (678) at (8.75, -3) {};
		\node [style=none] (679) at (8.75, -3.375) {};
		\node [style=none] (680) at (9, -3.2) {};
		\node [style=none] (681) at (8, -3) {};
		\node [style=none] (682) at (8, -3.375) {};
		\node [style=none] (683) at (8.75, -2.5) {};
		\node [style=none] (684) at (8.75, -2.875) {};
		\node [style=none] (685) at (9, -2.7) {};
		\node [style=none] (686) at (8, -2.5) {};
		\node [style=none] (687) at (8, -2.875) {};
		\node [style=none] (688) at (8.75, -2) {};
		\node [style=none] (689) at (8.75, -2.375) {};
		\node [style=none] (690) at (9, -2.2) {};
		\node [style=none] (691) at (8, -2) {};
		\node [style=none] (692) at (8, -2.375) {};
		\node [style=none] (693) at (8.75, -1.5) {};
		\node [style=none] (694) at (8.75, -1.875) {};
		\node [style=none] (695) at (9, -1.7) {};
		\node [style=none] (696) at (8, -1.5) {};
		\node [style=none] (697) at (8, -1.875) {};
		\node [draw, circle, cross, minimum width=0.5 cm, fill=white] (698) at (9, -0.275) {};
		\node [draw, circle, cross, minimum width=0.5 cm, fill=white] (699) at (9, 0.85) {};
		\node [style=none] (700) at (8, -4.375) {};
		\node [style=none] (701) at (2, -5.25) {};
		\node [style=none] (702) at (3, -5.25) {};
		\node [style=none] (703) at (4, -5.25) {};
		\node [style=none] (704) at (5, -5.25) {};
		\node [style=none] (705) at (6, -5.25) {};
		\node [style=none] (706) at (7, -5.25) {};
		\node [style=none] (707) at (8, -5.25) {};
		\node [style=gauge1] (708) at (-3.5, -7.75) {};
		\node [style=gauge1] (709) at (-2.5, -7.75) {};
		\node [style=gauge1] (710) at (-1.5, -7.75) {};
		\node [style=gauge1] (711) at (0, -7.75) {};
		\node [style=gauge1] (712) at (1.5, -7.75) {};
		\node [style=gauge1] (713) at (2.5, -7.75) {};
		\node [style=gauge1] (714) at (3.5, -7.75) {};
		\node [style=gauge1] (715) at (4.475, -7.75) {};
		\node [style=gauge1] (716) at (5.5, -7.75) {};
		\node [style=gauge1] (717) at (6.5, -7.75) {};
		\node [style=gauge1] (719) at (7.5, -7.75) {};
		\node [style=gauge1] (720) at (6.5, -6.5) {};
		\node [style=gauge1] (721) at (8.5, -7.75) {};
		\node [style=none] (722) at (-3.5, -8.25) {1};
		\node [style=none] (723) at (-2.5, -8.25) {2};
		\node [style=none] (724) at (-1.5, -8.25) {3};
		\node [style=none] (725) at (0, -8.25) {4};
		\node [style=none] (726) at (1.5, -8.25) {5};
		\node [style=none] (727) at (2.5, -8.25) {6};
		\node [style=none] (729) at (3.5, -8.25) {7};
		\node [style=none] (730) at (4.475, -8.25) {8};
		\node [style=none] (731) at (5.5, -8.25) {9};
		\node [style=none] (732) at (6.5, -8.25) {10};
		\node [style=none] (734) at (6.5, -6) {5};
		\node [style=none] (735) at (7.5, -8.25) {6};
		\node [style=none] (736) at (8.5, -8.25) {3};
		\node [style=none] (737) at (7, 1.25) {};
		\node [style=none] (738) at (8, 1.25) {};
		\node [style=none] (739) at (8, 0.75) {};
		\node [style=none] (740) at (7, 0.75) {};
		\node [style=none] (741) at (7, 0.25) {};
		\node [style=none] (742) at (8, 0.25) {};
		\node [style=none] (743) at (8, -0.25) {};
		\node [style=none] (744) at (7, -0.25) {};
		\node [style=none] (745) at (7, -0.75) {};
		\node [style=none] (746) at (8, -0.75) {};
	\end{pgfonlayer}
	\begin{pgfonlayer}{edgelayer}
		\draw (6.center) to (7.center);
		\draw [style=new edge style 1] (22.center) to (23.center);
		\draw (80.center) to (81.center);
		\draw (90.center) to (91.center);
		\draw (100.center) to (101.center);
		\draw (110.center) to (111.center);
		\draw (120.center) to (121.center);
		\draw (130.center) to (131.center);
		\draw (140.center) to (141.center);
		\draw (150.center) to (151.center);
		\draw (160.center) to (161.center);
		\draw (170.center) to (171.center);
		\draw (157.center) to (167.center);
		\draw (163.center) to (173.center);
		\draw (164.center) to (174.center);
		\draw (180.center) to (181.center);
		\draw (176.center) to (186.center);
		\draw (177.center) to (187.center);
		\draw (178.center) to (188.center);
		\draw (182.center) to (32.center);
		\draw (183.center) to (33.center);
		\draw (184.center) to (34.center);
		\draw (185.center) to (35.center);
		\draw (76.center) to (86.center);
		\draw (77.center) to (87.center);
		\draw (78.center) to (88.center);
		\draw (79.center) to (89.center);
		\draw (85.center) to (95.center);
		\draw (84.center) to (94.center);
		\draw (83.center) to (93.center);
		\draw (82.center) to (92.center);
		\draw (96.center) to (106.center);
		\draw (97.center) to (107.center);
		\draw (98.center) to (108.center);
		\draw (99.center) to (109.center);
		\draw (105.center) to (115.center);
		\draw (104.center) to (114.center);
		\draw (103.center) to (113.center);
		\draw (102.center) to (112.center);
		\draw (116.center) to (126.center);
		\draw (117.center) to (127.center);
		\draw (118.center) to (128.center);
		\draw (119.center) to (129.center);
		\draw (122.center) to (132.center);
		\draw (123.center) to (133.center);
		\draw (124.center) to (134.center);
		\draw (125.center) to (135.center);
		\draw (146.center) to (191.center);
		\draw [bend left=45] (191.center) to (194.center);
		\draw [bend right=315] (194.center) to (193.center);
		\draw (196.center) to (197.center);
		\draw [bend left=45] (197.center) to (199.center);
		\draw [bend right=315] (199.center) to (198.center);
		\draw (609.center) to (610.center);
		\draw (611.center) to (613.center);
		\draw (612.center) to (615.center);
		\draw (616.center) to (622.center);
		\draw (614.center) to (618.center);
		\draw (617.center) to (621.center);
		\draw (620.center) to (626.center);
		\draw (619.center) to (624.center);
		\draw (625.center) to (630.center);
		\draw (627.center) to (628.center);
		\draw (631.center) to (633.center);
		\draw (634.center) to (639.center);
		\draw (636.center) to (637.center);
		\draw (642.center) to (643.center);
		\draw (645.center) to (644.center);
		\draw (646.center) to (647.center);
		\draw (648.center) to (649.center);
		\draw (650.center) to (651.center);
		\draw (652.center) to (653.center);
		\draw (654.center) to (655.center);
		\draw (656.center) to (657.center);
		\draw (658.center) to (659.center);
		\draw (660.center) to (661.center);
		\draw (662.center) to (663.center);
		\draw (664.center) to (665.center);
		\draw (666.center) to (667.center);
		\draw (668.center) to (669.center);
		\draw (672.center) to (193.center);
		\draw [bend left=45] (678.center) to (680.center);
		\draw [bend right=315] (680.center) to (679.center);
		\draw (682.center) to (679.center);
		\draw (678.center) to (681.center);
		\draw [bend left=45] (683.center) to (685.center);
		\draw [bend right=315] (685.center) to (684.center);
		\draw (687.center) to (684.center);
		\draw (683.center) to (686.center);
		\draw [bend left=45] (688.center) to (690.center);
		\draw [bend right=315] (690.center) to (689.center);
		\draw (692.center) to (689.center);
		\draw (688.center) to (691.center);
		\draw [bend left=45] (693.center) to (695.center);
		\draw [bend right=315] (695.center) to (694.center);
		\draw (697.center) to (694.center);
		\draw (693.center) to (696.center);
		\draw (198.center) to (700.center);
		\draw (701.center) to (87.center);
		\draw (702.center) to (97.center);
		\draw (703.center) to (103.center);
		\draw (704.center) to (112.center);
		\draw (705.center) to (122.center);
		\draw (706.center) to (661.center);
		\draw (707.center) to (700.center);
		\draw (708) to (709);
		\draw (709) to (710);
		\draw (710) to (711);
		\draw (711) to (712);
		\draw (712) to (713);
		\draw (713) to (714);
		\draw (714) to (715);
		\draw (715) to (716);
		\draw (716) to (717);
		\draw (720) to (717);
		\draw (717) to (719);
		\draw (719) to (721);
		\draw (700.center) to (696.center);
		\draw (737.center) to (738.center);
		\draw (739.center) to (740.center);
		\draw (741.center) to (742.center);
		\draw (743.center) to (744.center);
		\draw (745.center) to (746.center);
		\draw (745.center) to (668.center);
	\end{pgfonlayer}
\end{tikzpicture}
\caption{Type IIA brane configuration and magnetic quiver for the $6$d $\mathrm{SU}(4)$ gauge theory with one 2nd rank antisymmetric and $12$ fundamentals.}
\label{fig:SU4_infinite}
\end{figure}

Likewise, one can consider the infinite coupling case for the $6$d $\mathrm{SU}(4)$ theory. Starting from the Type IIA brane configuration, it has been shown in \cite{Cabrera:2019izd} how to obtain the infinite coupling brane configuration and magnetic quiver via the small $E_8$ instanton transition. For convenience, the brane configuration and magnetic quiver are recalled in Figure \ref{fig:SU4_infinite}.

Subtracting from  magnetic quiver in Figure \ref{fig:SU4_infinite} the  magnetic quivers for elementary slices leads to the Hasse diagram shown in Figure \ref{tab:Hasse_SU4_infinite}. One observes that the infinite coupling Hasse diagram is simply an extension of the finite coupling Hasse diagram of Figure \ref{tab:Hasse_SU4_finite} by an $e_8$ transition.

\begin{figure}[t]
	\centering
	\begin{tabular}{m{2.6cm} m{5.7cm}}
	Hasse diagram  & Magnetic quiver 
	 \\ 
\begin{tikzpicture}[node distance=20pt]
	\tikzstyle{hasse} = [circle, fill,inner sep=2pt];
		\node at (-0.7,-0.5) [] (1a) [] {};
		\node [hasse] at (0,-0.5) [] (1b) [label=left:\footnotesize{$68$}] {}; %39
		\node at (0.7,-0.5) [] (1c) [] {};
		\node  (2a) [below of=1a] {};
		\node [] (2b) [below of=1b] {};
		\node (2c) [below of=1c] {};
		\node [] (3a) [below of=2a] {};
		\node  (3b) [below of=2b] {};
		\node (3c) [below of=2c] {};
		\node (4a) [below of=3a] {};
		\node [hasse] (4b)[label=left:\footnotesize{$39$},below of=3b] {}; %22
		\node (4c) [below of=3c] {};
		\node [] (5a) [below of=4a] {};
		\node (5b) [below of=4b] {};
		\node  (5c) [below of=4c] {};
		\node  (6a) [below of=5a] {};
		\node  (6b) [below of=5b] {};
		\node (6c) [below of=5c] {};
		\node  (7a) [below of=6a] {};
		\node [hasse] (7b) [label=right:\footnotesize{$22$},below of=6b] {};
		\node (7c) [below of=6c] {}; 
		\node (8a) [below of=7a] {};
		\node  (8b) [below of=7b] {};
	        \node (8c) [below of=7c] {};
	        \node (9a) [below of=8a] {};
		\node (9b) [below of=8b] {};
	        \node (9c) [below of=8c] {};
	        \node  (10a) [below of=9a] {};
		\node (10b) [below of=9b] {};
	        \node  (10c) [below of=9c] {};
	          \node (11a) [below of=10a] {};
		\node (11b) [below of=10b] {};
	        \node [hasse] (11c) [label=right:\footnotesize{$11$},below of=10c] {};
	         \node (12a) [below of=11a] {};
	         	\node (12b) [below of=11b] {};
	        \node (12c) [below of=11c] {};
	         \node (13a) [below of=12a] {};
	            \node (13b) [below of=12b] {};
	               \node (13c) [below of=12c] {};
	        \node [hasse] (13a) [label=left:\footnotesize{$1$},below of=13a] {};
		\node (14b) [below of=13b] {};
	        \node (14c) [below of=13c] {};
	        \node (14a) [below of=13a] {};
	          \node (15a) [below of=14a] {};
		\node [hasse] (15b) [label=right:\footnotesize{$0$},below of=14b] {};
	        \node (15c) [below of=14c] {};
	         \node (16a) [below of=15a] {};
	         \node (17a) [below of=16a] {}; 
		\node (12b) [below of=11b] {};
	        \node (12c) [below of=11c] {};
		\draw (1b) edge [] node[label=left:\footnotesize{$e_{8}$}] {} (4b)
	 (4b) edge [] node[label=left:\footnotesize{$d_{10}$}] {} (7b)
		        (7b) edge [] node[label=right:\footnotesize{$a_{11}$}] {} (11c)
		            (7b) edge [] node[label=left:\footnotesize{$d_{12}$}] {} (13a)
		             (13a) edge [] node[label=above:\footnotesize{$c_{1}$}] {} (15b)
			(11c) edge [] node[label=right:\footnotesize{$a_{11}$}] {} (15b);
	\end{tikzpicture}
	&
\thead{ $\node{}1 -\node{}2 - \node{}3 -\node{}4 -\node{}5-\node{}6-\node{}7-\node{}8-\node{}9-\node{\topnode{}5}{10}-\node{}6-\node{}3$  \\ \\ \\
$\node{}1 -\node{}2 - \node{}3 -\node{\topnode{}1}4 -\node{}4-\node{}4-\node{}4-\node{}4-\node{}4-\node{\topnode{}2}4-\node{}2-\node{}1$ 
\\ \\  
$\underbrace{\node{}1 -
 \overset{\overset{\displaystyle\overset 1 \circ}
	{\diagup~~~~~\diagdown}
	}
{\node{\topnode{}1}2  -\dots -\node{\topnode{}1}2}
-\node{}1}_{11}$  
 \\ \\ \\  \\ 
 $\underbrace{
 \overset{\overset{\displaystyle\overset 1 \circ}
	{\diagup~~~~~\diagdown}
	}
	{ \node{}{1}-\dots-\node{}{1} }
	}_{11}$ 
 \\ \\  \\
 $\node{}1 = \node{}1$ \\ \\
  $\node{}1$ \\ \\ \\ \\ \\ \\}
	\end{tabular}
	\caption{Hasse diagram for the Higgs branch of $6$d $\mathrm{SU}(4)$ with one 2nd rank antisymmetric and 12 fundamentals at infinite gauge coupling. Comparing to the finite coupling case of Figure \ref{tab:Hasse_SU4_finite}, we see that an $e_8$ line is added on top.}\label{tab:Hasse_SU4_infinite}
\end{figure}
\clearpage

\section{Results}
\label{sectionResults}
In this section, the symplectic leaf structure in the form of Hasse diagrams is provided for the following theories: 
\begin{itemize}
    \item Section \ref{subsectionSimpleGaugeGroup} contains theories with a simple gauge group of type $A$, $B$, $C$, $D$, and $G_2$ with fundamental matter.
    \item Section \ref{subsection5d} focuses on $5$d Higgs branches at infinite coupling, as obtained in \cite{Cabrera:2018jxt}. 
    \item In Section \ref{subsection6d}, Higgs branch of 6d SCFTs realised on $-1$ curves are considered. 
    \item Section \ref{subsectioncompletegraphs} considers examples of Argyres-Douglas theories with magnetic quivers in the form of complete graphs \cite{DelZotto:2014kka}.
\end{itemize}

\subsection{Simple Gauge Groups with Fundamentals}
\label{subsectionSimpleGaugeGroup}
In this subsection, we consider theories with a simple gauge group and enough fundamental matter. The motivation for studying these models is threefold: firstly, they are the most commonly used theories, in particular in the context of quiver gauge theories. Secondly, the results are purely classical, and consequently, valid for any number of dimensions $3 \leq d \leq 6$. Thirdly their partial Higgs mechanism can be studied using representation theory and the results are naturally expressed in terms of nilpotent orbits. One can then verify, as a consistency check, that the appropriate sections of the resulting Hasse diagrams are part of the nilpotent orbits Hasse diagrams. We elaborate below on what these appropriate sections are, in a case-by-case analysis. 

It is important to stress that the results presented below are valid for a number of flavours high enough so that complete Higgsing is possible. Below that threshold (which is indicated in tables of results), the Hasse diagram is modified. 

\begin{table}
	\centering
	\begin{tabular}{|l|c|c|}
	\hline
	Theory & $U(k)$ with $N \geq 2k$ flavours & $\mathrm{SU}(k)$ with $N \geq 2k$ flavours  
	 \\ \hline
	 Electric quiver & $ \node{\flav {\mathrm{SU}(N)}}{U(k)}  $ & $ \node{\flav {U(N)}}{\mathrm{SU}(k)}  $ \\ \hline
 Magnetic quiver & $\underbrace{\node{}1-\node{}2- \cdots -\overset{\overset{\displaystyle\overset 1 \circ}
	{\diagup~~~~~~\diagdown}
	}{\node{}{k}-\dots-\node{}{k} } - \cdots - \node{}2-\node{}1}_{N-1}$
  &
  $\underbrace{\node{}{1}-\node{}{2}-\cdots-\node{\topnode{}{1}}{k}-\cdots-\node{\topnode{}{1}}{k}-\cdots- \node{}{2}-\node{}{1}}_{N-1}$  \\ \hline
  Hasse diagram
  &
 \begin{tikzpicture}
		\tikzstyle{hasse} = [circle, fill,inner sep=2pt];
		\node [hasse] (1) [label=right:\footnotesize{$k(N-k)$}] {}; 
		\node [hasse] (2) [below of=1] {};
		\node [hasse] (3) [below of=2] {};
		\node [hasse] (4) [below of=3] {};
		\node [hasse] (5) [below of=4] {};
		\node [hasse] (6) [below of=5] {};
		\node [hasse] (7) [below of=6] {};
				\node at (0,-4.5) {$\vdots$};
		\draw (1) edge [] node[label=left:\footnotesize{$a_{N-2k+1}$}] {} (2)
			(2) edge [] node[label=left:\footnotesize{$a_{N-2k+3}$}] {} (3)
			(3) edge [] node[label=left:\footnotesize{$a_{N-2k+5}$}] {} (4)
			(4) edge [] node[label=left:\footnotesize{$a_{N-2k+7}$}] {} (5)
			(6) edge [] node[label=left:\footnotesize{$a_{N-1}$}] {} (7);
	\end{tikzpicture} &
	 \begin{tikzpicture}
		\tikzstyle{hasse} = [circle, fill,inner sep=2pt];
		\node [hasse] (1) [label=right:\footnotesize{$k(N-k)+1$}] {};
		\node [] (2) [below of=1] {};
		\node [hasse] (3) [below of=2] {};
		\node [hasse] (4) [below of=3] {};
		\node [hasse] (5) [below of=4] {};
		\node [hasse] (6) [below of=5] {};
		\node [hasse] (7) [below of=6] {};
				\node at (0,-4.5) {$\vdots$};
		\draw (1) edge [] node[label=left:\footnotesize{$d_{N -2k+4}$}] {} (3)
			(3) edge [] node[label=left:\footnotesize{$a_{N-2k+5}$}] {} (4)
			(4) edge [] node[label=left:\footnotesize{$a_{N-2k+7}$}] {} (5)
			(6) edge [] node[label=left:\footnotesize{$a_{N-1}$}] {} (7);
	\end{tikzpicture}\\ \hline
	
		\end{tabular}
	\caption{Hasse diagrams for gauge theories with unitary and special unitary gauge groups and enough fundamental matter.  Note that the only difference from left to right is that two $a$-transitions on the top have become a $d$-transition on the right, accompanied by an extra $\mathrm{U}(1)$ factor in the global symmetry of the right magnetic quiver.}\label{tab:hassetables1}
\end{table}

\paragraph{$U(k)$ gauge group. }
This case is very elementary, as the Higgs branch is a nilpotent orbit of height 2 of $\mathfrak{sl}(N,\mathbb{C})$. We assume $N \geq 2k$. The Hasse diagram displayed in Table \ref{tab:hassetables1} is a line, with elementary slices of type $a$. It is a subset of the Hasse diagram of the nilpotent orbits of $\mathfrak{sl}(N,\mathbb{C})$. 

When $N < 2k$, the theory might not be completely Higgsed. However, the Higgs branch can be seen as the Higgs branch of a theory with gauge group of smaller rank. 

\paragraph{$\mathrm{SU}(k)$ gauge group.}
This is a slight modification of the $U(k)$ case, and we still assume $N \geq 2k$. The Higgs branch is a baryonic extension of an $\mathfrak{sl}(N,\mathbb{C})$ nilpotent orbit.\footnote{This name comes from the fact that the generators of the nilpotent orbits can be identified with the mesons of the theory. In the $U(k)$ theory, there is no further gauge invariant operator, but in the $\mathrm{SU}(k)$ theory, there are baryons which provide additional generators. } It turns out the baryons affect only the top dimensional leaf of the diagram, and are eliminated after a transition of type $d$. Schematically, one can say that two transitions of type $a$ (namely $a_{N-2k+1}$ and $a_{N-2k+3}$, of combined dimension $2N-4k+4$) of the $U(k)$ theory are traded for one transition of type $d$ (namely $d_{N-2k+4}$, of dimension $2N-4k+5$). The Hasse diagram is shown in Table \ref{tab:hassetables1}. 

When the number of flavours is $N < 2k$, interesting phenomena occur, linked to the presence of nilpotent operators in the chiral ring. These nilpotent operators confer an additional structure to the Hasse diagram, giving non-trivial multiplicities to the leaves. We refer to \cite{Bourget:2019rtl} for in-depth analysis of this particular problem. 

\begin{table}
	\centering
	\begin{tabular}{|l|c|c|}
	\hline
         Theory &  $\mathrm{Sp}(k)$ with $\mathrm{SO}(2N), \; N \geq 2k$ &  $\mathrm{Sp}(k)$ with $\mathrm{SO}(2N+1)$, $N \geq 2k$
		 \\ \hline
      Electric quiver & $ \node{\flav {\mathrm{SO}(2N)}}{\mathrm{Sp}(k)} $ & $ \node{\flav {\mathrm{SO}(2N+1)}}{\mathrm{Sp}(k)} $ \\ \hline
      Magnetic quiver &
      $\node{}{1}-\node{}{2}-\cdots- \underbrace{  \node{\topnode{}{1}}{2k}-\cdots-\node{\topnode{}{k}}{2k}}_{N-2k-1} - \node{}{k}$
        &  $\node{}{1}-\node{}{2}-\cdots- \underbrace{  \node{\topnode{}{1}}{2k}-\cdots-\node{}{2k}}_{N-2k} => \node{}{k}$
         \\ \hline
      Hasse diagram &
	\begin{tikzpicture}
		\tikzstyle{hasse} = [circle, fill,inner sep=2pt];
		\node [hasse] (1) [label=right:\footnotesize{$2kN-2k^2-k$}] {};
		\node [hasse] (2) [below of=1] {};
		\node [hasse] (3) [below of=2] {};
		\node [hasse] (4) [below of=3] {};
		\node [hasse] (5) [below of=4] {};
		\node at (0,-1.5) {$\vdots$};
		\draw (1) edge [] node[label=left:\footnotesize{$d_{N-2(k-1)}$}] {} (2)
			(3) edge [] node[label=left:\footnotesize{$d_{N-2}$}] {} (4)
			(4) edge [] node[label=left:\footnotesize{$d_{N}$}] {} (5);
	\end{tikzpicture}
	&
				\begin{tikzpicture}
		\tikzstyle{hasse} = [circle, fill,inner sep=2pt];
		\node [hasse] (1) [label=right:\footnotesize{$2kN-2k^2$}] {};
		\node [hasse] (2) [below of=1] {};
		\node [hasse] (3) [below of=2] {};
		\node [hasse] (4) [below of=3] {};
		\node [hasse] (5) [below of=4] {};
		\node at (0,-1.5) {$\vdots$};
		\draw (1) edge [] node[label=left:\footnotesize{$b_{N-2(k-1)}$}] {} (2)
			(3) edge [] node[label=left:\footnotesize{$b_{N-2}$}] {} (4)
			(4) edge [] node[label=left:\footnotesize{$b_{N}$}] {} (5);
	\end{tikzpicture}
	
	 \\ \hline
	\end{tabular}
	\caption{Hasse diagrams for gauge theories with symplectic gauge groups and enough fundamental matter. }\label{tab:hassetables2}
\end{table}

\paragraph{$\mathrm{Sp}(k)$ gauge group.}
Here we have to distinguish between even and odd number of flavours. When the number of flavours is even, the global symmetry is $\mathrm{SO}(2N)$, and the Higgs branch is a nilpotent orbit of height 2 of $\mathfrak{so}(2N,\mathbb{C})$. This is very similar to the case studied before of $U(k)$ gauge group. The Hasse diagram is linear and consists of a chain of transitions of type $d$, see Table \ref{tab:hassetables2}. Note that for $k=1$, we have $\mathrm{Sp}(k)=\mathrm{Sp}(1)=\mathrm{SU}(2)$ and the results agree with our discussion of baryonic extensions above. 

More subtle is the case with global symmetry $B_N = \mathrm{SO}(2N+1)$. This corresponds to an odd number of half-hypermultiplets. We consider here only the classical theory. Quantum mechanically, this theory is anomalous in four dimensions, and requires a half-integer Chern-Simons level in three dimensions to be consistent. 
The Higgs branch is now a height 2 nilpotent orbit of $\mathfrak{so}(2N+1,\mathbb{C})$, and from that perspective we can read the elementary slices for the nilpotent orbits Hasse diagram. They are of type $b$, which means that the magnetic quivers involved will be non-simply laced. The rules of quiver subtraction of Appendix \ref{AppendixSubtraction} do not apply directly.

\begin{table}
	\centering
	\begin{tabular}{|l|c|c|}
	\hline
	Theory & $\mathrm{O}(k)$ with $\mathrm{Sp}(N)$, $N \geq k$  & $G_2$ with $\mathrm{Sp}(N)$, $N \geq 4$
	 \\ \hline
	 Electric quiver & $\node{\flav {\mathrm{Sp}(N)}}{\mathrm{O}(k)}$ & $ \node{\flav {\mathrm{Sp}(N)}}{G_2}$ \\ \hline
	 Magnetic quiver & $\node{}{1}-\node{}{2}-\cdots- \underbrace{  \node{\topnode{}{1}}{k}-\cdots-\node{}{k}}_{N-k} <= \node{}{k}$ & no known unitary \\ \hline
	 Hasse diagram &
	 	\begin{tikzpicture}
		\tikzstyle{hasse} = [circle, fill,inner sep=2pt];
		\node [hasse] (1) [label=right:\footnotesize{$kN+\frac{k}{2}(1-k)$}] {};
		\node [hasse] (2) [below of=1] {};
		\node [hasse] (3) [below of=2] {};
		\node [hasse] (4) [below of=3] {};
		\node [hasse] (5) [below of=4] {};
		\node at (0,-1.5) [] {$\vdots$};
		\draw (1) edge [] node[label=left:\footnotesize{$c_{N-(k-1)}$}] {} (2)
			(3) edge [] node[label=left:\footnotesize{$c_{N-1}$}] {} (4)
			(4) edge [] node[label=left:\footnotesize{$c_{N}$}] {} (5);
	\end{tikzpicture}
 &
  	\begin{tikzpicture}
		\tikzstyle{hasse} = [circle, fill,inner sep=2pt];
		\node [hasse] (1) [label=right:\footnotesize{$7N-14$}] {};
		\node [hasse] (2) [label=right:\footnotesize{$3(N-1)$},below of=1] {};
		\node [hasse] (3) [label=right:\footnotesize{$N$},below of=2] {};
		\node [hasse] (4) [below of=3] {};
		\draw (1) edge [] node[label=left:\footnotesize{$d_{2N-4}$}] {} (2)
			(2) edge [] node[label=left:\footnotesize{$a_{2N-3}$}] {} (3)
			(3) edge [] node[label=left:\footnotesize{$c_{N}$}] {} (4);
	\end{tikzpicture}
	 \\ \hline
	\end{tabular}
	\caption{Hasse diagrams for theories with orthogonal and $G_2$ gauge groups with enough fundamental matter.  }\label{tab:hassetables3}
\end{table}

\paragraph{$\mathrm{O}(k)$ gauge group.}
This case is similar to $\mathrm{Sp}(k)$ with symmetry $\mathrm{SO}(2N+1)$: we again find height 2 nilpotent orbits, this time of $\mathfrak{sp}(N,\mathbb{C})$, and again the transitions involve non-simply laced quivers. The Hasse diagram is displayed in Table \ref{tab:hassetables3}. In particular, note that it is important that the gauge group is $\mathrm{O}(k)$ and not $\mathrm{SO}(k)$.

\begin{figure}
    \centering
\begin{tikzpicture}
	\begin{pgfonlayer}{nodelayer}
		\node [style=miniG] (1) at (0, 4) {};
		\node [style=miniG] (2) at (0, 0) {};
		\node [style=miniG] (4) at (0, -8) {};
		\node [style=none] (6) at (-0.75, -2.25) {$a_{2N-3}$};
		\node [style=none] (9) at (-0.75, -6) {$c_N$};
		\node [style=none] (10) at (-0.75, 1.75) {$d_{2N-4}$};
		\node [style=none] (11) at (1.25, 4) {};
		\node [style=none] (12) at (1.75, 4) {};
		\node [style=none] (13) at (1.75, 0.075) {};
		\node [style=none] (14) at (1.25, 0.075) {};
		\node [style=none] (23) at (3.25, 4) {};
		\node [style=none] (24) at (3.75, 4) {};
		\node [style=none] (25) at (3.75, -3.875) {};
		\node [style=none] (26) at (3.25, -3.875) {};
		\node [style=none] (27) at (1.25, -0.075) {};
		\node [style=none] (28) at (1.75, -0.075) {};
		\node [style=none] (29) at (1.75, -8) {};
		\node [style=none] (30) at (1.25, -8) {};
		\node [style=none] (31) at (5.25, 4) {};
		\node [style=none] (32) at (5.75, 4) {};
		\node [style=none] (33) at (5.75, -8) {};
		\node [style=none] (34) at (5.25, -8) {};
		\node [style=gauge1] (47) at (2.5, 1.5) {};
		\node [style=flavour1] (48) at (2.5, 3) {};
		\node [style=none] (49) at (2.5, 1) {$\mathrm{SU}(2)$};
		\node [style=none] (50) at (2.5, 3.5) {$2N-4$};
		\node [style=gauge1] (59) at (4.5, -0.75) {};
		\node [style=flavour1] (60) at (4.5, 0.75) {};
		\node [style=none] (61) at (4.5, -1.25) {$\mathrm{SU}(3)$};
		\node [style=none] (62) at (4.5, 1.25) {$2N-2$};
		\node [style=gauge1] (67) at (6.75, -2.5) {};
		\node [style=flavour1] (68) at (6.75, -1) {};
		\node [style=none] (69) at (6.75, -3) {$G_2$};
		\node [style=none] (70) at (6.75, -0.5) {$C_N$};
		\node [style=gauge1] (86) at (2.5, -4.75) {};
		\node [style=flavour1] (87) at (2.5, -3.25) {};
		\node [style=none] (88) at (2.5, -5.25) {$\mathrm{SO}(3)$};
		\node [style=none] (89) at (2.5, -2.75) {$C_N$};
		\node [style=none] (90) at (3.25, -4.125) {};
		\node [style=none] (91) at (3.75, -4.125) {};
		\node [style=none] (92) at (3.75, -8) {};
		\node [style=none] (93) at (3.25, -8) {};
		\node [style=gauge1] (94) at (4.5, -6.75) {};
		\node [style=flavour1] (95) at (4.5, -5.25) {};
		\node [style=none] (96) at (4.5, -7.25) {$\mathrm{O}(1)$};
		\node [style=none] (97) at (4.5, -4.75) {$C_N$};
		\node [style=miniG] (98) at (0, -4) {};
		\node [style=none] (99) at (-2, 0) {};
		\node [style=none] (100) at (-2.5, 0) {};
		\node [style=none] (101) at (-2.5, -4) {};
		\node [style=none] (102) at (-2, -4) {};
		\node [style=gauge1] (103) at (-3.5, -2.75) {};
		\node [style=flavour1] (104) at (-3.5, -1.25) {};
		\node [style=none] (105) at (-3.5, -3.25) {$\mathrm{U}(1)$};
		\node [style=none] (106) at (-3.5, -0.75) {$2N-2$};
	\end{pgfonlayer}
	\begin{pgfonlayer}{edgelayer}
		\draw (1) to (2);
		\draw (11.center) to (12.center);
		\draw (11.center) to (12.center);
		\draw (12.center) to (13.center);
		\draw (13.center) to (14.center);
		\draw (23.center) to (24.center);
		\draw (23.center) to (24.center);
		\draw (24.center) to (25.center);
		\draw (25.center) to (26.center);
		\draw (27.center) to (28.center);
		\draw (27.center) to (28.center);
		\draw (28.center) to (29.center);
		\draw (29.center) to (30.center);
		\draw (31.center) to (32.center);
		\draw (31.center) to (32.center);
		\draw (32.center) to (33.center);
		\draw (33.center) to (34.center);
		\draw (47) to (48);
		\draw (59) to (60);
		\draw (67) to (68);
		\draw (86) to (87);
		\draw (90.center) to (91.center);
		\draw (90.center) to (91.center);
		\draw (91.center) to (92.center);
		\draw (92.center) to (93.center);
		\draw (94) to (95);
		\draw (2) to (98);
		\draw (98) to (4);
		\draw (99.center) to (100.center);
		\draw (100.center) to (101.center);
		\draw (101.center) to (102.center);
		\draw (103) to (104);
	\end{pgfonlayer}
\end{tikzpicture}
    \caption{Hasse diagram for $G_2$ with $N \geq 4$ fundamentals with electric quivers associated to every subdiagram between two points in the Hasse diagram.
    }
    \label{G2Hasse}
\end{figure}
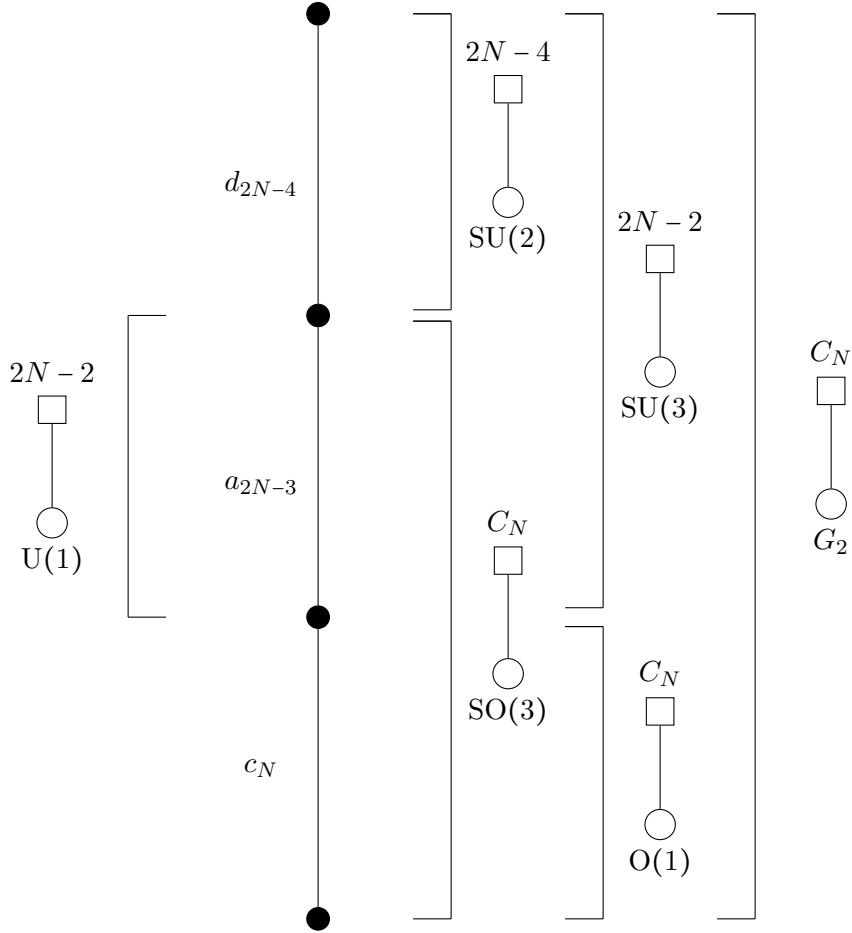

\paragraph{$G_2$ gauge group.} In this final example, we are forced out of the range of nilpotent orbits, and in addition we do not know of magnetic quivers to represent the Higgs branch. However, the Hasse diagram shown in Table \ref{tab:hassetables3} can be computed using group theory. Repeating the computation of Sections \ref{higgsforsu3} and \ref{higgsforsu4} for $G_2$ with $N$ fundamental hypermultiplets in Figure \ref{G2Hasse}, we find, that we can Higgs the theory to $\mathrm{SU}(3)$ with $2N-2$ fundamental hypermultiplets and that the elementary slice to the origin of the Higgs branch has dimension $N$. Since the fundamental representation of $G_2$ is real we expect an $\mathrm{Sp}(N)$ global symmetry. Hence the first line in the Hasse diagram is $c_N$. The rest of the Hasse diagram is easily computed as the Hasse diagram of $\mathrm{SU}(3)$ with $2N-2$ flavours, which can be read from Table \ref{tab:hassetables1} and the entire Hasse diagram is fixed. However, we are unable to give the magnetic quiver for this theory as quiver addition is not a unique operation, see Appendix \ref{quiveradd}.

\paragraph{Other exceptional gauge groups.} In principle, the technique used to derive the diagram of $G_2$ applies similarly for the other exceptional groups. However, the results are considerably more intricate (they involve both non-simple gauge groups and matter content in non-fundamental representations at various stages of the partial Higgsing process), and we postpone this computation for future work. 

\subsection{5d SQCD at Infinite Coupling}
\label{subsection5d}

\begin{table}[t]
\centering
$    \begin{array}{|c|c|} \hline 
        \textrm{First region} & \begin{array}{ccc}
             \begin{array}{c}
             |k| > \frac{1}{2}  \\
             \textrm{Cones Tab. } \ref{tab:region1}, \ref{tab:region1seq} \\ 
             \textrm{Tree Tab. } \ref{tab:hassetables5dGen1}
        \end{array}  & 
        \begin{array}{c}
             |k| = \frac{1}{2}  \\
             \textrm{Cones Tab. } \ref{tab:region1kHalf} \\ 
             \textrm{Tree Tab. } \ref{tab:hassetables5dGen1}
        \end{array} 
        & 
        \begin{array}{c}
             |k| = 0  \\
             \textrm{Cones Tab. } \ref{tab:region1k0} \\ 
             \textrm{Tree Tab. } \ref{tab:hassetables5dGen1}
        \end{array} 
        \end{array}    \\  \hline 
        \textrm{Second region} & \begin{array}{ccc}
             \begin{array}{c}
             |k| > 1  \\
             \textrm{Cones Tab. } \ref{tab:region2} \\ 
             \textrm{Tree Tab. } \ref{tab:hassetables5dGen1}
        \end{array}  & 
        \begin{array}{c}
             |k| =1  \\
             \textrm{Cones Tab. } \ref{tab:region2k1} \\ 
             \textrm{Tree Tab. } \ref{tab:hassetables5dGen1}
        \end{array} 
        \\ & \vspace{-1em} \\
        \begin{array}{c}
             |k| = \frac{1}{2}  \\
             \textrm{Cone Tab. } \ref{tab:region2kHalf} \\ 
             \textrm{Only one cone}
        \end{array} 
        & 
        \begin{array}{c}
             |k| =0  \\
             \textrm{Cone Tab. } \ref{tab:region2k0} \\ 
             \textrm{Only one cone}
        \end{array} 
        \end{array}    \\  \hline 
        \textrm{Third region} & \begin{array}{ccc}
        \begin{array}{c}
             |k| > \frac{3}{2}  \\
             \textrm{Cones Tab. } \ref{tab:region3} \\ 
             \textrm{Tree Tab. } \ref{tab:hassetables5dGen2}
        \end{array} &
        \begin{array}{c}
             |k| = \frac{3}{2}  \\
             \textrm{Cone Tab. } \ref{tab:region3k3Halves} \\ 
             \textrm{Only one cone}
        \end{array} 
             \begin{array}{c}
             |k| =1  \\
             \textrm{Cone Tab. } \ref{tab:region3k1} \\ 
             \textrm{Only one cone}
        \end{array}  \\  & & \vspace{-1em} \\ 
        \begin{array}{c}
             |k| = \frac{1}{2}  \\
             \textrm{Cone Tab. } \ref{tab:region3kHalf} \\ 
             \textrm{Only one cone}
        \end{array} 
        & 
        \begin{array}{c}
             |k| =0  \\
             \textrm{Cone Tab. } \ref{tab:region3k0} \\ 
             \textrm{Only one cone}
        \end{array} 
        \end{array}   \\   \hline 
        \textrm{Fourth region} & \begin{array}{ccc}
             \begin{array}{c}
             |k| > 2  \\
             \textrm{Cones Tab. } \ref{tab:region4} \\ 
             \textrm{Tree Tab. } \ref{tab:hassetables5dGen2}
        \end{array}  & 
        \begin{array}{c}
             |k| = 2  \\
             \textrm{Cone Tab. } \ref{tab:region4k2} \\ 
             \textrm{Only one cone}
        \end{array} 
        & 
        \begin{array}{c}
             |k| = \frac{3}{2}  \\
             \textrm{Cone Tab. } \ref{tab:region4k3Halves} \\ 
             \textrm{Only one cone}
        \end{array} 
         \\ & & \vspace{-1em} \\
        \begin{array}{c}
             |k| = 1   \\
             \textrm{Cone Tab. } \ref{tab:region4k1} \\ 
             \textrm{Only one cone}
        \end{array} 
        & 
        \begin{array}{c}
             |k| = \frac{1}{2}  \\
             \textrm{Cone Tab. } \ref{tab:region4kHalf} \\ 
             \textrm{Only one cone}
        \end{array} 
        & 
        \begin{array}{c}
             |k| = 0  \\
             \textrm{6d theory}
        \end{array} 
        \end{array}  \\  \hline 
    \end{array}$
    \caption{Summary of results of 5d SQCD at infinite coupling.}
    \label{tab:5dinf}
\end{table}

In this section, a three parameter family of 5d $\mathcal{N}=1$ SCFTs is considered. These are the strong coupling limits of SQCD with gauge group $\mathrm{SU}(N_c)$, $N_f$ flavours of fundamental matter and Chern-Simons level $k$. This level satisfies $k \in \mathbb{Z}$ when $N_f$ is even and $k \in \mathbb{Z} + \frac{1}{2}$ when $N_f$ is odd. The analysis closely follows \cite{Cabrera:2018jxt}, which employed 5-brane webs to derive the magnetic quivers for the Higgs branches at infinite coupling. This generally imposes the condition
\begin{equation}
    |k | \leq N_c - \frac{N_f}{2} + 2 \,. \label{eq:range_5d_para}
\end{equation}
(Some exceptional cases have been analysed in \cite{Cabrera:2018jxt}, but these are omitted here.)
In the range of parameters given in \eqref{eq:range_5d_para}, one can compute the magnetic quivers for the Higgs branches and use quiver subtractions to determine the Hasse diagrams. It is useful to divide this parameter space into the following four regions:
\begin{equation}
    \begin{array}{|c|c|} \hline 
        \textrm{First region} & |k| < N_c - \frac{N_f}{2}   \\  \hline 
        \textrm{Second region} & |k| = N_c - \frac{N_f}{2}   \\  \hline 
        \textrm{Third region} & |k| = N_c - \frac{N_f}{2} +1  \\  \hline 
        \textrm{Fourth region} & |k| = N_c - \frac{N_f}{2}  +2 \\  \hline 
    \end{array}
\end{equation}
The Higgs branches are unions of one or more cones. All the individual cones, which correspond to the magnetic quivers computed in \cite{Cabrera:2018jxt}, are tabulated in the Appendix \ref{full5d} and Table \ref{tab:5dinf} provides a summary of the results. The results of this section are organised in tables summarized in table \ref{tab:5dinf}

\begin{table}[t]
	\centering
	\begin{tabular}{|c|c|c|}
	\hline
	 First region, $N_c-\frac{N_f}{2}>1$  &  First region $N_c-\frac{N_f}{2}=1$  & Second region, $|k|>1$
	 \\ \hline
\begin{tikzpicture}[node distance=30pt]
	\tikzstyle{hasse} = [circle, fill,inner sep=2pt];
		\node at (-0.7,-0.5) [] (1a) [] {};
		\node at (0,-0.5) [hasse] (1b) [label=above:\footnotesize{$III$}] {};
		\node at (0.7,-0.5) [] (1c) [] {};
		\node [] (2a) [below of=1a] {};
		\node [hasse] (2b) [below of=1b] {};
		\node [] (2c) [below of=1c] {};
		\node  (3a) [below of=2a] {};
		\node [hasse] (3b) [below of=2b] {};
		\node  (3c) [below of=2c] {};
		\node [hasse] (4a) [label=above:\footnotesize{$I$},below of=3a] {};
		\node [hasse] (4b)[below of=3b] {};
		\node (4c) [below of=3c] {};
		\node (5a) [below of=4a] {};
		\node [hasse] (5b) [below of=4b] {};
		\node (5c) [below of=4c] {};
		\node (6a) [below of=5a] {};
		\node [hasse] (6b) [below of=5b] {};
		\node (6c) [below of=5c] {};
		\node [hasse] (7a) [label=above:\footnotesize{$II$},below of=6a] {};
		\node [hasse] (7b) [below of=6b] {};
		\node (7c) [below of=6c] {}; 
		\node (8a) [below of=7a] {};
		\node [hasse] (8b) [below of=7b] {};
	        \node (8c) [below of=7c] {};
	         \node (9a) [below of=8a] {};
	         \node [hasse] (9b) [below of=8b] {};
	          \node (9c) [below of=8c] {};
	          \node (10a) [below of=9a] {};
	          \node [hasse] (10b) [below of=9b] {};
	          \node (10c) [below of=9c] {};
	          \node (11a) [below of=10a] {};
	          \node [hasse] (11b) [below of=10b] {};
	          \node (11c) [below of=10c] {};
	          \node at (0,-3.1)[]{$\vdots$};
	          \node at (0,-6.2)[]{$\vdots$};
	           \node at (0,-9.4)[]{$\vdots$};
		\draw (1b) edge [] node[label=right:\footnotesize{$a_X$}] {} (2b)
			(2b) edge [] node[label=right:\footnotesize{$a_{X+2}$}] {} (3b)
			(5b) edge [] node[label=right:\footnotesize{$a_{2|k|+1}$}] {} (6b)
			(4b) edge [] node[label=right:\footnotesize{$a_{2|k|-1}$}] {} (5b)
			(4a) edge [] node[label=left:\footnotesize{$A_{N_c-\frac{N_f}{2}+|k|-1}$}] {} (5b)
			(7a) edge [] node[label=left:\footnotesize{$A_{2N_c-N_f-1}$}] {} (8b)
			(7b) edge [] node[label=right:\footnotesize{$a_{2N_c-N_f-1}$}] {} (8b)
			(8b) edge [] node[label=right:\footnotesize{$a_{2N_c-N_f+1}$}] {} (9b)
			(10b) edge [] node[label=right:$a_{N_f-1}$] {} (11b);
	\end{tikzpicture}
	&

\begin{tikzpicture}[node distance=30pt]
	\tikzstyle{hasse} = [circle, fill,inner sep=2pt];
		\node [hasse] (4b) [label=above:\footnotesize{$I$}] {};
		\node [hasse] (5b) [below of=4b] {};
		\node [hasse] (6b) [below of=5b] {};
		\node [hasse] (7b) [below of=6b] {};
		\node [hasse] (7a) [label=above:\footnotesize{$II$},left of=7b] {};
		\node (8a) [below of=7a] {};
		\node [hasse] (8b) [below of=7b] {};
	         \node [hasse] (9b) [below of=8b] {};
	          \node [hasse] (10b) [below of=9b] {};
	          \node [hasse] (11b) [below of=10b] {};
	          \node at (0,-2.6)[]{$\vdots$};
	           \node at (0,-5.8)[]{$\vdots$};
		\draw 
			(4b) edge [] node[label=right:\footnotesize{$a_{2}$}] {} (5b)
			(5b) edge [] node[label=right:\footnotesize{$a_{3}$}] {} (6b)
			(7a) edge [] node[label=left:\footnotesize{$A_{2N_c-N_f-1}$}] {} (8b)
			(7b) edge [] node[label=right:\footnotesize{$a_{2N_c-N_f-1}$}] {} (8b)
			(8b) edge [] node[label=right:\footnotesize{$a_{2N_c-N_f+1}$}] {} (9b)
			(10b) edge [] node[label=right:$a_{N_f-1}$] {} (11b);
	\end{tikzpicture}
	& \begin{tikzpicture}[node distance=20pt]
	\tikzstyle{hasse} = [circle, fill,inner sep=2pt];
		\node at (-0.7,-0.5) [hasse] (1a) [label=above:\footnotesize{$III$}] {};
		\node at (0,-0.5) [] (1b) [label=right:\footnotesize{$ $}] {};
		\node at (0.7,-0.5) [] (1c) [] {};
		\node  (2a) [below of=1a] {};
		\node [hasse] (2b) [below of=1b] {};
		\node (2c) [below of=1c] {};
		\node [hasse] (3a) [below of=2a] {};
		\node (3b) [below of=2b] {};
		\node (3c) [below of=2c] {};
		\node (4a) [below of=3a] {};
		\node [hasse] (4b)[below of=3b] {};
		\node (4c) [below of=3c] {};
		\node [hasse] (5a) [below of=4a] {};
		\node (5b) [below of=4b] {};
		\node (5c) [below of=4c] {};
		\node (6a) [below of=5a] {};
		\node [hasse] (6b) [below of=5b] {};
		\node (6c) [below of=5c] {};
		\node [hasse] (7a) [below of=6a] {};
		\node (7b) [below of=6b] {};
		\node (7c) [below of=6c] {}; 
		\node (8a) [below of=7a] {};
		\node [hasse] (8b) [below of=7b] {};
	    \node (8c) [below of=7c] {};
	        \node [hasse] (9a) [below of=8a] {};
		\node (9b) [below of=8b] {};
	        \node (9c) [below of=8c] {};
	        \node (10a) [below of=9a] {};
		\node [hasse] (10b) [below of=9b] {};
	        \node (10c) [below of=9c] {};
	        \node [hasse] (11a) [below of=10a] {};
		\node (11b) [below of=10b] {};
	        \node (11c) [below of=10c] {};
	        \node (12a) [below of=11a] {};
		\node [hasse] (12b) [below of=11b] {};
	        \node (12c) [below of=11c] {};
	        \node [hasse] (13a) [below of=12a] {};
		\node  (13b) [below of=12b] {};
	        \node (13c) [below of=12c] {};
	        \node(14a) [below of=13a] {};
		\node [hasse] (14b) [below of=13b] {};
	        \node (14c) [below of=13c] {};
            \node [hasse] (15a) [below of=14a] {};
		\node (15b) [below of=14b] {};
	        \node (15c) [below of=14c] {};
	        \node(16a) [below of=15a] {};
		\node [hasse] (16b) [below of=15b] {};
	        \node (16c) [below of=15c] {};
	          \node at (-0.4,-4.1)[]{$\vdots$};
	          \node at (-0.4,-8.3)[]{$\vdots$};
	          \node at (-3,-4.6) [hasse] (8aa) [label=above:\footnotesize{{I}}] {};
	          \node at (-1.3,-5.2) {\footnotesize{$a_{2|k|-1}$}}; 
	          \node at (-2.4,-5.5) {\footnotesize{$A_{2|k|-2}$}};
		\draw (1a) edge [] node[label=above:\footnotesize{$a_{1}$}] {} (2b)
			(1a) edge [] node[label=left:\footnotesize{$a_{X}$}] {} (3a)
			(2b) edge [] node[label=right:\footnotesize{$a_{X}$}] {} (4b)
			(3a) edge [] node[label=above:\footnotesize{$a_1$}] {} (4b)
			(4b) edge [] node[label=right:\footnotesize{$a_{X+2}$}] {} (6b)
			(5a) edge [] node[label=above:\footnotesize{$a_{1}$}] {} (6b)
			(3a) edge [] node[label=left:\footnotesize{$a_{X+2}$}] {} (5a)
			(7a) edge [] node[label=above:\footnotesize{$a_{1}$}] {} (8b)
			(7a) edge [] node[label=left:\footnotesize{$ $}] {} (9a)
			(8aa) edge [] node[label=left:\footnotesize{}] {} (9a)
			(9a) edge [] node[label=above:\footnotesize{$a_{1}$}] {} (10b)
			(8b) edge [] node[label=right:\footnotesize{$a_{2|k|-1}$}] {} (10b)
			(9a) edge [] node[label=left:\footnotesize{$a_{2|k|+1}$}] {} (11a)
			(11a) edge [] node[label=above:\footnotesize{$a_{1}$}] {} (12b)
			(10b) edge [] node[label=right:\footnotesize{$a_{2|k|+1}$}] {} (12b)
			(15a) edge [] node[label=above:\footnotesize{$a_{1}$}] {} (16b)
			(13a) edge [] node[label=left:\footnotesize{$a_{N_f-1}$}] {} (15a)
			(14b) edge [] node[label=right:\footnotesize{$a_{N_f-1}$}] {} (16b)
			(13a) edge [] node[label=above:\footnotesize{$a_{1}$}] {} (14b);
	\end{tikzpicture}
	\\ \hline
		\end{tabular}
	\caption{General Hasse diagrams of $5d$ SQCD: first and second regions. The roman numerals indicate the various cones which constitute the Higgs branch. For $N_f$ odd, $X=2$ and for $N_f$ even $X=1$.}\label{tab:hassetables5dGen1}
\end{table}

\begin{table}[t]
	\centering
	\begin{tabular}{|c|c|c|}
	\hline
	Third region, $|k|=1$ & Third region, $|k|>\frac{3}{2}$  & $4$-th region, $|k|>2$, $N_f$ even 
	 \\ \hline
	 \begin{tikzpicture}[node distance=20pt]
	\tikzstyle{hasse} = [circle, fill,inner sep=2pt];
		\node at (-0.7,-0.5) [hasse] (1a) [label=above:\footnotesize{$I$}] {};
		\node at (0,-0.5) [] (1b) [label=right:\footnotesize{$ $}] {};
		\node at (0.7,-0.5) [hasse] (1c) [label=above:\footnotesize{$III$}] {};
		\node  (2a) [below of=1a] {};
		\node [hasse] (2b) [below of=1b] {};
		\node (2c) [below of=1c] {};
		\node [hasse] (3a) [below of=2a] {};
		\node (3b) [below of=2b] {};
		\node (3c) [below of=2c] {};
		\node (4a) [below of=3a] {};
		\node [hasse] (4b)[below of=3b] {};
		\node (4c) [below of=3c] {};
		\node [hasse] (5a) [below of=4a] {};
		\node (5b) [below of=4b] {};
		\node (5c) [below of=4c] {};
		\node (6a) [below of=5a] {};
		\node [hasse] (6b) [below of=5b] {};
		\node (6c) [below of=5c] {};
		\node [hasse] (7a) [below of=6a] {};
		\node (7b) [below of=6b] {};
		\node (7c) [below of=6c] {}; 
		\node (8a) [below of=7a] {};
		\node [hasse] (8b) [below of=7b] {};
	        \node (8c) [below of=7c] {};
	        	\node [hasse] (9a) [below of=8a] {};
		\node (9b) [below of=8b] {};
	        \node (9c) [below of=8c] {};
	        	\node (10a) [below of=9a] {};
		\node [hasse] (10b) [below of=9b] {};
	        \node (10c) [below of=9c] {};
	        \node [hasse] (11a) [below of=10a] {};
		\node (11b) [below of=10b] {};
	        \node (11c) [below of=10c] {};
	        	\node (12a) [below of=11a] {};
		\node [hasse] (12b) [below of=11b] {};
	        \node (12c) [below of=11c] {};
	          \node at (-0.4,-5.5)[]{$\vdots$};
		\draw (1a) edge [] node[label=above:\footnotesize{$\;\;\;a_{2}$}] {} (2b)
			(1a) edge [] node[label=left:\footnotesize{$a_{4}$}] {} (3a)
				(1c) edge [] node[label=above:\footnotesize{$a_{1}\;$}] {} (2b)
			(2b) edge [] node[label=right:\footnotesize{$a_{3}$}] {} (4b)
			(3a) edge [] node[label=above:\footnotesize{$a_1$}] {} (4b)
				(7a) edge [] node[label=above:\footnotesize{$a_1$}] {} (8b)
			(4b) edge [] node[label=right:\footnotesize{$a_{5}$}] {} (6b)
			(6b) edge [] node[label=right:\footnotesize{$a_{7}$}] {} (8b)
			(3a) edge [] node[label=left:\footnotesize{$a_{5}$}] {} (5a)
			(5a) edge [] node[label=left:\footnotesize{$a_{7}$}] {} (7a)
			(5a) edge [] node[label=above:\footnotesize{$a_{1}$}] {} (6b)
			(9a) edge [] node[label=above:\footnotesize{$a_{1}$}] {} (10b)
			(9a) edge [] node[label=left:\footnotesize{$a_{N_f-1}$}] {} (11a)
			(10b) edge [] node[label=right:\footnotesize{$a_{N_f-1}$}] {} (12b)
			(11a) edge [] node[label=above:\footnotesize{$a_{1}$}] {} (12b);
	\end{tikzpicture} & 
\begin{tikzpicture}[node distance=30pt]
	\tikzstyle{hasse} = [circle, fill,inner sep=2pt];
		\node at (-0.7,-0.5) [] (1a) [] {};
		\node at (0,-0.5) [hasse] (1b) [label=above:\footnotesize{$III$}] {};
		\node at (0.7,-0.5) [] (1c) [] {};
		\node [] (2a) [below of=1a] {};
		\node [hasse] (2b) [below of=1b] {};
		\node [] (2c) [below of=1c] {};
		\node  (3a) [below of=2a] {};
		\node [hasse] (3b) [below of=2b] {};
		\node  (3c) [below of=2c] {};
		\node [hasse] (4a) [label=above:\footnotesize{$I$},below of=3a] {};
		\node [hasse] (4b)[below of=3b] {};
		\node (4c) [below of=3c] {};
		\node (5a) [below of=4a] {};
		\node [hasse] (5b) [below of=4b] {};
		\node (5c) [below of=4c] {};
		\node (6a) [below of=5a] {};
		\node [hasse] (6b) [below of=5b] {};
		\node (6c) [below of=5c] {};
		\node (7a) [below of=6a] {};
		\node [hasse] (7b) [below of=6b] {};
		\node (7c) [below of=6c] {}; 
		\node (8a) [below of=7a] {};
		\node [hasse] (8b) [below of=7b] {};
	        \node (8c) [below of=7c] {};
	          \node at (0,-3)[]{$\vdots$};
	          \node at (0,-6.2)[]{$\vdots$};
		\draw (1b) edge [] node[label=right:\footnotesize{$a_X$}] {} (2b)
			(2b) edge [] node[label=right:\footnotesize{$a_{X+2}$}] {} (3b)
			(5b) edge [] node[label=right:\footnotesize{$a_{2|k|}$}] {} (6b)
			(4b) edge [] node[label=right:\footnotesize{$a_{2|k|-2}$}] {} (5b)
			(4a) edge [] node[label=left:\footnotesize{$A_{2|k|-3}$}] {} (5b)
			(7b) edge [] node[label=right:\footnotesize{$a_{N_f}$}] {} (8b);
	\end{tikzpicture}
	
	&
	\begin{tikzpicture}[node distance=30pt]
	\tikzstyle{hasse} = [circle, fill,inner sep=2pt];
		\node at (-0.7,-0.5) [] (1a) [] {};
		\node at (0,-0.5) [] (1b) [] {};
		\node at (0.7,-0.5) [hasse] (1c) [label=above:\footnotesize{$V$}] {};
		\node [hasse] (2a) [label=above:\footnotesize{$IV$},below of=1a] {};
		\node (2b) [below of=1b] {};
		\node [hasse] (2c) [below of=1c] {};
		\node  (3a) [below of=2a] {};
		\node [hasse] (3b) [below of=2b] {};
		\node  (3c) [below of=2c] {};
		\node (4a) [below of=3a] {};
		\node [hasse] (4b)[below of=3b] {};
		\node (4c) [below of=3c] {};
		\node (5a) [below of=4a] {};
		\node [hasse] (5b) [below of=4b] {};
		\node (5c) [below of=4c] {};
		\node (6a) [below of=5a] {};
		\node [hasse] (6b) [below of=5b] {};
		\node (6c) [below of=5c] {};
		\node (7a) [below of=6a] {};
		\node [hasse] (7b) [below of=6b] {};
		\node (7c) [below of=6c] {}; 
	          \node at (0,-5.1)[]{$\vdots$};
		\draw (1c) edge [] node[label=right:\footnotesize{$A_{N_c-\frac{N_f}{2}-1}$}] {} (2c)
			(2a) edge [] node[label=left:\footnotesize{$A_{1}$}] {} (3b)
			(2c) edge [] node[label=right:\footnotesize{$A_{1}$}] {} (3b)
			
			(3b) edge [] node[label=right:\footnotesize{$d_{4}$}] {} (4b)
			(4b) edge [] node[label=right:\footnotesize{$d_{6}$}] {} (5b)
		       (6b) edge [] node[label=right:\footnotesize{$d_{N_f}$}] {} (7b);
	\end{tikzpicture}
		
	\\ \hline
		\end{tabular}
	\caption{General Hasse diagrams of $5d$ SQCD: third and fourth region. For $N_f$ odd, $X=1$ and for $N_f$ even $X=2$.}\label{tab:hassetables5dGen2}
\end{table}

When the Higgs branch is only one cone, there is nothing to add to the Hasse diagrams presented in the appendix. However, when the Higgs branch is a union of two or three cones, one has to specify their intersections. Magnetic quivers for these intersections are also given in \cite{Cabrera:2018jxt}, and one could compute the associated Hasse diagrams. However, the results of these computations are not included in the present paper, because the information they carry can be neatly encapsulated into a bigger Hasse diagram with more than one point of locally maximal dimension. 

For instance, consider the case $|k|>\frac{1}{2}$ in the first region. The Higgs branch is the union of three cones, called $I$, $II$ and $III$ in Tables \ref{tab:region1} and \ref{tab:region1seq}. These combine into the diagram on the left of Table \ref{tab:hassetables5dGen1}, which contains three points of locally maximal dimension. The Hasse diagram for the intersection $ I \cap III$ is the intersection of the branch that relates the origin to the extremity labeled $I$ on the one hand, and the branch that relates the origin to the extremity labeled $III$ on the other hand. Similarly, one reads off the diagrams for the intersections $I \cap II$ and $II \cap III$. Note that the diagram clarifies that $I \cap II = II \cap III$ holds.

\subsection{6d SCFTs}
\label{subsection6d}
Now, we turn our attention to six dimensional SCFTs. Cancellation of gauge anomalies imposes strong restrictions on the gauge groups and matter contents, giving a list of allowed theories \cite{Danielsson:1997kt}. This list has been reproduced from F-theory constructions \cite{Heckman:2015bfa}. The theories can be labeled by their rank, which by definition is the dimension of the tensor branch. Theories of rank $1$ can be realised on complex curves $\mathbb{P}^1$ with negative self-intersection. See, for instance, \cite{Heckman:2018jxk} for a review. 

In the present paper, no attempt is made to compute the Hasse diagrams for the Higgs branches of all these theories. Instead, focus is placed on a few examples as a proof of concept. Theories realised on so-called $-1$ curves experience a small $E_8$ instanton transition at the origin of the tensor branch, see \cite{Ganor:1996mu} and also \cite{Seiberg:1996vs,Intriligator:1997kq,Blum:1997mm,Hanany:1997gh}. This transition implies a jump by 29 quaternionic dimensions of the Higgs branch between a generic point and the origin of the tensor branch. (Higgs branches at the origin of the tensor branch have been addressed recently in \cite{Mekareeya:2016yal,Mekareeya:2017sqh,Hanany:2018uhm}.)
In the Hasse diagram this is expected to be manifested by a presence of an $e_8$ transition on the top of the Hasse diagram which describes the classical Higgs branch.

\begin{table}[t]
	\centering
	\begin{tabular}{|l|c|c|}
	\hline
	$6d$ SCFT & $\mathrm{SU}(2k)$ with $N{=}2k{+}8$ and $\Lambda^2$  & $\mathrm{SU}(2k{+}1)$ with $N {=} 2k{+}9$ and $\Lambda^2$
	 \\ \hline
 Magnetic quiver & $ \node{}1-\node{}2- \cdots -\node{\topnode{} {k+3}}{2k+6}-\node{}{k+4} - \node{}3$
   &
  $ \node{}1-\node{}2- \cdots -\node{\topnode{} {k+3}}{2k+7}-\node{}{k+5} - \node{}3$  \\ \hline
  Hasse diagram
   &
\begin{tikzpicture}[node distance=30pt]
	\tikzstyle{hasse} = [circle, fill,inner sep=2pt];
		\node at (-0.7,-0.5) [] (1a) [] {};
		\node at (0,-0.5) [hasse] (1b) [label=right:\footnotesize{$2k^2+15k+30$}] {};
		\node at (0.7,-0.5) [] (1c) [] {};
		\node [] (2a) [below of=1a] {};
		\node [hasse] (2b) [below of=1b] {};
		\node [] (2c) [below of=1c] {};
		\node  (3a) [below of=2a] {};
		\node [hasse] (3b) [below of=2b] {};
		\node  (3c) [below of=2c] {};
		\node (4a) [below of=3a] {};
		\node (4b)[below of=3b] {};
		\node (4c) [below of=3c] {};
		\node (5a) [below of=4a] {};
		\node (5b) [below of=4b] {};
		\node [hasse] (5c) [below of=4c] {};
		\node [hasse] (6a) [below of=5a] {};
		\node (6b) [below of=5b] {};
		\node (6c) [below of=5c] {};
		\node (7a) [below of=6a] {};
		\node [hasse] (7b) [below of=6b] {};
		\node (7c) [below of=6c] {}; 
		\node (8a) [below of=7a] {};
		\node (8b) [below of=7b] {};
	        \node (8c) [below of=7c] {};
	         \node (9a) [below of=8a] {};
	         \node (9b) [below of=8b] {};
	          \node [hasse] (9c) [below of=8c] {};
	          \node [hasse] (10a) [below of=9a] {};
	          \node (10b) [below of=9b] {};
	          \node (10c) [below of=9c] {};
	          \node (11a) [below of=10a] {};
	          \node [hasse] (11b) [below of=10b] {};
	          \node (11c) [below of=10c] {};
	          \node (12a) [below of=11a] {};
	          \node (12b) [below of=11b] {};
	          \node [hasse] (12c) [below of=11c] {};
	          \node [hasse] (13a) [below of=12a] {};
	          \node (13b) [below of=12b] {};
	          \node (13c) [below of=12c] {};
	          \node (14a) [below of=13a] {};
	          \node [hasse] (14b) [below of=13b] {};
	          \node (14c) [below of=13c] {};
	           \node (15a) [below of=14a] {};
	          \node (15b) [below of=14b] {};
	          \node (15c) [below of=14c] {};
	          \node at (0,-12)[]{$\vdots$};
		\draw (1b) edge [] node[label=left:\footnotesize{$e_8$}] {} (2b)
			(2b) edge [] node[label=left:\footnotesize{$d_{10}$}] {} (3b)
			(3b) edge [] node[label=left:\footnotesize{$d_{12}$}] {} (6a)
			(6a) edge [] node[label=above:\footnotesize{$\;\; A_1$}] {} (7b)
			(3b) edge [] node[label=right:\footnotesize{$a_{11}$}] {} (5c) 
			(5c) edge [] node[label=right:\footnotesize{$a_{11}$}] {} (7b)
			(7b) edge [] node[label=right:\footnotesize{$a_{12}$}] {} (9c)
			(9c) edge [] node[label=right:\footnotesize{$a_{13}$}] {} (11b)
			(6a) edge [] node[label=left:\footnotesize{$d_{14}$}] {} (10a)
			(10a) edge [] node[label=above:\footnotesize{$\;\;\;A_2$}] {} (11b)
			(10a) edge [] node[label=left:$ $] {} (11a)
			(12c) edge [] node[label=right:\footnotesize{$a_{2k+7}$}] {} (14b)
			(12a) edge [] node[label=left:\footnotesize{$d_{2k+8}$}] {} (13a)
			(13a) edge [] node[label=above:\footnotesize{$\;\;\;\;A_{k-1}$}] {} (14b);
	\end{tikzpicture}
	&
\begin{tikzpicture}[node distance=30pt]
	\tikzstyle{hasse} = [circle, fill,inner sep=2pt];
		\node at (-0.7,-0.5) [] (1a) [] {};
		\node at (0,-0.5) [hasse] (1b) [label=right:\footnotesize{$2k^2+17k+38$}] {};
		\node at (0.7,-0.5) [] (1c) [] {};
		\node [] (2a) [below of=1a] {};
		\node [hasse] (2b) [below of=1b] {};
		\node [] (2c) [below of=1c] {};
		\node  (3a) [below of=2a] {};
		\node [hasse] (3b) [below of=2b] {};
		\node  (3c) [below of=2c] {};
		\node (4a) [below of=3a] {};
		\node (4b)[below of=3b] {};
		\node (4c) [below of=3c] {};
		\node (5a) [below of=4a] {};
		\node (5b) [below of=4b] {};
		\node [hasse] (5c) [below of=4c] {};
		\node [hasse] (6a) [below of=5a] {};
		\node (6b) [below of=5b] {};
		\node (6c) [below of=5c] {};
		\node (7a) [below of=6a] {};
		\node [hasse] (7b) [below of=6b] {};
		\node (7c) [below of=6c] {}; 
		\node (8a) [below of=7a] {};
		\node (8b) [below of=7b] {};
	        \node (8c) [below of=7c] {};
	         \node (9a) [below of=8a] {};
	         \node (9b) [below of=8b] {};
	          \node [hasse] (9c) [below of=8c] {};
	          \node [hasse] (10a) [below of=9a] {};
	          \node (10b) [below of=9b] {};
	          \node (10c) [below of=9c] {};
	          \node (11a) [below of=10a] {};
	          \node [hasse] (11b) [below of=10b] {};
	          \node (11c) [below of=10c] {};
	          \node (12a) [below of=11a] {};
	          \node (12b) [below of=11b] {};
	          \node [hasse] (12c) [below of=11c] {};
	          \node [hasse] (13a) [below of=12a] {};
	          \node (13b) [below of=12b] {};
	          \node (13c) [below of=12c] {};
	          \node (14a) [below of=13a] {};
	          \node [hasse] (14b) [below of=13b] {};
	          \node (14c) [below of=13c] {};
	           \node (15a) [below of=14a] {};
	          \node[hasse] (15b) [below of=14b] {};
	          \node (15c) [below of=14c] {};
	          \node at (0,-12)[]{$\vdots$};
		\draw (1b) edge [] node[label=left:\footnotesize{$e_8$}] {} (2b)
			(2b) edge [] node[label=left:\footnotesize{$d_{10}$}] {} (3b)
			(3b) edge [] node[label=left:\footnotesize{$d_{12}$}] {} (6a)
			(6a) edge [] node[label=above:\footnotesize{$\;\; A_1$}] {} (7b)
			(3b) edge [] node[label=right:\footnotesize{$a_{11}$}] {} (5c) 
			(5c) edge [] node[label=right:\footnotesize{$a_{11}$}] {} (7b)
			(7b) edge [] node[label=right:\footnotesize{$a_{12}$}] {} (9c)
			(9c) edge [] node[label=right:\footnotesize{$a_{13}$}] {} (11b)
			(6a) edge [] node[label=left:\footnotesize{$d_{14}$}] {} (10a)
			(10a) edge [] node[label=above:\footnotesize{$\;\;\;A_2$}] {} (11b)
			(10a) edge [] node[label=left:$ $] {} (11a)
			(12c) edge [] node[label=right:\footnotesize{$a_{2k+7}$}] {} (14b)
			(12a) edge [] node[label=left:\footnotesize{$d_{2k+8}$}] {} (13a)
			(13a) edge [] node[label=above:\footnotesize{$\;\;\;A_{k-1}$}] {} (14b)
			(14b) edge [] node[label=right:\footnotesize{$a_{2k+8}$}] {} (15b);
	\end{tikzpicture}
	\\ \hline
	
		\end{tabular}
	\caption{Hasse diagrams of $6d$ SCFTs: $\mathrm{SU}(N)$ with $N+8$ fundamentals and one 2nd rank antisymmetric. Note that the two diagrams differ only at the bottom.   }\label{tab:hassetables6d1}
\end{table}
\clearpage

\begin{table}[t]
	\centering
	\begin{tabular}{|c|c|c|c|}
	\hline
	$6d$ SCFT & $\mathrm{Sp}(k)$ with $N = 4k+16$ flavours    & $G_2$ with $7$ flavours 
	 \\ \hline
 Magnetic quiver & $ \node{}1-\node{}2- \cdots -\node{\topnode{} {k+3}}{2k+6}-\node{}{k+4} - \node{}2$
   &  Not known  \\ \hline
  Hasse diagram  
   &
	\begin{tikzpicture}
		\tikzstyle{hasse} = [circle, fill,inner sep=2pt];
		\node [hasse] (1) [label=right:\footnotesize{$2k^2+15k+29$}] {};
		\node [hasse] (2) [below of=1] {};
		\node [hasse] (3) [below of=2] {};
		\node [hasse] (4) [below of=3] {};
		\node [hasse] (5) [below of=4] {};
		\node [hasse] (6) [below of=5] {};
		\node at (0,-3.5) {$\vdots$};
		\draw (1) edge [] node[label=left:$e_8$] {} (2)
			(2) edge [] node[label=left:$d_{10}$] {} (3)
			(3) edge [] node[label=left:$d_{12}$] {} (4)
			(5) edge [] node[label=left:$d_{2k+8}$] {} (6);
	\end{tikzpicture}
	&
	  	\begin{tikzpicture}
		\tikzstyle{hasse} = [circle, fill,inner sep=2pt];
		\node [hasse] (1) [label=right:\footnotesize{$64$}] {};
		\node [hasse] (2) [label=right:\footnotesize{$35$},below of=1] {};
		\node [hasse] (3) [label=right:\footnotesize{$18$},below of=2] {};
		\node [hasse] (4) [label=right:\footnotesize{$7$},below of=3] {};
		\node [hasse] (5) [below of=4] {};
		\draw (1) edge [] node[label=left:$e_8$] {} (2)
			(2) edge [] node[label=left:$d_{10}$] {} (3)
			(3) edge [] node[label=left:$a_{11}$] {} (4)
			(4) edge [] node[label=left:$c_{7}$] {} (5);
	\end{tikzpicture}
	\\ \hline
		\end{tabular}
		\caption{Hasse diagrams of $6d$ SCFTs: $\mathrm{Sp}(k)$ family and $G_2$ theory. }\label{tab:hassetables6d2}
\end{table}

As an illustration, we first consider two infinite families of rank one theories on a $-1$ curve. These are the following:  
\begin{itemize}
\item The $6$d $\mathrm{SU}(N)$ gauge theory with $N+8$ fundamental hypermultiplets and a $2$nd rank antisymmetric hypermultiplet, denoted by $\Lambda^2$. The Hasse diagrams for this theory is given in Table \ref{tab:hassetables6d1}, using the magnetic quiver of \cite[Sec.\ 3.6.2]{Cabrera:2019izd} for $\mathrm{SU}(2k)$ and of \cite[Sec.\ 3.6.4]{Cabrera:2019izd} for $\mathrm{SU}(2k+1)$. 
This family generalises the case $N=4$ studied in detail in Sections \ref{higgsforsu4}, \ref{magQuivforsu4}, and \ref{infinite_coupling_su4}. 
Note that the Hasse diagram of the $\mathrm{SU}(N_1)$ theory is entirely included into the Hasse diagram of the $\mathrm{SU}(N_2)$ provided $N_1 \leq N_2$. This means that one can Higgs the $\mathrm{SU}(N_2)$ theory with $N_2+8$ fundamentals and one $\Lambda^2$ to the $\mathrm{SU}(N_1)$ theory with exactly $N_1+8$ fundamentals and one $\Lambda^2$. Alternatively, this can be checked directly by decomposing the representations, see for instance \cite[Fig.\ 5]{DelZotto:2018tcj}. 
\item The $\mathrm{Sp}(k)$ with $N=4k+16$ fundamental 6d half-hypermultiplets, with magnetic quivers derived in \cite[Sec.\ 3.6.1 and 3.6.3]{Cabrera:2019izd}. The Hasse diagram for this family of theories is given in Table \ref{tab:hassetables6d2}. Again, theories defined by various $k$-values display Hasse diagrams included into one another.
\end{itemize}
As a consistency check, the global symmetry of the theories is reproduced by the bottom part of the diagrams, as per the rules of Section \ref{subsectionGlobalSymmetry}.

In addition, to these two infinite families of theories, there are $12$ other rank one theories defined by a $-1$ curve. When there is no ambiguity, we label these theories by their gauge group; when there is ambiguity, we give them a special name as specified below. We describe the 12 theories using the notation of \cite{DelZotto:2018tcj} as  
\begin{itemize}
    \item $\mathrm{SU}(6) \oplus \Lambda^{\oplus 15} \oplus \frac{1}{2} \Lambda^3$. This theory is denoted $\mathrm{SU}(6)_{\star}$. 
    \item $\mathrm{SO}(7)\oplus V^{\oplus 2} \oplus S^{\oplus 6}$ 
    \item $\mathrm{SO}(8)\oplus V^{\oplus 3} \oplus S_+^{\oplus 3}\oplus S_-^{\oplus 3}$ 
    \item $\mathrm{SO}(9)\oplus V^{\oplus 4} \oplus S^{\oplus 3}$ 
    \item $\mathrm{SO}(10)\oplus V^{\oplus 5} \oplus S^{\oplus 3}$ 
    \item $\mathrm{SO}(11)\oplus V^{\oplus 6} \oplus \frac{1}{2} S^{\oplus 3}$ 
    \item $\mathrm{SO}(12)\oplus V^{\oplus 7} \oplus \frac{1}{2} S_{\pm}^{\oplus 3}$. This theory is denoted $\mathrm{SO}(12)_{a}$.  
    \item $\mathrm{SO}(12)\oplus V^{\oplus 7} \oplus \frac{1}{2} S_{\pm} \oplus \frac{1}{2} S_{\mp}^{\oplus 2}$. This theory is denoted $\mathrm{SO}(12)_{b}$.  
    \item $G_2\oplus \mathbf{7}^{\oplus 7}$ 
    \item $F_4\oplus \mathbf{26}^{\oplus 4}$
    \item $E_6\oplus \mathbf{27}^{\oplus 5}$
    \item $E_7\oplus \frac{1}{2}\mathbf{56}^{\oplus 7}$
\end{itemize}
In all cases except the $\mathrm{SU}(6)_{\star}$ theory \cite{Zafrir:2015rga,Hayashi:2015zka} (more on this below), we are not aware of any brane system, and we therefore have to rely on more indirect arguments. 
Let us consider the $G_2$ theory as a warm up. We can rely on two principles: the classical part of the Hasse diagram is given by a representation theory computation as in Section \ref{subsectionSimpleGaugeGroup}, while the strong coupling limit is conjectured to be an $e_8$ transition. 
This leads directly to the diagram of Table \ref{tab:hassetables6d2}. We can repeat the process for the other theories. 

As mentioned above, for the $\mathrm{SU}(6)_{\star}$ theory a magnetic quiver is available, since it is part of an infinite family of theories \cite{Mekareeya:2017jgc,Hanany:2018vph}, for which the magnetic quivers have been given in \cite[Eq.\ 3.9]{Hanany:2018vph} and \cite[Sec.\ 3.6.5]{Cabrera:2019izd} . The magnetic quiver is 
\begin{equation}
    \node{}1-\node{}2-\node{}3-\node{}4-\node{}5-\node{}6-\node{}7-\node{}8-\node{}9-\node{}{10}-\node{}{11}-\node{\topnode{} {5}}{12}-\node{}{8} - \node{}4
\end{equation}
Quiver subtraction can be used to find the Hasse diagram, and we find agreement with the pattern of partial Higgsing. 

\begin{figure}
        \centering
\begin{tikzpicture}[node distance=20pt]
        \tikzstyle{hasse} = [circle, fill,inner sep=1.5pt];
                \node at (2,0) [hasse] (1a) [label=left:\footnotesize{$\textcolor{red}{\mathrm{Sp}(1)}, -17$}] {};
                \node at (2,3.4) [hasse] (0aa) [label=left:\footnotesize{$\textcolor{red}{\{1\}},0$}] {};
                \node at (2,9.2) [hasse] (0aaaa) [label=left:\footnotesize{$\textcolor{red}{\{\}},29$}] {};
                \node at (5,-9.6) [hasse] (7c) [label=below:\footnotesize{$\begin{array}{c} \textcolor{red}{\mathrm{SU}(6)_{\star}} \\ -65 \end{array}$}] {};
                \node at (12,-7) [hasse] (7g) [label=above:\footnotesize{$\textcolor{red}{F_4},-52$}] {};
                \node at (2,-10) [] {$\vdots$};
                \draw (2,-9.2) -- (2,-9.8);
                 \node at (4,-10.1) [] {$\vdots$};
                \draw (4,-9.3) -- (4,-9.9);
                \node at (2,-4.2) [hasse] (3a) [label=left:\footnotesize{$\textcolor{red}{\mathrm{Sp}(2)},-38$}] {};
                \node at (2,-9.2) [hasse] (4a) [label=left:\footnotesize{$\textcolor{red}{\mathrm{Sp}(3)},-63$}] {};
                \node at (4,-2.2) [hasse] (3b) [label=above:\footnotesize{$\textcolor{red}{\;\;\;\;\;\;\;\;\;\;\;\;\;\; \mathrm{SU}(3)},-28$}] {};
                \node at (4,-4.4) [hasse] (4b) [] {};
                \node at (4.9,-4.1) [] {\footnotesize{$\textcolor{red}{\mathrm{SU}(4)},-39$}};
                \node at (4,-6.8)  [hasse] (5b) [label=left:\footnotesize{$\textcolor{red}{\mathrm{SU}(5)},-51$}] {};
                \node at (4,-9.4)  [hasse] (7b) [label=below:\footnotesize{$\begin{array}{c} \textcolor{red}{\mathrm{SU}(6)} \\ -64 \end{array}$ \;\;\;\;\;\;\;\;\;\;}] {};
                \node at (8,-3.6) [hasse] (4e) [label=above:\footnotesize{$\textcolor{red}{G_2},-35$}] {};
                \node at (8,-4.8)  [hasse] (5e) [] {};
                \node at (7,-4.8) [hasse] (5e2) [] {}; 
                \node at (9,-4.8)  [hasse] (5e3) [label=right:\footnotesize{$\textcolor{red}{\mathrm{SO}(7)},-41$}] {};
                \node at (8,-5.44) [hasse] (6e) [label=right:\footnotesize{$\textcolor{red}{\mathrm{SO}(8)},-44$}] {};
                \node at (8,-6.2)  [hasse] (7e) [label=left:\footnotesize{$\textcolor{red}{\mathrm{SO}(9)},-48$}] {};
                \node at (8,-7.2) [hasse]  (8e) [] {};
               \node at (9,-7.05) [] {\footnotesize{$\textcolor{red}{\mathrm{SO}(10)},-53$}};  
                \node at (8,-8.4) [hasse] (9e) [label=left:\footnotesize{$\textcolor{red}{\mathrm{SO}(11)},-59$}] {};
                 \node at (9,-9.8) [hasse] (10e2) [label=below:\footnotesize{$\begin{array}{c} \textcolor{red}{\mathrm{SO}(12)_a} \\ -66 \end{array}$}] {};
                \node at (7,-9.8)  [hasse] (10e3) [label=below:\footnotesize{$\begin{array}{c} \textcolor{red}{\mathrm{SO}(12)_b} \\ -66 \end{array}$}] {};
                \node at (12,-8)  [hasse] (9g) [label=right:\footnotesize{$\textcolor{red}{E_6},-57$}] {};
                \node at (12,-9.2) [hasse] (11g) [label=right:\footnotesize{$\textcolor{red}{E_7},-63$}] {};
                                \draw (0aaaa) edge [] node[label=left:\footnotesize{$e_{8}$}] {} (0aa)
                                (0aa) edge [] node[label=left:\footnotesize{$d_{10}$}] {} (1a)
                                (1a) edge [] node[label=left:\footnotesize{$d_{12}$}] {} (3a)
                                (3a) edge [] node[label=left:\footnotesize{$d_{14}$}] {} (4a)
                                (3b) edge [] node[label=center:\footnotesize{$a_{11}\;\;\;\;\;\;$}] {} (4b)
                                (4b) edge [] node[label=center:\footnotesize{$a_{12}\;\;\;\;\;\;$}] {} (5b)
                                (5b) edge [] node[label=center:\footnotesize{$a_{13}\;\;\;\;\;\;\;$}] {} (7b)
                                (5b) edge [] node[label=right:\footnotesize{$a_{14}\;\;\;$}] {} (7c)
                                (3b) edge [] node[label=above:\footnotesize{$c_7$}] {} (4e)
                                (4e) edge [] (5e)
                                (4e) edge []  (5e2)
                                (4e) edge [] node[label=center:\footnotesize{$\;\;\;\;\;\; c_6$}] {} (5e3)
                                (5e) edge [] (6e)
                                (5e2) edge [] node[label=center:\footnotesize{$c_3 \;\;\;\;\;\;\;$}] {} (6e)
                                (5e3) edge [] node[label=center:\footnotesize{$c_3\;\;\;\;\;\;\;$}] {} (6e)
                                (6e) edge [] node[label=center:\footnotesize{$\;\;\;\;\;c_4$}] {} (7e)
                                (7e) edge [] node[label=center:\footnotesize{$\;\;\;\;\;c_5$}] {} (8e)
                                (8e) edge [] node[label=center:\footnotesize{$\;\;\;\;\;c_6$}] {} (9e)
                                (4b) edge [] (5e)
                                (4b) edge [] (5e2)
                                (4b) edge [] node[label=below:\footnotesize{$c_2$}] {} (5e2)
                                 (4b) edge [] node[] {} (5e3)
                                (5b) edge [] node[label=above:\footnotesize{$a_2$}] {} (8e)
                                (7e) edge [] node[label=above:\footnotesize{$\;\;c_4$}] {} (7g)
                                (7g) edge [] node[label=right:\footnotesize{$?$}] {} (9g)
                                (9g) edge [] node[label=right:\footnotesize{$?$}] {} (11g)
                                (1a) edge [] node[label=above:\footnotesize{$\;\; a_{11}$}] {} (3b)
                                (3a) edge [] node[label=above:\footnotesize{$A_{1}$}] {} (4b)
                                (4a) edge [] node[label=above:\footnotesize{$A_{2}$}] {} (7b)
                                (9e) edge [] node[label=above:\footnotesize{$c_{7}$}] {} (10e2)
                                (9e) edge [] node[label=above:\footnotesize{$c_{7}\;\;$}] {} (10e3)
                                (7c) edge [] node[label=below:\footnotesize{$a_{1}$}] {} (10e3)
                                (7b) edge [] node[label=above:\footnotesize{$a_{2}$}] {} (10e2)
                                (8e) edge [] node[label=below:\footnotesize{$a_{4}$}] {} (9g)
                                (9e) edge [] node[label=below:\footnotesize{$b_{3}$}] {} (11g);
        \end{tikzpicture}
        \caption{Conjectural Hasse diagram reproducing the E-string diagram which contains all the 6d Higgs branches for theories defined on $-1$ curves. A red label $\textcolor{red}{G}$ on a node X has the following meaning: the effective theory whose Higgs branch is the transverse slice between the top node and X is the 6d theory whose gauge group is $G$ (using the unambiguous labeling described in the text). 
        }\label{fig:Etree}
\end{figure}
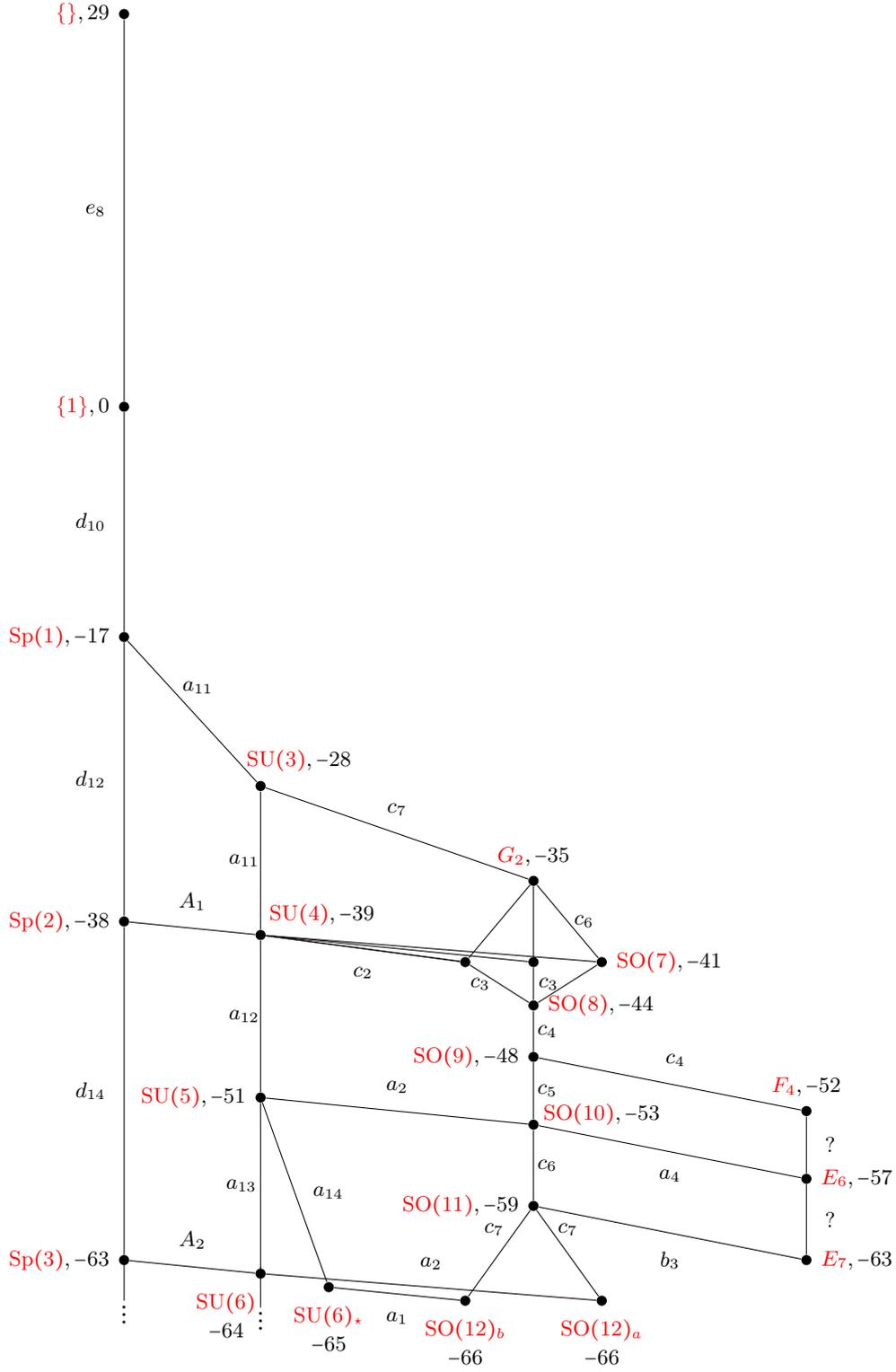

It turns out all the theories defined on $-1$ curves can be neatly encapsulated into a unique diagram, showing how they are all related by partial Higgsing, see Figure \ref{fig:Etree}. This partially reproduces what is called the E-string diagram in \cite{DelZotto:2018tcj}, see also the earlier works \cite{Bershadsky:1996nh,Distler:1996ub}. 

Comparison of our results with the E-string diagram \cite[Fig.\ 5]{DelZotto:2018tcj}
obtained from the Higgsing pattern, shows again that the Hasse diagram of the moduli space is intimately linked to the Higgs mechanism. Extending the E-string Higgsing diagram to a Hasse diagram of the Higgs branch is achieved by adding further minimal transitions, for instance there is an $A_k$ transition between the $Sp(k+1)$ theories and the $SU(2k+2)$ theory. These transitions are also a result of the known magnetic quivers for these theories.

On the other hand, the $SO$ gauge groups with spinor matter or exceptional gauge groups have no known magnetic quivers, but one can attempt a representation theory analysis. This would imply the following transitions on the right part of Figure \ref{fig:Etree}:
\begin{enumerate}
    \item We find a KP transition from the $\mathrm{SO}(9)$ theory to the $F_4$ theory. The transverse slice between $\mathrm{SO}(8)$ to $F_4$  can be seen from the diagram to be non-minimal. 
    \item We find a $c_2$ KP transition from $\mathrm{SU}(4)$ to $\mathrm{SO}(7)$.
    \item We find a $a_2$ KP transition from $\mathrm{SU}(5)$ to $\mathrm{SO}(10)$.
\end{enumerate}
While the global symmetry conjecture of Section \ref{subsectionGlobalSymmetry} is satisfied in most cases, some of the global symmetries listed in \cite[Tab.\ 1]{DelZotto:2018tcj} are difficult to interpret from partial Higgsing. As a consequence, our diagram is incomplete: 
\begin{enumerate}
\item We seem to need a family of elementary slices, denoted by $?_n$ in Figure \ref{fig:Etree}, which have isometry $\mathfrak{so}(n+1)$ and quaternionic dimension $n$. No closures of minimal nilpotent orbits have these properties. The elementary slice could be what is called $a_n^+$ in \cite{fu2017generic}. 
\item We also need an elementary slice, denoted $??_5$ of quaternionic dimension 5, from the $F_4$ to the $E_6$ theory. 
    \item For the $\mathrm{SO}(9)$ theory we could not find a $c_3$ transition. 
\end{enumerate}
We were not able to resolve these issues, and the reader should keep that in mind when considering the rightmost half of Figure \ref{fig:Etree}.

\subsection{Generalised Argyres-Douglas theories and Complete Graphs}
\label{subsectioncompletegraphs}
In this last section of results, we go back to four dimensions and focus on inherently non-Lagrangian theories. The $A_n$ Argyres-Douglas (AD) theory can be seen as a member of the more general family of $4$d $\mathcal{N}=2$ theories, denoted $(A_m,A_n)$, which was first introduced in \cite{Cecotti:2010fi}. We call these theories \emph{generalised Argyres-Douglas} theories. The $A_n$ AD theory corresponds to taking $m=1$. The $(A_m,A_n)$ theory can be realised as Type IIB string theory on a Calabi-Yau space defined by the equation 
\begin{equation}
    x^{n+1} + y^{m+1} + z^2 + w^2 = 0
\end{equation}
in $\mathbb{C}^4$. It was found in \cite{Xie:2012hs} that even though it is not Lagrangian, after compactification on a circle the $3$d theory admits in some cases a Lagrangian $3$d mirror given by a quiver. This quiver can be interpreted as the magnetic quiver for the Higgs branch of the generalised AD theory. Specifically, if one considers the $(A_{N-1},A_{kN-1})$ theory, this magnetic quiver is given by a quiver with $N$ $\mathrm{U}(1)$ nodes, where each pair of nodes is related by $k$ links \cite{boalch2008irregular,boalch2012simply,DelZotto:2014kka}. This is a two parameter family of quivers which we explore partially in this subsection. 
In particular, taking $k=1$, we obtain for $(A_{N-1},A_{N-1})$ the \emph{complete graph} with $N$ nodes. The complexity of computing the Hasse diagram for a high number of nodes is extremely high, as a multitude of slices of the same kind appear and the Hasse diagrams get more perplexed. Therefore, we only present the cases for $k=1$ with $N=2,3,4,$ and $5$ in Table \ref{tab:hassecompletegraphs1}. In Table \ref{tab:hassecompletegraphs2} we present the case $k>1$ for $N=2,3,4$.

\begin{table}
	\centering
	\begin{tabular}{|m{1.4cm}|m{1.8cm}|m{2.7cm}|m{5cm}|}
	\hline
	$N=2$  &  $N=3$ &  $N=4$ & $N=5$ 
	 \\ \hline
	 
$~~~~\mathbb{H}$
	&
	\begin{tikzpicture}[node distance=30pt]
	\tikzstyle{hasse} = [circle, fill,inner sep=1pt];
		\node at (-0.7,-0.5) [] (1a) [] {};
		\node at (0,-0.5) [hasse] (1b) [label=right:\footnotesize{$2$}] {};
		\node  [] (0b) [above of=1b] {};
			\node  [] (00b) [above of=0b] {};
		\node at (0.7,-0.5) [] (1c) [] {};
		\node [] (2a) [below of=1a] {};
	        \node  (3a) [below of=2a] {};%
		\node  (4a) [below of=3a] {};%
		\node [hasse] (2b) [label=right:\footnotesize{$0$},below of=1b] {};
		\draw (1b) edge [] node[label=right:\footnotesize{$a_2$}] {} (2b);
	\end{tikzpicture}
	
				&
	\begin{tikzpicture}[node distance=30pt]
	\tikzstyle{hasse} = [circle, fill,inner sep=1pt];
		\node at (-1.2,-0.5) [] (1a) [] {};
		\node at (-0.8,-0.5) [] (1b) [] {};
		\node at (-0.4,-0.5) [hasse] (1c) [label=above:\footnotesize{$3$}] {};
		\node at (0,-0.5) [] (1d) [] {};
		\node at (0.4,-0.5) [] (1e) [] {};	
\node [hasse] (2a) [below of=1a] {};
\node [hasse] (2b) [below of=1b] {};
\node [] (2c) [below of=1c] {};
\node [hasse] (2d) [below of=1d] {};
\node [hasse] (2e) [label=right:\footnotesize{$1$},below of=1e] {};
\node [] (3a) [below of=2a] {};
\node [] (3b) [below of=2b] {};
\node [] (3c) [below of=2c] {};
\node [] (3d) [below of=2d] {};
\node [] (3e) [below of=2e] {};
\node [hasse] (4c) [label=right:\footnotesize{$0$},below of=3c] {};
\draw 
%Ak-1
(1c) edge [] node[label=above:\footnotesize{$a_{2}$\;\;}] {} (2a)
(1c) edge [] node[] {} (2b)
(1c) edge [] node[] {} (2d)
(1c) edge [] node[] {} (2e)
%Ak-1
(2a) edge [] node[label=left:\footnotesize{$A_{2}$}] {} (4c)
(2b) edge [] node[] {} (4c)
(2d) edge [] node[] {} (4c)
(2e) edge [] node[] {} (4c);
	\end{tikzpicture}
	&
	\begin{tikzpicture}[node distance=30pt]
 	\tikzstyle{hasse} = [circle, fill,inner sep=1pt];
 		\node at (-2,-0.5) [] (1a) [] {};
 		\node at (-1.6,-0.5) [] (1b) [] {};
 		\node at (-1.2,-0.5) [] (1c) [] {};
 		\node at (-0.8,-0.5) [] (1d) [] {};
 		\node at (-0.4,-0.5) [] (1e) [] {};
 \node at (0,-0.5) [hasse] (1f) [label=above:\footnotesize{$4$}] {};
 \node at (0.4,-0.5) [] (1g) [] {};
 \node at (0.8,-0.5) [] (1h) [] {};
 \node at (1.2,-0.5) [] (1i) [] {};
 \node at (1.6,-0.5) [] (1j) [] {};
 \node at (2,-0.5) [] (1k) [] {};
 \node [hasse] (2a) [below of=1a] {};
 \node [hasse] (2b) [below of=1b] {};
 \node [hasse] (2c) [below of=1c] {};
 \node [hasse] (2d) [below of=1d] {};
 \node [hasse] (2e) [below of=1e] {};
 \node [] (2f) [below of=1f] {};
 \node [hasse] (2g) [below of=1g] {};
 \node [hasse] (2h) [below of=1h] {};
 \node [hasse] (2i) [below of=1i] {};
 \node [hasse] (2j) [below of=1j] {};
 \node [hasse] (2k) [label=right:\footnotesize{$2$},below of=1k] {};
 \node [] (3a) [below of=2a] {};
 \node [hasse] (3b) [below of=2b] {};
 \node [] (3c) [below of=2c] {};
 \node [hasse] (3d) [below of=2d] {};
 \node [] (3e) [below of=2e] {};
 \node [hasse] (3f) [below of=2f] {};
 \node [] (3g) [below of=2g] {};
 \node [hasse] (3h) [below of=2h] {};
 \node [] (3i) [below of=2i] {};
 \node [hasse] (3j) [label=right:\footnotesize{$1$},below of=2j] {};
 \node [] (3k) [below of=2k] {};
 \node [] (4a) [below of=3a] {};
 \node [] (4b) [below of=3b] {};
 \node [] (4c) [below of=3c] {};
 \node [] (4d) [below of=3d] {};
 \node [] (4e) [below of=3e] {};
 \node [] (4f) [below of=3f] {};
 \node [] (4g) [below of=3g] {};
 \node [] (4h) [below of=3h] {};
 \node [] (4i) [below of=3i] {};
 \node [] (4j) [below of=3j] {};
 \node [] (4k) [below of=3k] {};
 \node [] (5f) [below of=4f] {};
 \node [hasse] (6f) [label=right:\footnotesize{$0$},below of=5f] {};
 \draw 
 (2a) edge [] node[label=left:\footnotesize{$A_2$}] {} (3b)
 (2a) edge [] node[] {} (3d)
 (2b) edge [] node[] {} (3b)
 (2b) edge [] node[] {} (3f)
 (2c) edge [] node[] {} (3b)
 (2c) edge [] node[] {} (3h)
 (2d) edge [] node[] {} (3b)
 (2d) edge [] node[] {} (3j)
 (2e) edge [] node[] {} (3d)
 (2e) edge [] node[] {} (3f)
 (2g) edge [] node[] {} (3d)
 (2g) edge [] node[] {} (3h)
 (2h) edge [] node[] {} (3d)
 (2h) edge [] node[] {} (3j)
 (2i) edge [] node[] {} (3f)
 (2i) edge [] node[] {} (3h)
 (2j) edge [] node[] {} (3f)
 (2j) edge [] node[] {} (3j)
 (2k) edge [] node[] {} (3h)
 (2k) edge [] node[] {} (3j)
 (1f) edge [] node[label=left:\footnotesize{$a_2$}] {} (2a)
 (1f) edge [] node[] {} (2b)
  (1f) edge [] node[] {} (2c)
  (1f) edge [] node[] {} (2d)
  (1f) edge [] node[] {} (2e)
  (1f) edge [] node[] {} (2g)
  (1f) edge [] node[] {} (2h)
  (1f) edge [] node[] {} (2i)
  (1f) edge [] node[] {} (2j)
   (1f) edge [] node[] {} (2k)
   (3b) edge [] node[label=left:\footnotesize{$A_3$}] {} (6f)
     (3d) edge [] node[] {} (6f)
      (3f) edge [] node[] {} (6f)
       (3h) edge [] node[] {} (6f)
       (3j) edge [] node[] {} (6f)
 (1f) edge [] node[] {} (2b)
  (1f) edge [] node[] {} (2c)
  (1f) edge [] node[] {} (2d)
  (1f) edge [] node[] {} (2e)
  %(1f) edge [] node[] {} (2f)
  (1f) edge [] node[] {} (2g)
  (1f) edge [] node[] {} (2h)
  (1f) edge [] node[] {} (2i)
  (1f) edge [] node[] {} (2j)
   (1f) edge [] node[] {} (2k);	
 	\end{tikzpicture}
	\\ \hline
		\end{tabular}
	\caption{Hasse diagrams for complete graphs for $N=2,3,4,$ and $5$ with edges of multiplicity $k=1$.
	}\label{tab:hassecompletegraphs1}
\end{table}

 \begin{table}
 	\centering
 	\begin{tabular}{|m{1.1cm}|m{2.4cm}|m{5cm}|}
 	\hline
 	$N=2$ & $N=3$ & $N=4$ 
 	 \\ \hline
 	 	\begin{tikzpicture}[node distance=30pt]
 	\tikzstyle{hasse} = [circle, fill,inner sep=1pt];
 		\node at (-0.7,-0.5) [] (1a) [] {};
 		\node at (0,-0.5) [hasse] (1b) [label=right:\footnotesize{$1\;\;$}] {};
 	           \node   (0b) [above of=1b] {};
 		\node at (0.7,-0.5) [] (1c) [] {};
 		\node [] (2a) [below of=1a] {};
 	        \node  (3a) [below of=2a] {};%
 		\node  (4a) [below of=3a] {};%
 		\node [hasse] (2b) [label=right:\footnotesize{$0$},below of=1b] {};
 		\draw (1b) edge [] node[label=center:\footnotesize{$A_{k-1}\;\;\;\;\;\;\;\;\;$}] {} (2b);
 	\end{tikzpicture}
 	 &
 	 	\begin{tikzpicture}[node distance=30pt]
 	\tikzstyle{hasse} = [circle, fill,inner sep=1pt];
 		\node at (-0.7,-0.5) [] (1a) [] {};
 		\node at (0,-0.5) [hasse] (1b) [label=above:\footnotesize{$2$}] {};
 		\node at (0.7,-0.5) [] (1c) [] {};
 \node [hasse] (2a) [below of=1a] {};
 \node [hasse] (2b) [below of=1b] {};
 \node [hasse] (2c) [label=right:\footnotesize{$1$},below of=1c] {};
 \node [] (3a) [below of=2a] {};
 \node [hasse] (3b) [label=right:\footnotesize{$0$},below of=2b] {};
 \node [] (3c) [below of=2c] {};
 \draw 
 (1b) edge [] node[label=center:\footnotesize{$A_{k-1}\;\;\;\;\;\;\;\;\;\;\;$}] {} (2a)
 (2a) edge [] node[label=center:\footnotesize{$A_{2k-1}\;\;\;\;\;\;\;\;\;\;$}] {} (3b)
 (1b) edge [] node[] {} (2b)
 (2b) edge [] node[] {} (3b)
 (1b) edge [] node[] {} (2c)
 (2c) edge [] node[] {} (3b);
 	\end{tikzpicture}
 	 &
 	\begin{tikzpicture}[node distance=30pt]
 	\tikzstyle{hasse} = [circle, fill,inner sep=1pt];
 		\node at (-1.2,-0.5) [] (1a) [] {};
 		\node at (-0.8,-0.5) [] (1b) [] {};
 		\node at (-0.4,-0.5) [] (1c) [] {};
 		\node at (0,-0.5) [hasse] (1d) [label=above:\footnotesize{$3$}] {};
 		\node at (0.4,-0.5) [] (1e) [] {};	
 \node at (0.8,-0.5) [] (1f) [] {};
 \node at (1.2,-0.5) [] (1g) [] {};
 \node [hasse] (2a) [below of=1a] {};
 \node [hasse] (2b) [below of=1b] {};
 \node [hasse] (2c) [below of=1c] {};
 \node [] (2d) [below of=1d] {};
 \node [hasse] (2e) [below of=1e] {};
 \node [hasse] (2f) [below of=1f] {};
 \node [hasse] (2g) [label=right:\footnotesize{$2$},below of=1g] {};
 \node [hasse] (3a) [below of=2a] {};
 \node [hasse] (3b) [below of=2b] {};
 \node [hasse] (3c) [below of=2c] {};
 \node [hasse] (3d) [below of=2d] {};
 \node [hasse] (3e) [below of=2e] {};
 \node [hasse] (3f) [below of=2f] {};
 \node [hasse] (3g) [label=right:\footnotesize{$1$},below of=2g] {};
 \node [] (4d) [below of=3d] {};
 \node [hasse] (5d) [label=right:\footnotesize{$0$},below of=4d] {};
 \draw 
 (2a) edge [] node[label=left:\footnotesize{$A_{k-1}$}] {} (3a)
 (2a) edge [orange] node[] {} (3d)
 (2a) edge [orange] node[] {} (3e)
 (2b) edge [] node[] {} (3a)
 (2b) edge [orange] node[] {} (3f)
 (2b) edge [orange] node[] {} (3g)
 (2c) edge [] node[] {} (3b)
 (2c) edge [orange] node[] {} (3d)
 (2c) edge [orange] node[] {} (3f)
 (2e) edge [] node[] {} (3b)
 (2e) edge [orange] node[] {} (3e)
 (2e) edge [orange] node[] {} (3g)
 (2f) edge [] node[] {} (3c)
 (2f) edge [orange] node[] {} (3d)
 (2f) edge [orange] node[] {} (3g)
 (2g) edge [] node[] {} (3c)
 (2g) edge [orange] node[] {} (3e)
 (2g) edge [orange] node[label=right:\footnotesize{$\;\;A_{2k-1}$}] {} (3f)
 (1d) edge [] node[label=above:\footnotesize{$A_{k-1}\;\;$}] {} (2a)
 (1d) edge [] node[] {} (2b)
  (1d) edge [] node[] {} (2c)
   (1d) edge [] node[] {} (2e)
  (1d) edge [] node[] {} (2f)
  (1d) edge [] node[] {} (2g)
   (3b) edge [red] node[] {} (5d)
   (3c) edge [red] node[] {} (5d)
   (3d) edge [green] node[] {} (5d)
   (3e) edge [green] node[] {} (5d)
   (3f) edge [green] node[] {} (5d)
   (3g) edge [green] node[label=right:\footnotesize{$A_{3k-1}$}] {} (5d)
     (3a) edge [red] node[label=left:\footnotesize{$A_{4k-1}$}] {} (5d);
 	\end{tikzpicture}
 	\\ \hline
 		\end{tabular}
 	\caption{Hasse diagrams for complete graphs with $N=2,3,4$ with edges of multiplicity $k>1$.
 	}\label{tab:hassecompletegraphs2}
 \end{table}

\section{Conclusions and Outlook}
\label{conclusions}
The aim of this paper is to clarify the connection between the Higgs mechanism and the structure of the Higgs branch as a symplectic singularity. Both the partial order implied by partially Higgsing a gauge theory and the partial order of inclusion of closures of symplectic leaves in the Higgs branch are the same and are unified in one Hasse diagram. 

Since the different partially Higgsed theories correspond to different singular loci on the Higgs branch, it is natural to consider the singularity structure of the Higgs branch. As demonstrated in this paper, analysing the Higgs branch as an algebraic space and reducing the singularity only via Kraft-Procesi transitions leads to a Hasse diagram. We presented an extensive discussion for two examples in Section \ref{branesclass} and \ref{magQuivforsu4}.
There are two notable features:
\begin{compactenum}[(i)]
\item The partial Higgsing diagram has exactly the same structure as the Hasse diagram of the moduli space, and
\item using magnetic quivers and Kraft-Procesi transitions (realised via quiver subtractions) the nature of transitions themselves becomes clear.
\end{compactenum}
As elaborated on in Section \ref{sec:Hasse}, the Hasse diagram can be summarised as follows: 
The symplectic leaves are the spaces of VEVs along which the original gauge group is broken to one of its subgroups via the Higgs mechanism.
Symplectic leaves are mutally disjoint and not symplectic singularities themselves; only their closures are. The Higgs branch of the theory that contains the residual gauge group and only the hypermultiplets that are charged under it, on a given symplectic leaf, is the transverse slice of the leaf to the total space. No other transverse slices are accessible via the Higgs mechanism. The magnetic quiver approach allows to achieve: firstly, all minimal transitions (Kraft-Procesi transitions) have an immediate description in terms of spaces of dressed monopole operators, by assumption. Secondly, the closure of each symplectic leaf is described via a magnetic quiver.

Though the identification of the Higgsing pattern with the singularity structure of the Higgs branch via Hasse diagrams is remarkable, the method is not restricted to finite coupling or Lagrangian theories. As exemplified in Section \ref{sec:Higgs_infinite}, Hasse diagrams for Higgs branches at infinite coupling can be derived in the exact same fashion, provided a suitable brane construction or a magnetic quiver is known.

The Hasse diagrams for classical Higgs branches for theories with a single gauge group of type $A$, $B$, $C$, $D$, and $G_2$ with fundamental matter have been derived in Section \ref{subsectionSimpleGaugeGroup}. For $U(k)$ $\mathrm{Sp}(k)$, and $\mathrm{O}(k)$ gauge groups, the Higgs branches are known to be nilpotent orbit closures of height 2. The derived Hasse diagrams correctly reproduce the known Hasse diagrams of nilpotent orbit closures. Hence, these examples serve as non-trivial consistency check. 
Building on that, the Hasse diagrams for 5d SQCD at infinite gauge coupling are derived in Section \ref{subsection5d}. These results go beyond classical Higgs mechanism, as instanton operators enhance the Higgs branch at infinite coupling.  Nevertheless, the use of magnetic quivers and quiver subtraction allow to access the structure of these moduli spaces in a very computable fashion.
Moreover, Hasse diagrams for Higgs branches of 6d SCFTs were presented in Section \ref{subsection6d}. In contrast to Hasse diagrams of 5d Higgs branches at infinite coupling, the anomaly cancellation for 6d $\mathcal{N}=(1,0)$ theories imposes strong constraints. Consequently, the transition between finite and infinite coupling for theories on a ``$-1$" curve are realised by a small $E_8$ transition such that 6d Hasse diagrams at infinite couplings are extension of finite coupling Hasse diagrams by an additional $e_8$ transition.
In addition, Higgs branches of 4d $\mathcal{N}=2$ Argyres-Douglas theories provide another important testing ground for the proposed technique, as these are non-Lagrangian. Building on the magnetic quivers in the form of complete graphs, the Hasse diagram display an intricate structure of the moduli space as shown in Section \ref{subsectioncompletegraphs}.

Summarising the above, the use of the \emph{same} Hasse diagram for the pattern of partial Higgsing \emph{as well as} the singularity structure of the Higgs branch of classical theories is a surprising feature. Moreover, this realisation is not limited to classical cases, because recent developments allow to analyse Higgs branches at infinite coupling just as easily.

\paragraph{Outlook.}
The use of Hasse diagrams and Kraft-Procesi transitions to describe the features of hyper-K\"ahler moduli spaces appears as potent and computable tool. In light of the results of this paper, there are open questions that need to be addressed in future work. To name a few: 
\begin{itemize}
    \item Among other features, Hasse diagrams allow to read off the non-abelian part of the symmetry. However, the abelian part is less straightforward and deserves further studies.
    \item As a proof of concept, this paper has been restricted to single gauge groups with almost always fundamental matter. Extensions to different representations for the matter fields or products of simple gauge groups are expected to yield very intricate Hasse diagrams.

\begin{landscape}
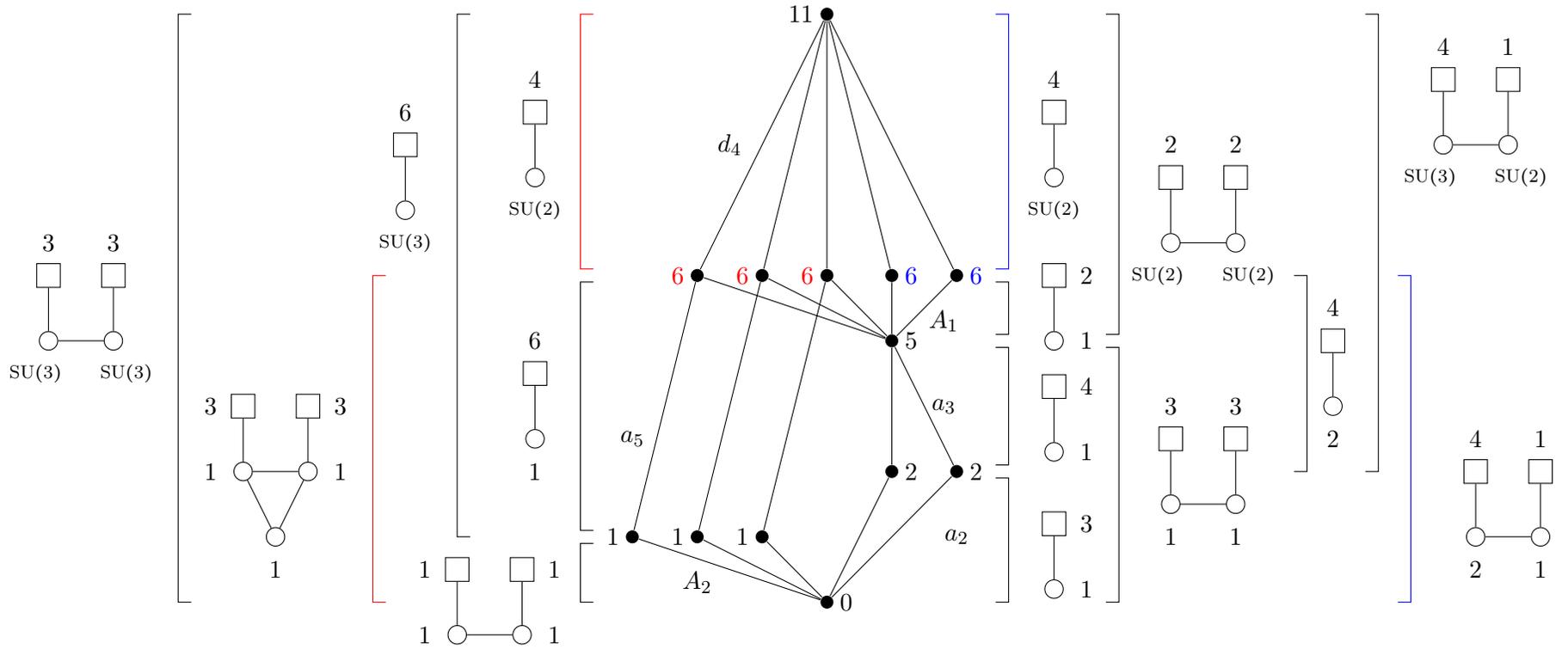
\begin{figure}[t]
    \centering
    \begin{tikzpicture}
		\tikzstyle{hasse} = [circle, fill,inner sep=2pt];
		\tikzstyle{gauge} = [inner sep=1mm,draw=none,fill=white,minimum size=2mm,circle, draw];
		\tikzstyle{flavour} = [draw=none,minimum size=0.3mm,fill=white, regular polygon,regular polygon sides=4,draw];
		
        \node[hasse] (1) at (0,0) {};
        \node[hasse] (2) at (-2,-4) {};
        \node[hasse] (3) at (-1,-4) {};
        \node[hasse] (4) at (0,-4) {};
        \node[hasse] (5) at (1,-4) {};
        \node[hasse] (6) at (2,-4) {};
        \node[hasse] (7) at (1,-5) {};
        \node[hasse] (8) at (-3,-8) {};
        \node[hasse] (9) at (-2,-8) {};
        \node[hasse] (10) at (-1,-8) {};
        \node[hasse] (11) at (1,-7) {};
        \node[hasse] (12) at (2,-7) {};
        \node[hasse] (13) at (0,-9) {};
		\draw (1)--(2)--(8)--(13) (1)--(3)--(9)--(13) (1)--(4)--(10)--(13) (2)--(7) (3)--(7) (4)--(7) (1)--(5)--(7)--(11)--(13) (1)--(6)--(7)--(12)--(13);
		\node at (-1.5,-2) {$d_4$};
		\node at (1.8,-4.7) {$A_1$};
		\node at (-3,-6.5) {$a_5$};
		\node at (1.8,-6) {$a_3$};
		\node at (2,-8) {$a_2$};
		\node at (-2,-8.7) {$A_2$};
		
		\node at (-0.4,0) {$11$};
		\node at (-2.3,-4) {\color{red}$6$};
		\node at (-1.3,-4) {\color{red}$6$};
		\node at (-0.3,-4) {\color{red}$6$};
		\node at (1.3,-4) {\color{blue}$6$};
		\node at (2.3,-4) {\color{blue}$6$};
		\node at (1.3,-5) {$5$};
		\node at (1.3,-7) {$2$};
		\node at (2.3,-7) {$2$};
		\node at (-3.3,-8) {$1$};
		\node at (-2.3,-8) {$1$};
		\node at (-1.3,-8) {$1$};
		\node at (0.3,-9) {$0$};
		
		\draw[red] (-3.6,0)--(-3.8,0)--(-3.8,-3.9)--(-3.6,-3.9);
		\draw (-3.6,-4.1)--(-3.8,-4.1)--(-3.8,-7.9)--(-3.6,-7.9) (-3.6,-8.1)--(-3.8,-8.1)--(-3.8,-9)--(-3.6,-9);
		
		\begin{scope}[shift={(-4.5,-2.5)}]
		\node[gauge] (14) at (0,0) {};
		\node[flavour] (15) at (0,1) {};
		\draw (14)--(15);
		\node at (0,-0.5) {$\scriptstyle{\mathrm{SU}(2)}$};
		\node at (0,1.5) {$4$};
		\end{scope}
		\begin{scope}[shift={(-4.5,-6.5)}]
		\node[gauge] (16) at (0,0) {};
		\node[flavour] (17) at (0,1) {};
		\draw (16)--(17);
		\node at (0,-0.5) {$1$};
		\node at (0,1.5) {$6$};
		\end{scope}
		\begin{scope}[shift={(-4.7,-9.5)}]
		\node[gauge] (18) at (0,0) {};
		\node[flavour] (19) at (0,1) {};
		\node[gauge] (20) at (-1,0) {};
		\node[flavour] (21) at (-1,1) {};
		\draw (19)--(18)--(20)--(21);
		\node at (0.5,0) {$1$};
		\node at (0.5,1) {$1$};
		\node at (-1.5,0) {$1$};
		\node at (-1.5,1) {$1$};
		\end{scope}
		
		\draw (-5.5,0)--(-5.7,0)--(-5.7,-8)--(-5.5,-8);
		
		\begin{scope}[shift={(-6.5,-3)}]
		\node[gauge] (22) at (0,0) {};
		\node[flavour] (23) at (0,1) {};
		\draw (22)--(23);
		\node at (0,-0.5) {$\scriptstyle{\mathrm{SU}(3)}$};
		\node at (0,1.5) {$6$};
		\end{scope}
		
		\draw[red] (-6.8,-4)--(-7,-4)--(-7,-9)--(-6.8,-9);
		
		\begin{scope}[shift={(-8,-7)}]
		\node[gauge] (24) at (0,0) {};
		\node[flavour] (25) at (0,1) {};
		\node[gauge] (26) at (-1,0) {};
		\node[flavour] (27) at (-1,1) {};
		\node[gauge] (28) at (-0.5,-1) {};
		\draw (25)--(24)--(26)--(27) (26)--(28)--(24);
		\node at (0.5,0) {$1$};
		\node at (0.5,1) {$3$};
		\node at (-1.5,0) {$1$};
		\node at (-1.5,1) {$3$};
		\node at (-0.5,-1.5) {$1$};
		\end{scope}
		
		\draw (-9.8,0)--(-10,0)--(-10,-9)--(-9.8,-9);
		
		\begin{scope}[shift={(-11,-5)}]
		\node[gauge] (60) at (0,0) {};
		\node[flavour] (61) at (0,1) {};
		\node[gauge] (62) at (-1,0) {};
		\node[flavour] (63) at (-1,1) {};
		\draw (61)--(60)--(62)--(63);
		\node at (0.2,-0.5) {$\scriptstyle{\mathrm{SU}(3)}$};
		\node at (0,1.5) {$3$};
		\node at (-1.2,-0.5) {$\scriptstyle{\mathrm{SU}(3)}$};
		\node at (-1,1.5) {$3$};
		\end{scope}
		
		\draw[blue] (2.6,0)--(2.8,0)--(2.8,-3.9)--(2.6,-3.9);
		\draw (2.6,-4.1)--(2.8,-4.1)--(2.8,-4.9)--(2.6,-4.9) (2.6,-5.1)--(2.8,-5.1)--(2.8,-6.9)--(2.6,-6.9) (2.6,-7.1)--(2.8,-7.1)--(2.8,-9)--(2.6,-9);
		
		\begin{scope}[shift={(3.5,-1.5)}]
		\node[gauge] (29) at (0,-1) {};
		\node[flavour] (30) at (0,0) {};
		\draw (29)--(30);
		\node at (0,-1.5) {$\scriptstyle{\mathrm{SU}(2)}$};
		\node at (0,0.5) {$4$};
		\end{scope}
		\begin{scope}[shift={(3.5,-5)}]
		\node[gauge] (31) at (0,0) {};
		\node[flavour] (32) at (0,1) {};
		\draw (31)--(32);
		\node at (0.5,0) {$1$};
		\node at (0.5,1) {$2$};
		\end{scope}
		\begin{scope}[shift={(3.5,-6.7)}]
		\node[gauge] (33) at (0,0) {};
		\node[flavour] (34) at (0,1) {};
		\draw (33)--(34);
		\node at (0.5,0) {$1$};
		\node at (0.5,1) {$4$};
		\end{scope}
		\begin{scope}[shift={(3.5,-8.8)}]
		\node[gauge] (35) at (0,0) {};
		\node[flavour] (36) at (0,1) {};
		\draw (35)--(36);
		\node at (0.5,0) {$1$};
		\node at (0.5,1) {$3$};
		\end{scope}
		
		\draw (4.3,0)--(4.5,0)--(4.5,-4.9)--(4.3,-4.9) (4.3,-5.1)--(4.5,-5.1)--(4.5,-9)--(4.3,-9);
		
		\begin{scope}[shift={(6.3,-3.5)}]
		\node[gauge] (37) at (0,0) {};
		\node[flavour] (38) at (0,1) {};
		\node[gauge] (39) at (-1,0) {};
		\node[flavour] (40) at (-1,1) {};
		\draw (38)--(37)--(39)--(40);
		\node at (0.2,-0.5) {$\scriptstyle{\mathrm{SU}(2)}$};
		\node at (0,1.5) {$2$};
		\node at (-1.2,-0.5) {$\scriptstyle{\mathrm{SU}(2)}$};
		\node at (-1,1.5) {$2$};
		\end{scope}
		\begin{scope}[shift={(6.3,-7.5)}]
		\node[gauge] (41) at (0,0) {};
		\node[flavour] (42) at (0,1) {};
		\node[gauge] (43) at (-1,0) {};
		\node[flavour] (44) at (-1,1) {};
		\draw (42)--(41)--(43)--(44);
		\node at (0,-0.5) {$1$};
		\node at (0,1.5) {$3$};
		\node at (-1,-0.5) {$1$};
		\node at (-1,1.5) {$3$};
		\end{scope}
		
		\draw (7.2,-4)--(7.4,-4)--(7.4,-7)--(7.2,-7);
		
		\begin{scope}[shift={(7.8,-6)}]
		\node[gauge] (45) at (0,0) {};
		\node[flavour] (46) at (0,1) {};
		\draw (45)--(46);
		\node at (0,-0.5) {$2$};
		\node at (0,1.5) {$4$};
		\end{scope}
		
		\draw (8.3,0)--(8.5,0)--(8.5,-7)--(8.3,-7);
		
		\begin{scope}[shift={(10.5,-2)}]
		\node[gauge] (47) at (0,0) {};
		\node[flavour] (48) at (0,1) {};
		\node[gauge] (49) at (-1,0) {};
		\node[flavour] (50) at (-1,1) {};
		\draw (48)--(47)--(49)--(50);
		\node at (0.2,-0.5) {$\scriptstyle{\mathrm{SU}(2)}$};
		\node at (0,1.5) {$1$};
		\node at (-1.2,-0.5) {$\scriptstyle{\mathrm{SU}(3)}$};
		\node at (-1,1.5) {$4$};
		\end{scope}
		
		\draw[blue] (8.8,-4)--(9,-4)--(9,-9)--(8.8,-9);
		\begin{scope}[shift={(11,-8)}]
		\node[gauge] (51) at (0,0) {};
		\node[flavour] (52) at (0,1) {};
		\node[gauge] (53) at (-1,0) {};
		\node[flavour] (54) at (-1,1) {};
		\draw (52)--(51)--(53)--(54);
		\node at (0,-0.5) {$1$};
		\node at (0,1.5) {$1$};
		\node at (-1,-0.5) {$2$};
		\node at (-1,1.5) {$4$};
		\end{scope}
    \end{tikzpicture}
\caption{Hasse diagram for the Higgs branch of the linear $6$d quiver with two $\mathrm{SU}(3)$ gauge nodes, a product of two $\mathrm{SU}(3)$ flavour symmetries, and bifundamental matter. While the symplectic leaves denoted {\color{red}$6$} correspond to the brane set up where one NS5 is moving along a Higgs branch direction. The symplectic leaves denoted {\color{blue}$6$} correspond to set ups where all NS5 branes are connected to D6 branes.
In order words, the $d_4$ transition leading to {\color{red}$6$} involves two NS5 branes transitioning away from the Higgs branch, while the $d_4$ transition leading to {\color{blue}$6$} involves all three NS5 branes. 
Note that in both cases the transverse slices are the Higgs branch of $\mathrm{SU}(2)$ with $4$ flavours, but the difference arises in the leaves {\color{red}$6$} and {\color{blue}$6$}, respectively. }
\label{fig:example_3-3-3-3}
\end{figure}
\end{landscape}
    
As an illustration, we demonstrate the Hasse diagram for the $6$d $\mathcal{N}=(1,0)$ theory describing three M$5$ branes transverse to $\mathbb{C}^2\slash \mathbb{Z}_3 $ in Figure \ref{fig:example_3-3-3-3}. Comparing to studies like \cite{Heckman:2016ssk,Hassler:2019eso}, the full Hasse diagram is much more intricate than the linear Hasse diagram for $\mathfrak{sl}(3,\mathbb{C})$, which only displays the $A_2$ and $a_2$ transition at the bottom of Figure \ref{fig:example_3-3-3-3}. Physically, the reason is clear, while Higgsing processes in \cite{Heckman:2016ssk} are limited to a subset of all possibilities, namely boundary conditions of D$6$ on D$8$ branes; the process of subsequently removing elementary slices explores all processes along the Higgs branch. As an example, the $d_4$ transition at the top of the Hasse diagram involves either two or three NS$5$ branes in the Type IIA description. Considering an arbitrary number of M$5$ branes results in a significantly increased complexity of the Hasse diagram. 
    \item As hinted on in the Introduction, when describing the Higgs branch as a union of symplectic singularities one describes it as an algebraic variety, ignoring nilpotent elements in the Higgs branch chiral ring. However, nilpotent operators may be present in which case the Higgs branch may be seen as a non-reduced affine scheme. Consequently, the structure of the Hasse diagram needs to be reconsidered \cite{Bourget:2019rtl}.
    \item The techniques of this paper rely on theories which either have a brane realisation without an orientifold or have known magnetic quiver with unitary gauge groups only. In order to analyse models with multiple ortho-symplectic gauge groups or hypermultiplets transforming in spinor representations, the techniques need to be extended to this set-up.
    \item Finally, it is known that \emph{non-simply laced quivers} can be used to describe certain symplectic singularities, as was mentioned in Section \ref{subsectionSimpleGaugeGroup} when the gauge group is $\mathrm{Sp}(k)$. It would be useful to generalise quiver subtraction to such quivers, as only partial results are available to date. The subtlety comes from the fact that for a non-simply laced quiver, there are various inequivalent ways of ungauging the global $\mathrm{U}(1)$ -- there exist various \emph{ungauging schemes}. As an illustration, the quiver 
    \begin{equation}
         \node{}1-\node{}2-\node{}3<=\node{}4-\nodesq{}2
    \end{equation}
    corresponds to the $E_6$ minimal nilpotent orbit (and can be used to represent an $e_6$ elementary slice) while the quiver 
    \begin{equation}
        \nodesq{}{1}-\node{}2-\node{}3<=\node{}4-\node{}{2}
    \end{equation}
    corresponds to the next-to-minimal nilpotent orbit closure of $F_4$. In the last two quivers the square-circle $$\boxcircle$$ denotes a particular choice of ungauging scheme which is adopted. Ungauging schemes are studied in an ongoing work \cite{Anton}. Using this, we can for instance compute the Hasse diagram for the family of quivers 
    \begin{equation}
    \label{quiverc}
         \node{}1-\node{}2-\node{}3- \cdots - \node{}{n-1}<=\node{}{n}-\nodesq{}{2}
    \end{equation}
    The result is shown in Figure \ref{fig:HasseQuiverc}. 
\end{itemize}
We leave these questions for further investigation.

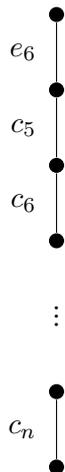
\begin{figure}[t]
    \centering
    	  	\begin{tikzpicture}
		\tikzstyle{hasse} = [circle, fill,inner sep=2pt];
		\node at (0,0) [hasse] (0) [] {};
		\node at (0,-1) [hasse] (1) [] {};
		\node at (0,-2) [hasse] (2) [] {};
		\node at (0,-3) [hasse] (3) [] {};
		\node at (0,-4) [] (4) [] {$\vdots$};
		\node at (0,-5) [hasse] (5) [] {};
		\node at (0,-6) [hasse] (6) [] {};
		\draw (0) edge [] node[label=left:$e_6$] {} (1);
		\draw (1) edge [] node[label=left:$c_5$] {} (2);
		\draw (2) edge [] node[label=left:$c_6$] {} (3);
		\draw (5) edge [] node[label=left:$c_n$] {} (6);
	\end{tikzpicture}
    \caption{Hasse diagram for the family of quivers \eqref{quiverc}. The number of minimal transverse slices is $n-3$. }
    \label{fig:HasseQuiverc}
\end{figure}

\section*{Acknowledgements}
We are indebted to Gabi Zafrir for many insightful discussions and clarifications. 
We would like to thank Rudolph Kalveks, Paul Levy, Hiraku Nakajima, Travis Schedler and Ben Webster for useful discussions. 
This work was supported by STFC grant ST/P000762/1 and STFC Consolidated Grant ST/J0003533/1.  
MS was supported by the National Thousand-Young-Talents Program of China and the China Postdoctoral Science Foundation (grant no.\ 2019M650616).
This work benefited from the 2019 Pollica summer workshop, which was supported in part by the Simons Foundation (Simons Collaboration on the Non-perturbative Bootstrap) and in part by the INFN. AB, AH and AZ are grateful for this support.
AB, JFG, AH, MS, AZ and ZZ gratefully acknowledge support from the Simons Center for Geometry and Physics, Stony Brook University where part of the research for this paper was performed.

\clearpage

\appendix

\section{Branes, Magnetic Quivers, Kraft-Procesi Transitions and Quiver Subtraction}
\label{app:Branes_Quivers}

\subsection{Brane Constructions and Magnetic Quivers}
\label{AppendixBranes}

The theories considered in this paper can be realised as $p$-dimensional world-volume theories on intersecting D$p$, D$(p+2)$, and NS$5$ brane configurations in Type IIA or Type IIB. In particular, $p=5$ and $6$ are the most relevant to this work and are briefly summarised below. The world-volume theories are read off from the suspension pattern of the fundamental string and the arising massless degrees of freedom are electrically charged. Therefore, the $p$-dimensional theories are called \emph{electric theories}.

\paragraph{5-brane web configuration.}
For $p=5$ the simplest theories are derived from D$5$ branes ending on NS$5$ in Type IIB; however, as extensively discussed in \cite{Aharony:1997ju,Aharony:1997bh}, the D$5$s ending on NS$5$ lead to brane bending and the notion of $(p,q)$ 5-branes, i.e.\ bound states of $p$ D$5$ and $q$ NS$5$. Consequently, the entire brane configuration becomes a 5-brane web. In addition, a $(p,q)$ 5-brane can terminate only on a $[p,q]$ 7-brane. The space time occupation of the different branes is summarised in Table \ref{tab:branes}. The electric theory at finite gauge coupling is read off when all NS$5$ branes are well-separated. 

\begin{table}[t]
    \centering
    \begin{tabular}{c|cccccccccc}
    \toprule
         Type IIB & $x_0$ &  $x_1$ & $x_2$ & $x_3$ & $x_4$ & $x_5$ & $x_6$ & $x_7$ & $x_8$ & $x_9$    \\ \midrule
         NS$5$ & $\times$ &  $\times$ & $\times$ & $\times$ & $\times$ & $\times$ & \\
         D$5$ & $\times$ &  $\times$ & $\times$ & $\times$ & $\times$ &  & $\times$ & \\
         $(p,q)$5-brane & $\times$ &  $\times$ & $\times$ & $\times$ & $\times$ & \multicolumn{2}{c}{$\times$}  & \\ 
         $[p,q]$7-brane & $\times$ &  $\times$ & $\times$ & $\times$ & $\times$ &  &  & $\times$ & $\times$ &$\times$  \\
         \midrule
Type IIA & $x_0$ &  $x_1$ & $x_2$ & $x_3$ & $x_4$ & $x_5$ & $x_6$ & $x_7$ & $x_8$ & $x_9$    \\ \midrule
         NS$5$ & $\times$ &  $\times$ & $\times$ & $\times$ & $\times$ & $\times$ & \\
         D$6$ & $\times$ &  $\times$ & $\times$ & $\times$ & $\times$ & $\times$ & $\times$ & \\
         D$8$, O8 & $\times$ &  $\times$ & $\times$ & $\times$ & $\times$ & $\times$  &  & $\times$ & $\times$ & $\times$\\
         \bottomrule
    \end{tabular}
    \caption{Occupation of space-times directions by the D$p$, D$(p{+}2)$, and NS $5$ branes. Upper part: branes for the 5-brane webs ending on 7-branes in Type IIB. Lower part: brane configuration for $6$d theories in Type IIA.}
    \label{tab:branes}
\end{table}

\paragraph{6d brane configuration.}
Next, for $p=6$ the D$6$, D$8$, and NS$5$ branes intersect as indicated in Table \ref{tab:branes}. The $6$-dimensional $\mathcal{N}=(1,0)$ world-volume theories have been first discussed in \cite{Hanany:1997gh,Brunner:1997gk,Hanany:1997sa}. Analogous to the above, the electric theory at finite gauge coupling is most easily derived from the phase of the brane configuration where all NS$5$ branes are well-separated and D$6$ branes are suspended only between NS$5$s. Contrary to the 5-branes webs, there exist restrictions on the 6d brane configurations in form of charge conservation. This charge conservation is equivalent to anomaly cancellation conditions in the resulting low-energy field theory. 

\paragraph{Higgs branches at different phases of the theory.}
In the spirit of \cite{Cabrera:2019izd,Cabrera:2018jxt}, given a $5$d or $6$d electric theory, one may consider the inverse gauge coupling $ 1\slash g^2 $ as the order parameter of the various phases of the theory, each with an associated moduli space. For a given phase, the moduli space generically decomposes into various mixed branches. Starting from the maximal Higgs branch, i.e.\ any assignment of vacuum expectation values that breaks the (electric) gauge group maximally, one can move onto singular loci of the Higgs branch towards mixed branches. This is because along the singularity the gauge group is not fully broken anymore and Coulomb branch directions may open up.
Depending on the phase, the Higgs branches can behave quite differently. For instance, the Higgs branch at finite coupling is a hyper-K\"ahler quotient thanks to the amount of supersymmetry \cite{Hitchin:1986ea}. Moving to the phase where the gauge coupling becomes infinite, the corresponding Higgs branch may enhance due to BPS gauge instantons that become massless. In the 5d setting, this enhancement of the Higgs branch is due to instanton operators \cite{Cremonesi:2015lsa,Ferlito:2017xdq,Cabrera:2018jxt}, while the enhancement in 6d is due to tensionless strings \cite{Mekareeya:2017jgc,Hanany:2018uhm,Hanany:2018vph,Cabrera:2019izd}. 

These BPS states are not captured by conventional F and D-terms; hence, no hyper-K\"ahler quotient description is available for Higgs branches over singular loci. Fortunately, these Higgs branches are still symplectic singularities (or unions thereof) and can be described as a space of dressed monopole operators by formally computing the $3$d Coulomb branch $\mathcal{C}$ of the associated \emph{magnetic quivers} \cite{Cabrera:2019izd}.
To be explicit, given an electric theory in $d=3,4,5,6$ with $8$ supercharges in a phase $\mathcal{P}$ there exists a (finite, non-empty) set of associated magnetic quivers $\mathsf{Q}^{\alpha}_{\mathcal{P}}$, such that
\begin{align}
    \mathcal{H}^{d} \left( \text{phase } \mathcal{P} \text{ of electric theory}\right) 
    =
    \bigcup_{\alpha} \mathcal{C} \left( \text{magnetic quiver } \mathsf{Q}^{\alpha}_{\mathcal{P}} \right) ,
\end{align}
holds as an \emph{equality of moduli spaces}. Here,  $\mathcal{H}^{d}$ is the Higgs branch of the $d$-dimensional theory in a given phase and $\alpha$ labels the (finite number of) symplectic singularities the Higgs branch is a union of. 

The procedure on how to derive a magnetic quiver from a brane configuration is now recapitulated. The main idea is to change the phase of the brane system to the state where all D$p$ branes are suspended between D$(p{+}2)$ branes. For every inequivalent decomposition of splitting D$p$ between D$(p{+}2)$ branes, one can read off a magnetic quiver. Contrary to $3$d brane configurations \cite{Hanany:1996ie} involving D$3$, D$5$, and NS$5$ branes, in brane constructions for higher dimensional theories the NS$5$ branes contribute dynamical degrees of freedom to the magnetic quiver.

As a remark, the magnetic quiver for a $3$d $\mathcal{N}=4$ theory constructed in Type IIB set-up of \cite{Hanany:1996ie} with D$3$, D$5$, and NS$5$ branes is nothing else than the magnetic theory determined from the suspension pattern of the D1-string in the D$3$-D$5$-NS$5$ brane configuration. 

\paragraph{Magnetic quiver for 5d brane webs.}
Following \cite{Cabrera:2018jxt}, each semi-infinite $(p,q)$ $5$-brane in the $5$-brane web terminates on a $[p,q]$ 7-brane. Next, the brane configuration is transitioned to the origin of the Coulomb branch and the web is subsequently divided into sub-webs, which respect charge conservation at every vertex as well as the s-rule. The magnetic quiver is then associated via
\begin{compactenum}[(i)]
\item $n$ copies of an independent sub-web give rise to a magnetic vector multiplet of $U(n)$.
\item The intersection numbers between the sub-webs determine which magnetic gauge nodes are connected via bi-fundamtental matter as well as the edge multiplicity.
\end{compactenum}
\paragraph{Magnetic quiver for 6d brane configurations.}
As explained in \cite{Cabrera:2019izd}, the magnetic phase of the Type IIA brane is reached by pulling in D$8$ branes from infinity such that all D$6$ branes can be suspended between D$8$ branes. Then, there are four rules to derive the magnetic quiver
\begin{compactenum}[(i)]
\item A stack of $n$ D$6$ branes between two adjacent D$8$ branes yields a magnetic $U(n)$ gauge node.
\item Two stacks of $n_1$ and $n_2$ D$6$ branes in adjacent D$8$ intervals have magnetic gauge nodes connected by a bi-fundamental hypermultiplets of $U(n_1) \times U(n_2)$.
\item A stack of $k$ coincident NS$5$ branes yields a magnetic $U(k)$ gauge node together with an additional hypermultiplet in the adjoint representation if the stack is free to move in the $x_6$ direction. In other words, NS$5$ stuck on the O$8^-$ do not have an adjoint hypermultiplet.
\item A stack of $k$ NS$5$s in the same D$8$ interval as $n$ D$6$s gives rise to magnetic bi-fundamentals that connect the $U(k)$ and $U(n)$ gauge nodes.
\end{compactenum}

\begin{figure}[t]
\centering
\begin{tikzpicture}
	\begin{pgfonlayer}{nodelayer}
		\node [style=gauge1] (1) at (3.5, -7) {};
		\node [style=gauge1] (2) at (2, -7) {};
		\node [style=gauge1] (3) at (0.5, -7) {};
		\node [style=gauge1] (4) at (4.5, -5) {};
		\node [style=gauge1] (5) at (6.5, -5) {};
		\node [style=gauge1] (6) at (4.5, -9) {};
		\node [style=gauge1] (7) at (6.5, -9) {};
		\node [style=gauge1] (8) at (7.5, -7) {};
		\node [style=gauge1] (9) at (8.75, -7) {};
		\node [style=gauge1] (10) at (10.25, -7) {};
		\node [style=none] (11) at (2, -6.875) {};
		\node [style=none] (12) at (2, -7.125) {};
		\node [style=none] (13) at (3.5, -6.875) {};
		\node [style=none] (14) at (3.5, -7.125) {};
		\node [style=none] (15) at (3.5, -6.875) {};
		\node [style=none] (16) at (3.5, -7.125) {};
		\node [style=none] (17) at (3.5, -7) {};
		\node [style=none] (18) at (7.5, -6.875) {};
		\node [style=none] (19) at (7.5, -7) {};
		\node [style=none] (20) at (7.5, -7.125) {};
		\node [style=none] (21) at (7.5, -6.875) {};
		\node [style=none] (22) at (7.5, -7.125) {};
		\node [style=none] (23) at (8.75, -6.875) {};
		\node [style=none] (24) at (8.75, -7.125) {};
		\node [style=none] (25) at (4.5, -6) {};
		\node [style=none] (26) at (6.5, -6) {};
		\node [style=none] (27) at (7, -6.875) {};
		\node [style=none] (28) at (7, -7.125) {};
		\node [style=none] (29) at (6.5, -8) {};
		\node [style=none] (30) at (4, -7.125) {};
		\node [style=none] (31) at (4.5, -8) {};
		\node [style=none] (32) at (4, -6.875) {};
	\end{pgfonlayer}
	\begin{pgfonlayer}{edgelayer}
		\draw [style=cyane] (19.center) to (17.center);
		\draw [style=rede] (22.center) to (24.center);
		\draw [style=rede] (12.center) to (14.center);
		\draw [style=bluee] (9) to (10);
		\draw [style=bluee] (3) to (2);
		\draw [style=greene] (25.center) to (4);
		\draw [style=greene] (5) to (26.center);
		\draw [style=greene] (30.center) to (16.center);
		\draw [style=greene] (30.center) to (31.center);
		\draw [style=greene] (31.center) to (6);
		\draw [style=greene] (29.center) to (7);
		\draw [style=greene] (29.center) to (28.center);
		\draw [style=greene] (28.center) to (22.center);
		\draw [style=greene] (27.center) to (21.center);
		\draw [style=greene] (26.center) to (27.center);
		\draw [style=greene] (32.center) to (15.center);
		\draw [style=greene] (25.center) to (32.center);
		\draw [style=greene] (25.center) to (26.center);
		\draw [style=greene] (29.center) to (31.center);
		\draw [style=greene] (15.center) to (11.center);
		\draw [style=greene] (21.center) to (23.center);
	\end{pgfonlayer}
\end{tikzpicture}
    \caption{Brane web construction of 5d $\mathrm{SU}(3)$ with 6 flavours at finite coupling on the mixed branch where the theory is broken to $\mathrm{SU}(2)$ with 4 flavours and 5 neutral hypermultiplets. Coloured branes are assumed to be on different positions along the 7-branes.}
    \label{fig:MixedSU(3)}
\end{figure}
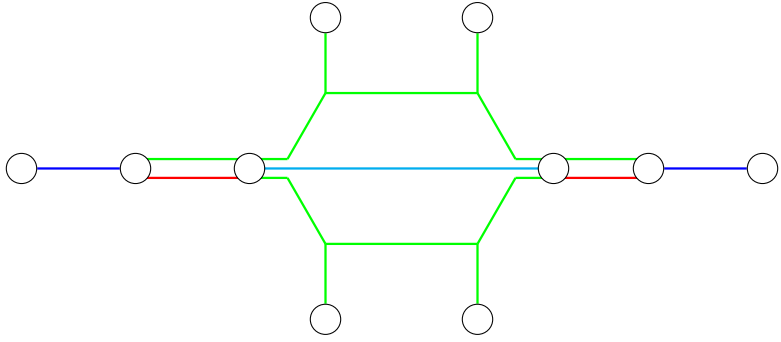

\paragraph{Mixed branches from branes.}
Here we present a new result. The notion of magnetic quivers allows not only for the computation of the Higgs branch of a theory, but it can be combined with reading electric quivers for the effective gauge theory in the brane picture to gain information about mixed branches. The $5$-brane set-up of $5$d $\mathrm{SU}(3)$ with 6 flavours on the mixed branch where the theory is broken to $\mathrm{SU}(2)$ with 4 flavours and 5 neutral hypermultiplets is shown in Figure \ref{fig:MixedSU(3)}. We find the singular loci of the Higgs branch by opening up Coulomb branch directions in the brane web (more generally in any brane system for $d<6$ dim theories) and using the rules to read magnetic quivers for the Higgs branch at this phase. Both the effective theory for the residual gauge group and the magnetic quiver for the transverse slice can be identified. This way, if one opens up only one Coulomb branch modulus at a time, one can obtain the elementary slice between two symplectic leaves in the Higgs branch. This had never been done before and results from a natural combination of the brane realisation of Kraft-Procesi transitions \cite{Cabrera:2016vvv,Cabrera:2017njm} with the proposal of magnetic quivers of 5-brane webs \cite{Cabrera:2018jxt}. See Figure \ref{fig:SU(3)6branesclass} for the example of $\mathrm{SU}(3)$ with 6 flavours.

The same procedure extends to any brane system, with a caveat in 6 dimensions, where no Coulomb branch moduli exists in the brane picture and it is more difficult to identify singular loci of the Higgs branch. Nevertheless, it is possible by identifying all phases where different electric theories are realised on the tensor branch. See Figure \ref{fig:SU(4)12Branes} and Figure \ref{fig:SU(4)Magnetic} for $\mathrm{SU}(4)$ with $12$ fundamental and one $2$nd rank antisymmetric hypermultiplet.

\subsection{Quiver Subtraction}
\label{AppendixSubtraction}

Subtracting quivers in a systematic way to obtain information about symplectic leaves (see Appendix \ref{SymplecticLeaves} for more on symplectic leaves) and transverse slices was first discussed in \cite{Cabrera:2018ann}. They take the magnetic quivers representing the closures of two symplectic leaves in a symplectic singularity and compute the magnetic quiver for the transverse slice. Here, we aim to give an algorithm that computes the quiver representing the closure of the lower symplectic leaf in the Hasse diagram by subtracting the quiver representing the transverse slice from the quiver representing the closure of the symplectic leaf higher in the Hasse diagram. The minus sign of any quiver subtraction cannot be compared to the arithmetic operation which for $a,b,c\in\mathbb{R}:$ $a-b=a+(-1)b$ and $a-b=c\Leftrightarrow a-c=b$, both generally untrue for quiver subtraction. Therefore, the algorithm given here looks different from the algorithm presented in \cite{Cabrera:2018ann}, even though the idea behind it is the same (i.e. a way to express the work of Kraft and Procesi \cite{kraft1980minimal,Kraft1982} in terms of quivers).

By construction, the Higgs branch of a theory with 8 supercharges is a symplectic singularity (or union thereof, this is elaborated in Section \ref{subsectionMultiple}) foliated into symplectic leaves with an associated Hasse diagram $\mathfrak{H}$.  The closure of a symplectic leaf $\mathcal{L}$ in a symplectic singularity $\mathcal{S}$ is obtained by the union of all the symplectic leaves which are lower in the Hasse diagram of $\mathcal{S}$. The magnetic quiver we read for the local Higgs branch on various mixed branches and hence on different points on the Hasse diagram then represents the closure of a symplectic leaf.

Let $\mathcal{L}'$ be a symplectic leaf just below $\mathcal{L}$ in $\mathfrak{H}$ and $\mathsf{Q}'$ its associated magnetic quiver. Then, there is a transverse slice (which is a symplectic singularity itself) to $\mathcal{L}'$ in $\mathcal{L}$ with an associated magnetic quiver $\mathsf{D}$. The process of \emph{quiver subtraction of an elementary slice} takes the input of a) magnetic quiver $\mathsf{Q}$ of a symplectic leaf inside a Higgs branch and b) magnetic quiver $\mathsf{D}$ representing an elementary slice and returns the magnetic quiver $\mathsf{Q}'$ representing the closure of the symplectic leaf lower in the Hasse diagram with $\mathsf{D}$ being the corresponding transverse slice. Diagrammatically it is defined as an operation on the quivers:
\begin{equation}
    \mathsf{Q} - \mathsf{D} = \mathsf{Q}'\,,
\end{equation}

If a brane construction exists, all three quivers, $\mathsf{Q}$, $\mathsf{D}$ and $\mathsf{Q}'$ can be read from it by moving between mixed branches. This allows us to deduce general rules of quiver subtraction for unitary quivers even for cases where no brane description is known. If we move to a singular locus in a minimal way, i.e. moving as few branes as possible, this translates to magnetic quivers as subtracting an elementary slice. In order to obtain all points in the Hasse diagram, by starting with the magnetic quiver representing the entire symplectic singularity we have to be careful to subtract only elementary slices, which are transverse slices between two neighbouring symplectic leaves. Following the analysis of \cite{Kraft1982} for nilpotent orbit closures, which has been applied to $3$d $\mathcal{N}=4$ Higgs and Coulomb branches in \cite{Cabrera:2016vvv,Cabrera:2017njm}, the elementary slices are generally \emph{assumed} to be either minimal nilpotent orbit closures or Kleinian singularities. An assumption which is backed up by manipulating brane constructions. With this assumption, one can start from a given Higgs branch, with its corresponding magnetic quiver $\mathsf{Q}$, then one finds all subgraphs\footnote{Note that the edge multiplicity between the nodes of the quiver of the elementary slice has to fit the edge multiplicity in the original magnetic quiver.} $\mathsf{D}$ which are elementary slices (see Appendix \ref{minimaltransitions}) and performs the quiver subtraction using the following prescription.

\begin{enumerate}
    \item [\underline{Quiver Subtraction of an Elementary Slice:}$\qquad\mathsf{Q}-\mathsf{D}=\mathsf{Q}'$]
    \item Align quivers $\mathsf{Q}$ and $\mathsf{D}$ (this may be done in different ways, not necessarily related by action of the automorphism group of the quiver $\mathsf{Q}$).
    \item Subtract the ranks of $\mathsf{D}$ from the ranks of $\mathsf{Q}$ to obtain a quiver $\mathsf{S}$.
    \item Restore the balance of the nodes: Add a $\mathrm{U}(1)$ node to $\mathsf{S}$ and connect it, with possible edge multiplicity, to the remaining non-zero nodes of $\mathsf{S}$ to create a new quiver $\mathsf{Q}'$  such that the balance\footnote{Note that this does not imply that the resulting quiver should be balanced. Rather that the balance of each individual node is the same before and after subtraction.} of the nodes in $\mathsf{Q}'$ and $\mathsf{Q}$ match.
\end{enumerate}
\paragraph{Example.}
Consider $\mathrm{SU}(3)$ with $6$ fundamentals at finite coupling, the magnetic quiver is the first quiver in \eqref{qsexample}. Upon inspection, one finds that it contains an affine $\hat{D}_4$ sub-graph. Therefore, it is suggestive to subtract the affine $\hat{D}_4$ quiver such that the closure of the minimal nilpotent orbit of $\mathrm{SO}(8)$ is an elementary slice. The algorithmic evaluation proceeds by aligning the $\hat{D}_4$ quiver with the original magnetic quiver and then subtracting the ranks and rebalancing the nodes. In particular, one has
\be
\left(  \node{}1-\node{}2-\node{\overset{\displaystyle\overset 1 \circ~~} \diagdown  \overset{\displaystyle\overset{~~1} {~~ \circ}} \diagup}3 -\node{}2-\node{}1 \right) ~ - ~ \left(  \node{}1-\node{\overset{\displaystyle\overset 1 \circ~~} \diagdown  \overset{\displaystyle\overset{~~1} {~~ \circ}} \diagup}2 -\node{}1 \right) ~= ~\left( \overset{\overset{\displaystyle\overset 1 \circ}{\diagup~~~~~~~~~~\diagdown}}{\node{}{1}-\node{}1-\node{}1-\node{}1-\node{}{1}}\right) \,,
\label{qsexample}
\ee
where it is apparent that the two $\mathrm{U}(1)$ gauge nodes, which the original quiver and the affine $\hat{D}_4$ have in common, disappear as their ranks are identical. Also, the  $\mathrm{U}(2)$ and $\mathrm{U}(3)$ nodes of the original magnetic quiver have been reduced to $\mathrm{U}(1)$ nodes due to the subtraction of their ranks. Lastly, the resulting quiver has to be rebalanced via an additional $\mathrm{U}(1)$ gauge node that has been added on top. This node is connected to the leftmost and rightmost $\mathrm{U}(1)$ gauge nodes such that their balance remains the same (which is zero). The balance of the other three nodes on the bottom is unchanged by the subtraction. One observes that the subtraction yields an affine $\hat{A}_5$ quiver whose Coulomb branch is the closure of the minimal nilpotent orbit of $\mathrm{SU}(6)$.

The same algorithm applies to subtracting any suitable transverse slice. However, identifying which transverse slices are suitable is highly non-trivial.

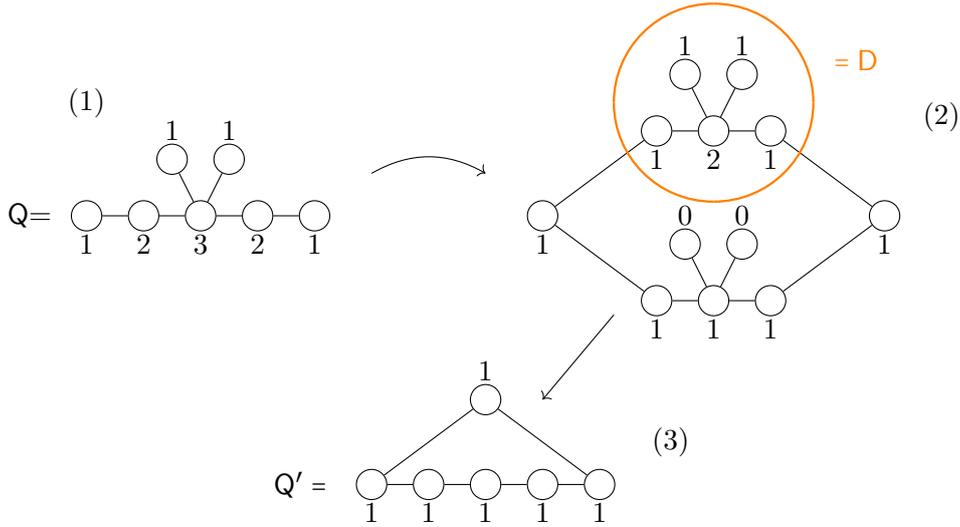
\begin{figure}[t]
\makebox[\textwidth][c]{
\begin{tikzpicture}[scale=0.75]
	\begin{pgfonlayer}{nodelayer}
		\node [style=gauge1] (0) at (-6, 1.5) {};
		\node [style=gauge1] (1) at (-5, 1.5) {};
		\node [style=gauge1] (2) at (-4, 1.5) {};
		\node [style=gauge1] (3) at (-3, 1.5) {};
		\node [style=gauge1] (4) at (-2, 1.5) {};
		\node [style=gauge1] (5) at (-3.5, 2.5) {};
		\node [style=gauge1] (6) at (-4.5, 2.5) {};
		\node [style=none] (7) at (-6, 1) {1};
		\node [style=none] (8) at (-5, 1) {2};
		\node [style=none] (9) at (-4, 1) {3};
		\node [style=none] (10) at (-3, 1) {2};
		\node [style=none] (11) at (-2, 1) {1};
		\node [style=none] (12) at (-3.5, 3) {1};
		\node [style=none] (13) at (-4.5, 3) {1};
		\node [style=none] (14) at (-1, 2.25) {};
		\node [style=none] (15) at (1, 2.25) {};
		\node [style=gauge1] (17) at (4, 3) {};
		\node [style=gauge1] (18) at (5, 3) {};
		\node [style=gauge1] (19) at (6, 3) {};
		\node [style=gauge1] (20) at (2, 1.5) {};
		\node [style=gauge1] (21) at (4, 0) {};
		\node [style=gauge1] (22) at (5, 0) {};
		\node [style=gauge1] (23) at (6, 0) {};
		\node [style=gauge1] (24) at (8, 1.5) {};
		\node [style=gauge1] (25) at (4.5, 4) {};
		\node [style=gauge1] (26) at (5.5, 4) {};
		\node [style=gauge1] (27) at (4.5, 1) {};
		\node [style=gauge1] (28) at (5.5, 1) {};
		\node [style=none] (29) at (4.5, 4.5) {1};
		\node [style=none] (30) at (5.5, 4.5) {1};
		\node [style=none] (31) at (5, 2.5) {2};
		\node [style=none] (32) at (4, 2.5) {1};
		\node [style=none] (33) at (6, 2.5) {1};
		\node [style=none] (34) at (2, 1) {1};
		\node [style=none] (35) at (8, 1) {1};
		\node [style=none] (36) at (4, -0.5) {1};
		\node [style=none] (37) at (5, -0.5) {1};
		\node [style=none] (38) at (6, -0.5) {1};
		\node [style=none] (39) at (3.25, -0.25) {};
		\node [style=none] (40) at (4.5, 1.5) {0};
		\node [style=none] (41) at (5.5, 1.5) {0};
		\node [style=none] (42) at (2, -1.75) {};
		\node [style=gauge1] (43) at (1, -1.75) {};
		\node [style=gauge1] (44) at (-1, -3.25) {};
		\node [style=gauge1] (45) at (0, -3.25) {};
		\node [style=gauge1] (46) at (1, -3.25) {};
		\node [style=gauge1] (47) at (2, -3.25) {};
		\node [style=gauge1] (48) at (3, -3.25) {};
		\node [style=none] (49) at (1, -1.25) {1};
		\node [style=none] (51) at (-1, -3.75) {1};
		\node [style=none] (52) at (0, -3.75) {1};
		\node [style=none] (53) at (1, -3.75) {1};
		\node [style=none] (54) at (2, -3.75) {1};
		\node [style=none] (55) at (3, -3.75) {1};
		\node [style=none] (57) at (5, 5.25) {};
		\node [style=none] (58) at (6.75, 3.5) {};
		\node [style=none] (59) at (5, 1.75) {};
		\node [style=none] (60) at (3.25, 3.5) {};
		\node [style=none] (61) at (-7, 1.5) {$\mathsf{Q}$=};
		\node [style=none] (62) at (7.5, 4.25) {$\color{orange}{=\mathsf{D}}$};
		\node [style=none] (63) at (-2.25, -3.25) {$\mathsf{Q}^\prime=$};
		\node [style=none] (64) at (-6, 3.5) {(1)};
		\node [style=none] (65) at (9, 3.25) {(2)};
		\node [style=none] (66) at (4.25, -2.5) {(3)};
	\end{pgfonlayer}
	\begin{pgfonlayer}{edgelayer}
		\draw (0) to (1);
		\draw (1) to (2);
		\draw (2) to (3);
		\draw (3) to (4);
		\draw (2) to (5);
		\draw (6) to (2);
		\draw [style=->, bend left] (14.center) to (15.center);
		\draw (17) to (20);
		\draw (17) to (18);
		\draw (18) to (19);
		\draw (19) to (24);
		\draw (24) to (23);
		\draw (23) to (22);
		\draw (22) to (21);
		\draw (20) to (21);
		\draw (25) to (18);
		\draw (18) to (26);
		\draw (27) to (22);
		\draw (22) to (28);
		\draw [style=->] (39.center) to (42.center);
		\draw (43) to (44);
		\draw (44) to (45);
		\draw (45) to (46);
		\draw (47) to (48);
		\draw (46) to (47);
		\draw (43) to (48);
		\draw [style=orangee, in=90, out=0] (57.center) to (58.center);
		\draw [style=orangee, in=0, out=-90] (58.center) to (59.center);
		\draw [style=orangee, in=-90, out=-180] (59.center) to (60.center);
		\draw [style=orangee, in=-180, out=90] (60.center) to (57.center);
	\end{pgfonlayer}
\end{tikzpicture}}
    \caption{Example of quiver subtraction of an elementary slice quiver $\mathsf{Q}-\mathsf{D}=\mathsf{Q}'$: Start with the quiver $\mathsf{Q}$ in (1) and identify the $d_4$ centered in the $U(3)$ node. (2) split this quiver such that the nodes containing the elementary slice quiver are split in two, conserving the rank and revealing the elementary slice quiver $\mathsf{D}$. (3) replace $\mathsf{D}$ with a single $\mathrm{U}(1)$ node to obtain the quiver $\mathsf{Q'}$. The subtraction $\mathsf{Q}-\mathsf{Q}'=D$ in \cite{Cabrera:2018ann} is achieved by replacing what is not in the orange circle in (2) by an empty node.}
    \label{blowingdown}
\end{figure}

The reason the quivers before and after subtraction have the same balance, with exception of the $\mathrm{U}(1)$ node added in step 3, is clear from the brane picture and can be visualised in the quiver. For the case where the quiver is that of an elementary slice itself one replaces it with a $\mathrm{U}(1)$ node. For a general quiver first one identifies the quiver of an elementary slice, then splits all nodes into two conserving overall rank, such that the elementary slice appears with the correct rank. Now one replaces the elementary slice quiver with a $\mathrm{U}(1)$ node. One is left with a quiver which has the same balance as the previous one, except for the $\mathrm{U}(1)$ node that replaced the elementary slice quiver. The previous example is reconsidered with this method in Figure \ref{blowingdown}.

In cases where edges with multiplicities appear, it is important to keep in mind that quivers can only be subtracted from one another if the corresponding edge multiplicities are identical. One has to be careful in identifying the possible elementary slices. For example:
\begin{equation}
    \overset{\overset{\displaystyle\overset 1 \circ}{\diagup~\diagdown}}{\node{}{1} = \node{}{1}}
\qquad \textnormal{only has an }a_1 \;(\textnormal{or equivalently}\; A_1)  \textnormal{and not an }a_2\textnormal{ elementary slice}
\end{equation}

\begin{equation}
\node{~~\diagup}1- \node{\topnode{}1}1-~ \node{\diagdown~}1
\qquad \textnormal{only has an }a_2\textnormal{ not an }a_3\textnormal{ elementary slice}\,.
\end{equation}

So far we have not mentioned what happens in the presence of flavour nodes. Given a flavourless quiver, any node can be ungauged. If the node we ungauge is an $\mathrm{U}(1)$, then we are left with a quiver with flavour node(s). Conversely, when we come across a quiver with flavours, we can always absorb them into a $\mathrm{U}(1)$ node which is connected to all nodes with edge multiplicity equal to the number of flavours they are connected to. In order to see all possible elementary slices for a quiver, we have to turn it into a flavourless quiver. Note that all of the above holds only for simply laced unitary quivers.

In the brane constructions used in this paper, we only come across elementary slices which are closures of minimal nilpotent orbits of ADE type or Kleinian singularities of type A. D-type Kleinian singularities appear as 3d $\mathcal{N}=4$ Coulomb branches of orthosymplectic quivers and are not considered here. No Coulomb branch construction of an E-type singularity is known. Closures of minimal nilpotent orbits of groups with non-simply laced Dynkin diagrams have Coulomb branch representations and can be used for quiver subtraction. However, the process of ungauging is non-trivial, see \cite{Anton}, and a detailed study remains to be done.

\subsection{Minimal Transitions}
\label{minimaltransitions}

Inspired by the analysis of minimal transitions in the study of nilpotent orbit closures \cite{Kraft1982}, an underlying assumption of the present work is that the minimal transitions are either minimal nilpotent orbit closures or Kleinian singularities. The relevant magnetic quivers for minimal nilpotent orbit closures are provided in Table \ref{tab:minimalNOtransitions}. The magnetic quiver for the $A_{N-1}$ singularity is 
\begin{equation}
    \displaystyle\overset 1 \circ\xlongequal{N}\overset 1 \circ \,.
\end{equation}

\begin{table}
	\centering
	\begin{tabular}{|c|c|c|}
	\hline
	Minimal NO transition & Quiver & Dimension \\ \hline
	$a_n$ & $\underbrace{\overset{\overset{\displaystyle\overset 1 \circ}
	{\diagup~~~~~\diagdown}
	}{\node{}{1}-\dots-\node{}{1}}}_{n}$ & $n$ \\ \hline
	$b_n$ & $\underbrace{\node{}1-\node{\topnode{}1}2-{\node{}{2}-\dots-\node{}{2}}-\node{}2 => \node{}1
	}_{n}$ & $2n-2$ \\ \hline
	$c_n$  & $\underbrace{\node{}1=> \node{}1-{\node{}{1}-\dots-\node{}{1}}-\node{}1 <= \node{}1
	}_{n+1}$ & $n$\\ \hline
	$d_n$  & $\underbrace{\node{}1-\node{\topnode{}1}2-
	\node{}{2}-\dots-\node{}{2}-\node{\topnode{}1}2-\node{}1
	}_{n-1}$ & $2n-3$\\ \hline
	$e_6$  & $\node{}1-\node{}2-\node{\topnode{\topnode{}1}2}3-
	\node{}{2}-\node{}{1} $ & $11$\\ \hline
		$e_7$  & $\node{}1-\node{}2-\node{}3-\node{\topnode{}2}4-
	\node{}3-\node{}2-\node{}1 $ & $17$\\ \hline
		$e_8$  & $\node{}1-\node{}2-\node{}3-\node{}4-\node{}5-\node{\topnode{}3}6-\node{}4-\node{}2 $ & $29$\\ \hline
	\end{tabular}
	\caption{Minimal nilpotent orbit transitions, the corresponding quivers and dimensions.}\label{tab:minimalNOtransitions}
\end{table}

\subsection{Addition of an Elementary Slice Quiver}
\label{quiveradd}

In Figure \ref{fig:two_cones} one obtains the magnetic quiver representing the intersection by subtracting an $a_1$ from either the magnetic quiver for the mesonic or the baryonic branch. Clearly both quivers are different. Hence, we see that a notion of \emph{quiver addition} is not unique. Adding an elementary slice quiver $\mathsf{D}$ to a quiver $\mathsf{Q}$ may be defined as choosing a quiver $\widetilde{\mathsf{Q}}$ such that
\begin{equation}
    \widetilde{\mathsf{Q}} - \mathsf{D}=\mathsf{Q}\,.
\end{equation}
For a specific $\mathsf{Q}$ and $\mathsf{D}$ a finite set of such quivers $\widetilde{\mathsf{Q}}$ may be found. If there is a $\mathrm{U}(1)$ node present in the quiver, one can always add a elementary slice quiver by replacing the $\mathrm{U}(1)$ node with that quiver. However, it may not always be possible to add an elementary slice quiver to a given quiver.

\section{Symplectic Leaves}
\label{SymplecticLeaves}

In this subsection, we provide a brief guide to the mathematics of Hasse diagrams employed in the main text, referring the reader to references for the details. 

In classical mechanics, a central tool is the Poisson bracket $\{ f , g \}$ for $f$ and $g$ two functions on the phase space. This bracket satisfies 
\begin{enumerate}
    \item Skew-symmetry $\{f,g\} + \{g,f\}= 0$
    \item The Jacobi identity $\{f,\{g,h\}\} + \{g,\{h,f\}\} + \{h,\{f,g\}\} = 0$
    \item The Leibniz rule $\{h,f g\} = \{h,f\} g + f \{h,g\} $
\end{enumerate}
More generally, a Poisson variety is any variety whose coordinate ring possesses a bracket satisfying the above three properties. Note that the Poisson bracket can be seen as a two-vector defined on the variety, called the \emph{Poisson bivector}. 

Dualizing the Poisson bivector, one obtains a two-form, and vice versa. Recalling that a symplectic variety is a variety with a non-degenerate two-form, we immediately see that any symplectic variety is a Poisson variety. However, the converse is not true, because the Poisson bivector can be degenerate. For reviews of Poisson geometry, we refer to \cite{vanhaecke2001integrable,vaisman2012lectures,fernandes2014lectures}. In this sense, Poisson geometry generalises symplectic geometry, by ``allowing" the symplectic form to be degenerate.

A Poisson structure on a smooth manifold gives rise to a foliation by symplectic leaves, and these leaves can have jumps in dimensionality. 
This extends to symplectic singularities \cite{beauville2000symplectic,namikawa2000extension}, see also \cite{fu2006survey,bellamy} for a review. Given an affine normal variety $X$ over $\mathbb{C}$ of even complex dimension with a non-degenerate closed two-form, we say that $X$ is a symplectic singularity if the two-form extends to a two-form on a resolution of $X$. We stress that this two form does not have to be non-degenerate in general. 

According to \cite[Thm.\ 2.3]{kaledin2006symplectic}, every (normal) symplectic singularity admits a finite stratification $\{0 \} =X_0 \subset X_1 \subset \cdots \subset X_n = X$ such that 
\begin{compactenum}[(i)]
    \item the singular part of $X_i$ is $X_{i-1}$, and
    \item the normalization of any irreducible component of $X_i$ is a symplectic singularity. 
\end{compactenum}
In general, there exists more than one such stratification. The set of all the spaces involved in these stratifications are nevertheless partially ordered by the operation of taking the singular part. This partial order is represented by a Hasse diagram. 

Questions about the properties of the symplectic leaves of a $3$d $\mathcal{N}=4$ Coulomb branch have already been raised in \cite[Sec.\ 2]{Nakajima:2015gxa}.

\section{Results for 5d SQCD Theories}
\label{full5d}
In this appendix, we provide the Hasse diagrams for each of the components of the Higgs branch $\mathcal{H}_\infty$ at infinite coupling for the theories discussed in Section \ref{subsection5d}. The results are shown in Tables \ref{tab:region1}--\ref{tab:region4kHalf} (see Table \ref{tab:5dinf} for an overview), which closely follows the tables in \cite{Cabrera:2018jxt}. Note that the intersections are not represented, as they can be read directly from the Hasse diagrams shown in Section \ref{subsection5d}. 

In each case, one can check that the non-Abelian part of the global symmetry, as read using the general principle of Section \ref{subsectionGlobalSymmetry}, agrees with what is shown in \cite[Tab.\ 1]{Cabrera:2018jxt}.

\begin{table}
	\makebox[\textwidth][c]{ 
\begin{tabular}{| m{1cm} |m{6.8cm}| m{3cm} |}
		\hline
		Phase & Quiver & Hasse diagram  \\ \hline
		I &  $\underbrace{\node{}1-\dots -\overset{\overset{\displaystyle\overset 1 \circ\xlongequal{N_c - \frac{N_f}{2}+|k|}\overset 1 \circ}{\diagup~~~~~~~~~~~~~~~~~~\diagdown}}{\node{}{\frac{N_f}{2}-|k|}-~~~~\dots~~~~-\node{}{\frac{N_f}{2}-|k|}}-\dots-\node{}1}_{N_f - 1}$ &
		\begin{tikzpicture}
		\tikzstyle{hasse} = [circle, fill,inner sep=2pt];
		\node [hasse] (1) [label=right:\footnotesize{$ $}] {};
		\node [hasse] (2) [below of=1] {};
		\node [hasse] (3) [below of=2] {};
		\node [hasse] (4) [below of=3] {};
		\node [hasse] (5) [below of=4] {};
		\node [hasse] (6) [below of=5] {};
		 \node at (0,-3.5)[]{$\vdots$};
		\draw (1) edge [] node[label=left:\footnotesize{$A_{N_c-\frac{N_f}{2} +|k|-1}$}] {} (2)
		(2) edge [] node[label=left:\footnotesize{$a_{2|k|+1}$}] {} (3)
		(3) edge [] node[label=left:\footnotesize{$a_{2|k|+3}$}] {} (4)
		(5) edge [] node[label=left:\footnotesize{$a_{N_f-1}$}] {} (6);
	         \end{tikzpicture}
		\\ \hline
		II &  $~\underbrace{\node{}1-\dots -\overset{\overset{\displaystyle\overset 1 \circ\xlongequal{2N_c - N_f}\overset 1 \circ}{\diagup~~~~~~~~~~~~~~~\diagdown}}{\node{}{N_f-N_c}-~~\dots~~-\node{}{N_f-N_c}}-\dots-\node{}1}_{N_f - 1}$ &
		\begin{tikzpicture}
		\tikzstyle{hasse} = [circle, fill,inner sep=2pt];
		\node [hasse] (1) [label=right:\footnotesize{$ $}] {};
		\node [hasse] (2) [below of=1] {};
		\node [hasse] (3) [below of=2] {};
		\node [hasse] (4) [below of=3] {};
		\node [hasse] (5) [below of=4] {};
		\node [hasse] (6) [below of=5] {};
		 \node at (0,-3.5)[]{$\vdots$};
		\draw (1) edge [] node[label=left:\footnotesize{$\;\;\;\;A_{2N_c-N_f-1}$}] {} (2)
		(2) edge [] node[label=left:\footnotesize{$a_{2N_c-N_f+1}$}] {} (3)
		(3) edge [] node[label=left:\footnotesize{$a_{2N_c-N_f+3}$}] {} (4)
		(5) edge [] node[label=left:\footnotesize{$a_{N_f-1}$}] {} (6);
	         \end{tikzpicture}
		  \\ \hline
	\end{tabular}
	}
	\caption{Components of $\mathcal{H}_\infty$ for  $\frac{1}{2}<|k|<N_c - \frac{N_f}{2}$, part 1.  
	Component I is present for $\frac {N_f}{2} \geq |k|$, Component II is present for $N_f \geq N_c$. The three dots denote a chain of balanced gauge nodes.}
	\label{tab:region1}
\end{table}

\begin{table}
	\makebox[\textwidth][c]{ 
\begin{tabular}{| m{3cm} |m{5.8cm}| m{2.8cm} |}
		\hline
		Phase & Quiver & Hasse diagram  \\ \hline
		III ($N_f$ even) &  $~~~~~\underbrace{\node{}1-\node{}2-\dots-\node{\overset{\displaystyle \overset{1}{\circ}}{\parallel}}{\frac{N_f }2}-\dots-\node{}2-\node{}1}_{N_f - 1}$ &
		\begin{tikzpicture}
		\tikzstyle{hasse} = [circle, fill,inner sep=2pt];
		\node [hasse] (1) [label=right:\footnotesize{$ $}] {};
		\node [hasse] (2) [below of=1] {};
		\node [hasse] (3) [below of=2] {};
		\node [hasse] (4) [below of=3] {};
		\node [hasse] (5) [below of=4] {};
		 \node at (0,-2.5)[]{$\vdots$};
		\draw (1) edge [] node[label=left:\footnotesize{$a_{1}$}] {} (2)
		(2) edge [] node[label=left:\footnotesize{$a_{3}$}] {} (3)
		(4) edge [] node[label=left:\footnotesize{$a_{N_f-1}$}] {} (5);
	         \end{tikzpicture}
		  \\ \hline
		III ($N_f$ odd) & $\underbrace{\node{}1-\node{}2-\dots -\overset{\overset{\displaystyle\overset 1 \circ}{\diagup~~\diagdown}}{\node{}{\frac{N_f-1}2}-\node{}{\frac{N_f-1}2}}-\dots-\node{}2-\node{}1}_{N_f - 1}$ &
		\begin{tikzpicture}
		\tikzstyle{hasse} = [circle, fill,inner sep=2pt];
		\node [hasse] (1) [label=right:\footnotesize{$ $}] {};
		\node [hasse] (2) [below of=1] {};
		\node [hasse] (3) [below of=2] {};
		\node [hasse] (4) [below of=3] {};
		\node [hasse] (5) [below of=4] {};
		 \node at (0,-2.5)[]{$\vdots$};
		\draw (1) edge [] node[label=left:\footnotesize{$a_{2}$}] {} (2)
		(2) edge [] node[label=left:\footnotesize{$a_{4}$}] {} (3)
		(4) edge [] node[label=left:\footnotesize{$a_{N_f-1}$}] {} (5);
	         \end{tikzpicture}
		 \\ \hline
	\end{tabular}
	}
	\caption{Components of $\mathcal{H}_\infty$ for  $\frac{1}{2}<|k|<N_c - \frac{N_f}{2}$, part 2. Component III is present for $N_f \ge 2$. The three dots denote a chain of balanced gauge nodes.}
	\label{tab:region1seq}
\end{table}

\begin{table}
	\centering
\begin{tabular}{| m{1cm} |m{6.7cm}| m{3cm} |}
		\hline
		Phase & Quiver & Hasse diagram
		\\ \hline
		I & $~~~~~~\underbrace{\node{}1-\dots -\overset{\overset{\displaystyle\overset 1 \circ\xlongequal{N_c - \frac{N_f}{2}+\frac 1 2}\overset 1 \circ}{\diagdown~~~~~~~~~~~\diagup}}{\node{}{\frac{N_f-1}{2}}-\node{}{\frac{N_f-1}{2}}}-\dots-\node{}1}_{N_f - 1}$  &
		\begin{tikzpicture}
		\tikzstyle{hasse} = [circle, fill,inner sep=2pt];
		\node [hasse] (1) [label=right:\footnotesize{$ $}] {};
		\node [hasse] (2) [below of=1] {};
		\node [hasse] (3) [below of=2] {};
		\node [hasse] (4) [below of=3] {};
		\node [hasse] (5) [below of=4] {};
		\node [hasse] (6) [below of=5] {};
		 \node at (0,-3.5)[]{$\vdots$};
		\draw (1) edge [] node[label=left:\footnotesize{$\;A_{N_c-\frac{N_f+1}{2}}$}] {} (2)
		(2) edge [] node[label=left:\footnotesize{$a_{2}$}] {} (3)
		(3) edge [] node[label=left:\footnotesize{$a_{4}$}] {} (4)
		(5) edge [] node[label=left:\footnotesize{$a_{N_f-1}$}] {} (6);
	         \end{tikzpicture}
		  \\ \hline
		II & $~\underbrace{\node{}1-\dots -\overset{\overset{\displaystyle\overset 1 \circ\xlongequal{2N_c - N_f}\overset 1 \circ}{\diagup~~~~~~~~~~~~~~~\diagdown}}{\node{}{N_f-N_c}-~~\dots~~-\node{}{N_f-N_c}}-\dots-\node{}1}_{N_f - 1}$  &
		\begin{tikzpicture}
		\tikzstyle{hasse} = [circle, fill,inner sep=2pt];
		\node [hasse] (1) [label=right:\footnotesize{$ $}] {};
		\node [hasse] (2) [below of=1] {};
		\node [hasse] (3) [below of=2] {};
		\node [hasse] (4) [below of=3] {};
		\node [hasse] (5) [below of=4] {};
		\node [hasse] (6) [below of=5] {};
		 \node at (0,-3.5)[]{$\vdots$};
		\draw (1) edge [] node[label=left:\footnotesize{$A_{2N_c-N_f-1}$}] {} (2)
		(2) edge [] node[label=left:\footnotesize{$a_{2N_c-N_f+1}$}] {} (3)
		(3) edge [] node[label=left:\footnotesize{$a_{2N_c-N_f+3}$}] {} (4)
		(5) edge [] node[label=left:\footnotesize{$a_{N_f-1}$}] {} (6);
	         \end{tikzpicture}
		  \\ \hline
	\end{tabular}
	\caption{Components of $\mathcal{H}_\infty$ for  $\frac 1 2 = |k|<N_c - \frac{N_f}{2}$.
	Component I appears for $N_f \geq 1$, Component II appears for $N_f \geq N_c$.
	}
	\label{tab:region1kHalf}
\end{table}

\begin{table}
	\centering
\begin{tabular}{| m{1cm} |m{6.7cm}| m{3cm} |m{3cm} |}
		\hline
		Phase & Quiver & Hasse diagram, $N_c \neq \frac{N_f+2}{2}$ & Hasse diagram, $N_c = \frac{N_f+2}{2}$  \\ \hline
		I &  $~~~\underbrace{\node{}1-\node{}2-\dots -\node{\overset{\displaystyle\overset{1}{\circ}\xlongequal{N_c - \frac{N_f}{2}}\overset{1}{\circ}}{\diagdown~~~~~~\diagup}}{\frac{N_f}{2}}-\dots-\node{}2-\node{}1}_{N_f - 1}$ &
		\begin{tikzpicture}
		\tikzstyle{hasse} = [circle, fill,inner sep=2pt];
		\node [hasse] (1) [label=right:\footnotesize{$ $}] {};
		\node [hasse] (2) [below of=1] {};
		\node [hasse] (3) [below of=2] {};
		\node [hasse] (4) [below of=3] {};
		\node [hasse] (5) [below of=4] {};
		\node [hasse] (6) [below of=5] {};
		\node [hasse] (7) [below of=6] {};
		 \node at (0,-4.5)[]{$\vdots$};
		\draw (1) edge [] node[label=left:\footnotesize{$A_{N_c-\frac{N_f}{2}-1}$}] {} (2)
		(2) edge [] node[label=left:\footnotesize{$a_{1}$}] {} (3)
		(3) edge [] node[label=left:\footnotesize{$a_{3}$}] {} (4)
		(4) edge [] node[label=left:\footnotesize{$a_{5}$}] {} (5)
		(6) edge [] node[label=left:\footnotesize{$a_{N_f-1}$}] {} (7);
	         \end{tikzpicture}
	         &
		\begin{tikzpicture}
		\tikzstyle{hasse} = [circle, fill,inner sep=2pt];
		\node [hasse] (1) [label=right:\footnotesize{$ $}] {};
		\node (2a) [below of=1] {};
		\node [hasse] (2) [below of=2a] {};
		\node [hasse] (3) [below of=2] {};
		\node [hasse] (4) [below of=3] {};
		\node [hasse] (5) [below of=4] {};
		\node [hasse] (6) [below of=5] {};
		 \node at (0,-4.5)[]{$\vdots$};
		\draw (1) edge [] node[label=left:\footnotesize{$a_2$}] {} (2)
		(2) edge [] node[label=left:\footnotesize{$a_{3}$}] {} (3)
		(3) edge [] node[label=left:\footnotesize{$a_{5}$}] {} (4)
		(5) edge [] node[label=left:\footnotesize{$\;\;\;a_{N_f-1}$}] {} (6);
	         \end{tikzpicture}
		 \\
\hline
	\end{tabular}
	\caption{Components of $\mathcal{H}_\infty$ for  $0=|k|<N_c - \frac{N_f}{2}$. 
	Component II is present for $N_f \geq N_c$
	}
	\label{tab:region1k0}
\end{table}

\begin{table}
	\centering
\begin{tabular}{| m{2.3cm} |m{7.1cm}| m{3cm} |}
		\hline
		Phase & Quiver & Hasse diagram \\ \hline
		I & $\underbrace{\node{}1-\dots -\overset{\overset{\overset{\overset{\displaystyle \overset 1 \circ}{\diagup~~~\diagdown}}{\displaystyle\overset 1 \circ\xlongequal{2|k|-1}\overset 1 \circ}}{\diagup~~~~~~~~~~~~~~~~~~~\diagdown}}{\node{}{\frac{N_f}2-|k|}-~~~~~\dots~~~~~-\node{}{\frac{N_f}2-|k|}}-\dots-\node{}1}_{N_f - 1}$ &
      \begin{tikzpicture}[node distance=20pt]
	\tikzstyle{hasse} = [circle, fill,inner sep=2pt];
		\node at (-0.7,-0.5) [hasse] (1a) [] {};
		\node at (0,-0.5) [] (1b) [label=right:\footnotesize{$ $}] {};
		\node at (0.7,-0.5) [] (1c) [] {};
		\node  (2a) [below of=1a] {};
		\node (2b) [below of=1b] {};
		\node (2c) [below of=1c] {};
		\node [hasse] (3a) [below of=2a] {};
		\node (3b) [below of=2b] {};
		\node (3c) [below of=2c] {};
		\node (4a) [below of=3a] {};
		\node [hasse] (4b)[below of=3b] {};
		\node (4c) [below of=3c] {};
		\node [hasse] (5a) [below of=4a] {};
		\node (5b) [below of=4b] {};
		\node (5c) [below of=4c] {};
		\node (6a) [below of=5a] {};
		\node [hasse] (6b) [below of=5b] {};
		\node (6c) [below of=5c] {};
		\node [hasse] (7a) [below of=6a] {};
		\node [hasse] (7b) [below of=6b] {};
		\node (7c) [below of=6c] {}; 
		\node (8a) [below of=7a] {};
		\node [hasse] (8b) [below of=7b] {};
	        \node (8c) [below of=7c] {};
	          \node at (-0.4,-4.1)[]{$\vdots$};
		\draw (1a) edge [] node[label=left:\footnotesize{$A_{2|k|-2}$}] {} (3a)
			(3a) edge [] node[label=right:\footnotesize{$a_1$}] {} (4b)
			(4b) edge [] node[label=right:\footnotesize{$a_{2|k|+1}$}] {} (6b)
			(5a) edge [] node[label=above:\footnotesize{$a_{1}$}] {} (6b)
			(3a) edge [] node[label=left:\footnotesize{$a_{2|k|+1}$}] {} (5a)
			(7a) edge [] node[label=below:\footnotesize{$a_{1}$}] {} (8b)
			(7b) edge [] node[label=right:\footnotesize{$a_{N_f-1}$}] {} (8b);
	\end{tikzpicture}
		  \\
\hline
	\end{tabular}
	\caption{Components of $\mathcal{H}_\infty$ for $1<|k|=N_c - \frac{N_f}{2}$.
	Component I is present for $N_f \geq N_c$, which means $\frac{N_f}{2} \ge |k|$, and Component III appears for $N_f \ge 1$. 
	}
	\label{tab:region2}
\end{table}

\begin{table}
	\centering
\begin{tabular}{| m{2.3cm} |m{6.8cm}| m{2.8cm} |}
		\hline
		Phase & Quiver & Hasse diagram \\
\hline
	\end{tabular}
	\caption{Components of $\mathcal{H}_\infty$ for  $1=|k|=N_c - \frac{N_f}{2}$.}
	\label{tab:region2k1}
\end{table}

\begin{table}
	\centering
\begin{tabular}{| m{1cm} |m{6.6cm}| m{2.7cm} |}
		\hline
		Phase & Quiver & Hasse diagram \\
\hline
	\end{tabular}
	\caption{Component of $\mathcal{H}_\infty$ for  $\frac 1 2=|k|=N_c - \frac{N_f}{2}$.}
	\label{tab:region2kHalf}
\end{table}

\begin{table}
	\centering
\begin{tabular}{| m{1cm} |m{5.7cm}| m{5cm} |}
		\hline
		Phase & Quiver & Hasse diagram \\
\hline
	\end{tabular}
	\caption{The component of $\mathcal{H}_\infty$ for  $0=|k|=N_c - \frac{N_f}{2}$. Note that for $N_f=2$ there is no $e_6$ nor $e_7$ elementary slice, as expected for the $\mathrm{SU}(2)$ theory with $4$ flavours. For $N_f=3$ there is no $e_7$ elementary slice, and we recover the Hasse diagram of Figure \ref{fig:Hasse_SU3_infinite}. Also note that since the quiver has an $\mathbb{Z}_2$ automorphism symmetry, there is branching into two $e_6$ transitions. As a consequence, the non-Abelian part of the global symmetry is $A_{N_f-1} \times A_1 \times A_1$. }
	\label{tab:region2k0}
\end{table}

\clearpage

\begin{table}
	\centering
\begin{tabular}{| m{2.3cm} |m{7.7cm}| m{2.7cm} |}
		\hline
		Phase & Quiver & Hasse diagram \\ \hline
		I & $\underbrace{\node{}1-\dots -\overset{\overset{\displaystyle\overset 1 \circ\xlongequal{2|k|-2}\overset 1 \circ}{\diagup~~~~~~~~~~~~~~~~~~\diagdown}}{\node{}{\frac{N_f}{2}-|k|+1}-~~~~\dots~~~~-\node{}{\frac{N_f}{2}-|k|+1}}-\dots-\node{}1}_{N_f}$ & 
		\begin{tikzpicture}
		\tikzstyle{hasse} = [circle, fill,inner sep=2pt];
		\node [hasse] (1) [label=right:\footnotesize{$ $}] {};
		\node [hasse] (2) [below of=1] {};
		\node [hasse] (3) [below of=2] {};
		\node [hasse] (4) [below of=3] {};
		\node [hasse] (5) [below of=4] {};
		\node [hasse] (6) [below of=5] {};
		 \node at (0,-3.5)[]{$\vdots$};
		\draw (1) edge [] node[label=left:\footnotesize{$A_{2|k|-3}$}] {} (2)
		(2) edge [] node[label=left:\footnotesize{$a_{2|k|}$}] {} (3)
		(3) edge [] node[label=left:\footnotesize{$a_{2|k|+2}$}] {} (4)
		(5) edge [] node[label=left:\footnotesize{$a_{N_f}$}] {} (6);
	         \end{tikzpicture}
		\\ \hline
		III ($N_f$ even) & $~~~~~~~~\underbrace{\node{}1-\node{}2-\dots -\overset{\overset{\displaystyle\overset 1 \circ}{\diagup~~\diagdown}}{\node{}{\frac{N_f}2}-\node{}{\frac{N_f}2}}-\dots-\node{}2-\node{}1}_{N_f}$ &
		\begin{tikzpicture}
		\tikzstyle{hasse} = [circle, fill,inner sep=2pt];
		\node [hasse] (1) [label=right:\footnotesize{$ $}] {};
		\node [hasse] (2) [below of=1] {};
		\node [hasse] (3) [below of=2] {};
		\node [hasse] (4) [below of=3] {};
		\node [hasse] (5) [below of=4] {};
		\node [hasse] (6) [below of=5] {};
		 \node at (0,-3.5)[]{$\vdots$};
		\draw (1) edge [] node[label=left:\footnotesize{$a_{2}$}] {} (2)
		(2) edge [] node[label=left:\footnotesize{$a_{4}$}] {} (3)
		(3) edge [] node[label=left:\footnotesize{$a_{6}$}] {} (4)
		(5) edge [] node[label=left:\footnotesize{$\;\;\;\;\;a_{N_f}$}] {} (6);
	         \end{tikzpicture}
			 \\ \hline
		III ($N_f$ odd) & $~~~~~~~~~~\underbrace{\node{}1-\node{}2-\dots-\node{\overset{\displaystyle \overset{1}{\circ}}{\parallel}}{\frac{N_f+1 }2}-\dots-\node{}2-\node{}1}_{N_f}$  &
		\begin{tikzpicture}
		\tikzstyle{hasse} = [circle, fill,inner sep=2pt];
		\node [hasse] (1) [label=right:\footnotesize{$ $}] {};
		\node [hasse] (2) [below of=1] {};
		\node [hasse] (3) [below of=2] {};
		\node [hasse] (4) [below of=3] {};
		\node [hasse] (5) [below of=4] {};
		\node [hasse] (6) [below of=5] {};
		 \node at (0,-3.5)[]{$\vdots$};
		\draw (1) edge [] node[label=left:\footnotesize{$a_{1}$}] {} (2)
		(2) edge [] node[label=left:\footnotesize{$a_{3}$}] {} (3)
		(3) edge [] node[label=left:\footnotesize{$a_{5}$}] {} (4)
		(5) edge [] node[label=left:\footnotesize{$\;\;\;\;\;a_{N_f}$}] {} (6);
	         \end{tikzpicture}
		  \\ \hline
	\end{tabular}
	\caption{Components of $\mathcal{H}_\infty$ for  $\frac 3 2<|k|=N_c - \frac{N_f}{2}+1$. Component I is present for $N_f \ge N_c$, which means $\frac{N_f}{2} \ge |k|-1$, and Component III is present for $N_f \ge 1$}
	\label{tab:region3}
\end{table}
\clearpage

\begin{table}
	\centering
\begin{tabular}{| m{1cm} |m{6.8cm}| m{2.6cm} |}
		\hline
		Phase & Quiver & Hasse diagram \\ \hline
		I & $\underbrace{\node{}1-\node{}2-\dots -\overset{\overset{\displaystyle\overset 1 \circ~ -~ \overset 1 \circ}{\diagup~~~~~~~~~\diagdown}}{\node{}{\frac{N_f-1}{2}}-\node{}{\frac{N_f-1}{2}}-\node{}{\frac{N_f-1}{2}}}-\dots-\node{}2-\node{}1}_{N_f}$ & 
		\begin{tikzpicture}
		\tikzstyle{hasse} = [circle, fill,inner sep=2pt];
		\node [hasse] (1) [label=right:\footnotesize{$ $}] {};
		\node [hasse] (2) [below of=1] {};
		\node [hasse] (3) [below of=2] {};
		\node [hasse] (4) [below of=3] {};
		\node [hasse] (5) [below of=4] {};
		\node [hasse] (6) [below of=5] {};
		 \node at (0,-3.5)[]{$\vdots$};
		\draw (1) edge [] node[label=left:\footnotesize{$a_{4}$}] {} (2)
		(2) edge [] node[label=left:\footnotesize{$a_{5}$}] {} (3)
		(3) edge [] node[label=left:\footnotesize{$a_{7}$}] {} (4)
		(5) edge [] node[label=left:\footnotesize{$\;\;\;\;a_{N_f}$}] {} (6);
	         \end{tikzpicture}
		\\ \hline
	\end{tabular}
	\caption{The component of $\mathcal{H}_\infty$ for  $\frac 3 2 = |k|=N_c - \frac{N_f}{2}+1$.}
	\label{tab:region3k3Halves}
\end{table}

\begin{table}
	\centering
\begin{tabular}{| m{1cm} |m{5cm}| m{2.6cm} |}
		\hline
		Phase & Quiver & Hasse diagram \\ \hline
		I & $~~~\underbrace{\node{}1-\dots -\node{\upnode 1}{\frac{N_f}{2}}-\node{\upnode 1}{\frac{N_f}{2}}-\dots-\node{}1}_{N_f}$ & 
		\begin{tikzpicture}
		\tikzstyle{hasse} = [circle, fill,inner sep=2pt];
		\node [hasse] (1) [label=right:\footnotesize{$ $}] {};
		\node [hasse] (2) [below of=1] {};
		\node [hasse] (3) [below of=2] {};
		\node [hasse] (4) [below of=3] {};
		\node [hasse] (5) [below of=4] {};
		\node [hasse] (6) [below of=5] {};
		 \node at (0,-3.5)[]{$\vdots$};
		\draw (1) edge [] node[label=left:\footnotesize{$d_{5}$}] {} (2)
		(2) edge [] node[label=left:\footnotesize{$a_{6}$}] {} (3)
		(3) edge [] node[label=left:\footnotesize{$a_{8}$}] {} (4)
		(5) edge [] node[label=left:\footnotesize{$\;\;\;a_{N_f}$}] {} (6);
	         \end{tikzpicture}
		 \\ \hline
	\end{tabular}
	\caption{The component of $\mathcal{H}_\infty$ for  $1=|k|=N_c - \frac{N_f}{2}+1$.}
	\label{tab:region3k1}
\end{table}

\begin{table}
	\centering
\begin{tabular}{| m{1cm} |m{6.1cm}| m{3.1cm} |}
		\hline
		Phase & Quiver & Hasse diagram \\ \hline
		I$'$ & $~~~~\underbrace{\node{}1-\node{}2-\dots -\node{\overset{\displaystyle \overset{\overset{\displaystyle {\circ}\rlap{\,\,$\scriptstyle 1$}}{\vert}}{\circ}\rlap{\,\,$\scriptstyle 2$}}{\vert}}{\frac{N_f+1 }2}-\dots-\node{}2-\node{}1}_{N_f }$ &
\begin{tikzpicture}[node distance=30pt]
	\tikzstyle{hasse} = [circle, fill,inner sep=2pt];
		\node at (-0.7,-0.5) [] (1a) [] {};
		\node at (0,-0.5) [hasse] (1b) [label=right:\footnotesize{$dim$}] {};
		\node at (0.7,-0.5) [] (1c) [] {};
		\node [hasse] (2a) [below of=1a] {};
		\node (2b) [below of=1b] {};
		\node (2c) [below of=1c] {};
		\node  (3a) [below of=2a] {};
		\node (3b) [below of=2b] {};
		\node [hasse] (3c) [below of=2c] {};
		\node [hasse] (4a) [below of=3a] {};
		\node (4b)[below of=3b] {};
		\node (4c) [below of=3c] {};
		\node (5a) [below of=4a] {};
		\node (5b) [below of=4b] {};
		\node [hasse] (5c) [below of=4c] {};
		\node [hasse] (6a) [below of=5a] {};
		\node (6b) [below of=5b] {};
		\node (6c) [below of=5c] {};
		\node (7a) [below of=6a] {};
		\node (7b) [below of=6b] {};
		\node [hasse] (7c) [below of=6c] {}; 
		\node [hasse] (8a) [below of=7a] {};
		\node (8b) [below of=7b] {};
	        \node (8c) [below of=7c] {};
	          \node at (0,-6.2)[]{$\vdots$};
		\draw (1b) edge [] node[label=left:$e_6$] {} (2a)
			(1b) edge [] node[label=right:$e_7$] {} (3c)
			(2a) edge [] node[label=left:$a_7$] {} (4a)
			(4a) edge [] node[label=left:$a_9$] {} (6a)
			(3c) edge [] node[label=right:$a_9$] {} (5c)
			(3c) edge [] node[label=above:$a_1$] {} (4a)
			(5c) edge [] node[label=above:$a_1$] {} (6a)
			(5c) edge [] node[label=right:$ $] {} (6c)
			(7a) edge [] node[label=left:$a_{N_f}$] {} (8a)
			(7c) edge [] node[label=above:$a_1$] {} (8a);
	\end{tikzpicture}
		   \\ \hline
	\end{tabular}
	\caption{The component of $\mathcal{H}_\infty$ for  $\frac 1 2 = |k|=N_c - \frac{N_f}{2}+1$.}
	\label{tab:region3kHalf}
\end{table}

\begin{table}
	\centering
\begin{tabular}{| m{1cm} |m{5.9cm}| m{2.6cm} |}
		\hline
		Phase & Quiver & Hasse diagram \\ \hline
		I$'$ & $~~~~\underbrace{\node{}{1}-\node{}{2}-\dots -\node{\upnode{2}}{\frac{N_f+2}{2}} -\dots -\node{}{2}-\node{}{1}}_{N_f + 1}$ & 
		\begin{tikzpicture}
		\tikzstyle{hasse} = [circle, fill,inner sep=2pt];
		\node [hasse] (1) [label=right:\footnotesize{$ $}] {};
		\node [hasse] (2) [below of=1] {};
		\node [hasse] (3) [below of=2] {};
		\node [hasse] (4) [below of=3] {};
		\node [hasse] (5) [below of=4] {};
		\node [hasse] (6) [below of=5] {};
		 \node at (0,-3.5)[]{$\vdots$};
		\draw (1) edge [] node[label=left:\footnotesize{$e_{7}$}] {} (2)
		(2) edge [] node[label=left:\footnotesize{$a_{9}$}] {} (3)
		(3) edge [] node[label=left:\footnotesize{$a_{11}$}] {} (4)
		(5) edge [] node[label=left:\footnotesize{$\;a_{N_f+1}$}] {} (6);
	         \end{tikzpicture}
		 \\ \hline
	\end{tabular}
	\caption{The component of $\mathcal{H}_\infty$ for  $0=|k|=N_c - \frac{N_f}{2}+1$.}
	\label{tab:region3k0}
\end{table}

\begin{table}
	\centering
\begin{tabular}{| m{2.3cm} |m{6.6cm}| m{2.7cm} |}
		\hline
		Phase & Quiver & Hasse diagram \\ \hline
		IV ($N_f$ even) & 
		 $~~~~~\node{}{1}-\node{}2-\dots-\node{}{N_f-3}-\node{\upnode{\frac{N_f-2}2}}{N_f-2}-\node{}{\frac{N_f}2}=\node{}{1}$ &
		\begin{tikzpicture}
		\tikzstyle{hasse} = [circle, fill,inner sep=2pt];
		\node [hasse] (1) [label=right:\footnotesize{$ $}] {};
		\node [hasse] (2) [below of=1] {};
		\node [hasse] (3) [below of=2] {};
		\node [hasse] (4) [below of=3] {};
		\node [hasse] (5) [below of=4] {};
		\node [hasse] (6) [below of=5] {};
		 \node at (0,-3.5)[]{$\vdots$};
		\draw (1) edge [] node[label=left:\footnotesize{$A_{1}$}] {} (2)
		(2) edge [] node[label=left:\footnotesize{$d_{4}$}] {} (3)
		(3) edge [] node[label=left:\footnotesize{$d_{6}$}] {} (4)
		(5) edge [] node[label=left:\footnotesize{$\;\;\;\;\;\;\;\;\;\;d_{N_f}$}] {} (6);
	         \end{tikzpicture}
		   \\ \hline
		V ($N_f$ even) & $~~\node{}{1}-\node{}2-\dots-\node{}{N_f - 3}-\node{\upnode{\frac{N_f-2}2}}{N_f - 2}-\overset{\overset{\displaystyle\overset 1 \circ}{\diagup~\Nwseline\rlap{\,\,$\scriptstyle N_c - \frac{N_f}2$}}}{\node{}{\frac{N_f}2}-\node{}{1}}~~~~~~~~~$ &
		\begin{tikzpicture}
		\tikzstyle{hasse} = [circle, fill,inner sep=2pt];
		\node [hasse] (1) [label=right:\footnotesize{$ $}] {};
		\node [hasse] (2) [below of=1] {};
		\node [hasse] (3) [below of=2] {};
		\node [hasse] (4) [below of=3] {};
		\node [hasse] (5) [below of=4] {};
		\node [hasse] (6) [below of=5] {};
		\node [hasse] (7) [below of=6] {};
		 \node at (0,-4.5)[]{$\vdots$};
		\draw (1) edge [] node[label=left:\footnotesize{$A_{N_c -\frac{N_f}{2}-1}$}] {} (2)
		(2) edge [] node[label=left:\footnotesize{$A_{1}$}] {} (3)
		(3) edge [] node[label=left:\footnotesize{$d_{4}$}] {} (4)
		(4) edge [] node[label=left:\footnotesize{$d_{6}$}] {} (5)
		(6) edge [] node[label=left:\footnotesize{$d_{N_f}$}] {} (7);
	         \end{tikzpicture}
		   \\ \hline
		V ($N_f$ odd) & $~~\node{}{1}-\node{}2-\dots-\node{}{N_f - 3}-\overset{\overset{\displaystyle\overset {\frac{N_f-1}2} \circ-~\overset {1} \circ}{~\diagup~~~~~~~~~\Nwseline\rlap{\,\,$\scriptstyle N_c - \frac{N_f-1}2$}}}{\node{}{N_f - 2}-\node{}{\frac{N_f - 1}2}-\node{}{1}} ~~~~~~~~~$  &
		\begin{tikzpicture}
		\tikzstyle{hasse} = [circle, fill,inner sep=2pt];
		\node [hasse] (1) [label=right:\footnotesize{$ $}] {};
		\node [hasse] (2) [below of=1] {};
		\node [hasse] (3) [below of=2] {};
		\node [hasse] (4) [below of=3] {};
		\node [hasse] (5) [below of=4] {};
		\node [hasse] (6) [below of=5] {};
		\node [hasse] (7) [below of=6] {};
		 \node at (0,-4.5)[]{$\vdots$};
		\draw (1) edge [] node[label=left:\footnotesize{$A_{N_c -\frac{N_f+1}{2}}$}] {} (2)
		(2) edge [] node[label=left:\footnotesize{$a_{3}$}] {} (3)
		(3) edge [] node[label=left:\footnotesize{$d_{5}$}] {} (4)
		(4) edge [] node[label=left:\footnotesize{$d_{7}$}] {} (5)
		(6) edge [] node[label=left:\footnotesize{$d_{N_f}$}] {} (7);
	         \end{tikzpicture}
		   \\ \hline
	\end{tabular}
	\caption{Components of $\mathcal{H}_\infty$ for  $2<|k|=N_c - \frac{N_f}{2}+2$.
	Component IV is present for $N_f \ge 2$ with $N_f$ even. Component V ($N_f$ even) is present for $N_f \ge 0$ if $k$ is even and is present for $N_f \ge 2$ if $k$ is odd. Component V ($N_f$ odd) is present for $N_f \ge 1$.}
	\label{tab:region4}
\end{table}
\clearpage

\begin{table}
	\centering
\begin{tabular}{| m{2.3cm} |m{6.1cm}| m{2.3cm} |}
		\hline
		Phase & Quiver & Hasse diagram \\ \hline
		V ($N_f$ even) &   $~~\node{}{1}-\node{}2-\dots-\node{}{N_f-3}-\node{\upnode{\frac{N_f-2}2}}{N_f-2}-\node{\upnode{1}}{\frac{N_f}2}-\node{}{1}$ &
		\begin{tikzpicture}
		\tikzstyle{hasse} = [circle, fill,inner sep=2pt];
		\node [hasse] (1) [label=right:\footnotesize{$ $}] {};
		\node [hasse] (2) [below of=1] {};
		\node [hasse] (3) [below of=2] {};
		\node [hasse] (4) [below of=3] {};
		\node [hasse] (5) [below of=4] {};
		\node [hasse] (6) [below of=5] {};
		\node [hasse] (7) [below of=6] {};
		 \node at (0,-4.5)[]{$\vdots$};
		\draw (1) edge [] node[label=left:\footnotesize{$d_5$}] {} (2)
		(2) edge [] node[label=left:\footnotesize{$d_{6}$}] {} (3)
		(3) edge [] node[label=left:\footnotesize{$d_{8}$}] {} (4)
		(4) edge [] node[label=left:\footnotesize{$d_{10}$}] {} (5)
		(6) edge [] node[label=left:\footnotesize{$\;\;\;d_{N_f}$}] {} (7);
	         \end{tikzpicture}
		\\ \hline
	\end{tabular}
	\caption{The component of $\mathcal{H}_\infty$ for  $2=|k|=N_c - \frac{N_f}{2}+2$.}
	\label{tab:region4k2}
\end{table}

\begin{table}
	\centering
\begin{tabular}{| m{2.2cm} |m{6.2cm}| m{2.6cm} |}
		\hline
		Phase & Quiver & Hasse diagram \\ \hline
		V ($N_f$ odd) &   $~~\node{}{1}-\node{}2-\dots-\node{}{N_f - 3}-\overset{\overset{\displaystyle\overset {\frac{N_f-1}2} \circ-~\overset {1} \circ}{~\diagup~~~~~~~~~}}{\node{}{N_f - 2}-\node{}{\frac{N_f - 1}2}-\node{}{1}} $ &
		\begin{tikzpicture}
		\tikzstyle{hasse} = [circle, fill,inner sep=2pt];
		\node [hasse] (1) [label=right:\footnotesize{$ $}] {};
		\node [hasse] (2) [below of=1] {};
		\node [hasse] (3) [below of=2] {};
		\node [hasse] (4) [below of=3] {};
		\node [hasse] (5) [below of=4] {};
		\node [hasse] (6) [below of=5] {};
		\node [hasse] (7) [below of=6] {};
		 \node at (0,-4.5)[]{$\vdots$};
		\draw (1) edge [] node[label=left:\footnotesize{$e_6$}] {} (2)
		(2) edge [] node[label=left:\footnotesize{$d_{7}$}] {} (3)
		(3) edge [] node[label=left:\footnotesize{$d_{9}$}] {} (4)
		(4) edge [] node[label=left:\footnotesize{$d_{11}$}] {} (5)
		(6) edge [] node[label=left:\footnotesize{$\;\;\;\;d_{N_f}$}] {} (7);
	         \end{tikzpicture}  
		\\ \hline
	\end{tabular}
	\caption{The component of $\mathcal{H}_\infty$ for  $\frac 3 2 = |k|=N_c - \frac{N_f}{2}+2$.}
	\label{tab:region4k3Halves}
\end{table}

\begin{table}
	\centering
\begin{tabular}{| m{2.3cm} |m{6.5cm}| m{2.7cm} |}
		\hline
		Phase & Quiver & Hasse diagram \\ \hline
		V$'$ ($N_f$ even) &   $~~\node{}{1}-\node{}2-\dots-\node{}{N_f-3}-\node{\upnode{\frac{N_f-2}2}}{N_f-2}-\node{}{\frac{N_f}2}-\node{}{2}-\node{}{1}$  &
\begin{tikzpicture}[node distance=20pt]
	\tikzstyle{hasse} = [circle, fill,inner sep=2pt];
		\node at (-0.7,-0.5) [hasse] (1a) [] {};
		\node at (0,-0.5) [] (1b) [label=right:\footnotesize{$ $}] {};
		\node at (0.7,-0.5) [] (1c) [] {};
		\node  (2a) [below of=1a] {};
		\node [hasse] (2b) [below of=1b] {};
		\node (2c) [below of=1c] {};
		\node [hasse] (3a) [below of=2a] {};
		\node (3b) [below of=2b] {};
		\node (3c) [below of=2c] {};
		\node (4a) [below of=3a] {};
		\node [hasse] (4b)[below of=3b] {};
		\node (4c) [below of=3c] {};
		\node [hasse] (5a) [below of=4a] {};
		\node (5b) [below of=4b] {};
		\node (5c) [below of=4c] {};
		\node (6a) [below of=5a] {};
		\node [hasse] (6b) [below of=5b] {};
		\node (6c) [below of=5c] {};
		\node [hasse] (7a) [below of=6a] {};
		\node [hasse] (7b) [below of=6b] {};
		\node (7c) [below of=6c] {}; 
		\node (8a) [below of=7a] {};
		\node [hasse] (8b) [below of=7b] {};
	        \node (8c) [below of=7c] {};
	          \node at (-0.4,-4.1)[]{$\vdots$};
		\draw (1a) edge [] node[label=above:\footnotesize{$e_{7}$}] {} (2b)
			(1a) edge [] node[label=left:\footnotesize{$e_{8}$}] {} (3a)
			(2b) edge [] node[label=right:\footnotesize{$d_{8}$}] {} (4b)
			(3a) edge [] node[label=above:\footnotesize{$a_1$}] {} (4b)
			(4b) edge [] node[label=right:\footnotesize{$d_{10}$}] {} (6b)
			(5a) edge [] node[label=above:\footnotesize{$a_{1}$}] {} (6b)
			(3a) edge [] node[label=left:\footnotesize{$d_{10}$}] {} (5a)
			(7a) edge [] node[label=above:\footnotesize{$a_{1}$}] {} (8b)
			(7b) edge [] node[label=right:\footnotesize{$d_{N_f}$}] {} (8b);
	\end{tikzpicture}
		\\ \hline
	\end{tabular}
	\caption{The component of $\mathcal{H}_\infty$ for  $1=|k|=N_c - \frac{N_f}{2}+2$.}
	\label{tab:region4k1}
\end{table}

\begin{table}
	\centering
\begin{tabular}{| m{2.3cm} |m{6.5cm}| m{2.7cm} |}
		\hline
		Phase & Quiver & Hasse diagram \\ \hline
		V$'$ ($N_f$ odd) &  $~~\node{}{1}-\node{}2-\dots-\node{}{N_f-2}-\node{\upnode{\frac{N_f-1}2}}{N_f-1}-\node{}{\frac{N_f+1}{2}}-\node{}{2}$ &
		\begin{tikzpicture}
		\tikzstyle{hasse} = [circle, fill,inner sep=2pt];
		\node [hasse] (1) [label=right:\footnotesize{$ $}] {};
		\node [hasse] (2) [below of=1] {};
		\node [hasse] (3) [below of=2] {};
		\node [hasse] (4) [below of=3] {};
		\node [hasse] (5) [below of=4] {};
		\node [hasse] (6) [below of=5] {};
		 \node at (0,-3.5)[]{$\vdots$};
		\draw (1) edge [] node[label=left:\footnotesize{$e_8$}] {} (2)
		(2) edge [] node[label=left:\footnotesize{$d_{10}$}] {} (3)
		(3) edge [] node[label=left:\footnotesize{$d_{12}$}] {} (4)
		(5) edge [] node[label=left:\footnotesize{$\;\;\;\;\;d_{N_f}$}] {} (6);
	         \end{tikzpicture}  	 
		\\ \hline
	\end{tabular}
	\caption{The component of $\mathcal{H}_\infty$ for  $\frac 1 2 = |k|=N_c - \frac{N_f}{2}+2$.}
	\label{tab:region4kHalf}
\end{table}
\clearpage

\bibliographystyle{JHEP}
\bibliography{bibli.bib}

\end{document}